\documentclass[times,final]{elsarticle}

\usepackage{jcomp}
\usepackage{framed,multirow}

\usepackage{amssymb}
\usepackage{latexsym}

\usepackage[fleqn]{amsmath}

\usepackage{url}
\usepackage{hyperref}
\usepackage[dvipsnames]{xcolor}
\usepackage{subfig}
\usepackage[normalem]{ulem}
\def\myclosedcircle{$\bullet$}
\def\mycircle{$\circ$}
\def\mystar{\raise 1pt\hbox{$\star$}}
\def\mytriangle{\raise 1pt\hbox{$\triangle$}}
\def\mytriangledown{\raise 0.1pt\hbox{$\bigtriangledown$}}
\def\mysquare{\raise 1pt\hbox{$\scriptstyle\Box$}}
\def\mydiamond{\raise 1pt\hbox{$\diamond$}}
\def\dashed{\bdash\bdash\bdash\nobreak}

\def\spacce#1{\hskip #1pt}
\def\drawline#1#2{\raise 2.5pt\vbox{\hrule width #1pt height #2pt}}
\def\solid{\drawline{24}{.5}\nobreak}

\def\bdash{\hbox{\drawline{5.8}{.5}\spacce{2}}}

\def\bdot{\hbox{\drawline{1}{.5}\spacce{2}}}

\def\dotted{\hbox{\leaders\bdot\hskip 24pt}\nobreak}

\definecolor{newcolor}{rgb}{.8,.349,.1}

\hypersetup{
  colorlinks=true,
  linkcolor=blue,
  citecolor=blue
}

\journal{Journal of Computational Physics}

\begin{document}

\verso{A. Lozano-Dur\'an and H.~J. Bae}

\begin{frontmatter}

\title{Error scaling of large-eddy simulation in the outer region of
wall-bounded turbulence}%

\author[1]{Adri\'an \snm{Lozano-Dur\'an}\corref{cor1}}
\cortext[cor1]{Corresponding author: 
  Tel.: +0-000-000-0000;  
  fax: +0-000-000-0000;}
\ead{adrianld@stanford.edu}
\author[2]{Hyunji Jane  \snm{Bae}}

\address[1]{Center for Turbulence Research, Stanford University, Stanford, California 94305, USA}
\address[2]{Graduate Aerospace Laboratories, California Institute of Technology, Pasadena, California, 91125}

\received{1 May 2018}
\finalform{10 May 2018}
\accepted{13 May 2018}
\availableonline{15 May 2018}

\begin{abstract}
We study the error scaling properties of large-eddy simulation (LES)
in the outer region of wall-bounded turbulence at moderately high
Reynolds numbers.  In order to avoid the additional complexity of
wall-modeling, we perform LES of turbulent channel flows in which the
no-slip condition at the wall is replaced by a Neumann condition
supplying the exact mean wall-stress.  The statistics investigated are
the mean velocity profile, turbulence intensities, and kinetic energy
spectra. The errors follow $(\Delta/L)^{\alpha}Re_\tau^{\gamma}$,
where $\Delta$ is the characteristic grid resolution, $Re_\tau$ is the
friction Reynolds number, and $L$ is the meaningful length-scale to
normalize $\Delta$ in order to collapse the errors across the
wall-normal distance. We show that $\Delta$ can be expressed as the
$L_2$-norm of the grid vector and that $L$ is well represented by the
ratio of the friction velocity and mean shear.  The exponent $\alpha$
is estimated from theoretical arguments for each statistical quantity
of interest and shown to roughly match the values computed by
numerical simulations. For the mean profile and kinetic energy
spectra, $\alpha\approx1$, whereas the turbulence intensities converge
at a slower rate $\alpha<1$.  The exponent $\gamma$ is approximately
$0$, i.e. the LES solution is independent of the Reynolds number. The
expected behavior of the turbulence intensities at high Reynolds
numbers is also derived and shown to agree with the classic log-layer
profiles for grid resolutions lying within the inertial range.
Further examination of the LES turbulence intensities and spectra
reveals that both quantities resemble their filtered counterparts from
direct numerical simulation (DNS) data, but that the mechanism
responsible for this similarity is related to the balance between the
input power and dissipation rather than to filtering.
\end{abstract}


\end{frontmatter}



\section{Introduction} 
%
%

Most turbulent flows cannot be calculated by DNS of the Navier-Stokes
equations because the range of scales of motions is so large that the
computational cost becomes prohibitive. In LES, only the large eddies
are resolved, and the effect of the small scales on the larger scales
is modeled through an SGS model. The approach enables a reduction of
the computational cost by several orders of magnitude while still
capturing the statistical quantities of interest.  However, the
solutions provided by most LES approaches are grid-dependent, and
multiple computations are required in order to faithfully assess the
quality of the LES results. This brings the fundamental question of
what is the expected LES error as a function of Reynolds number and
grid resolution.
The necessity of assessing the impact of grid resolution on both the
accuracy and convergence properties of SGS models and flow statistics
has been highlighted in the NASA Vision 2030 \citep{Slotnick2014} as a
pacing item for computational fluid mechanics. The issue was also
remarked by \citet{Pope2004} as a central problem concerning the
foundations of LES.  Therefore, LES should not be framed as the result
of one single solution, but instead as a convergence study using
multiple grid resolutions. It is then pertinent to determine the grid
requirements in order to deem LES as a cost-saving approach compared
to DNS.  In the present work, we analyze the LES error scaling of the
mean velocity profile, turbulence intensities, and energy spectra in
the outer region of wall-bounded flows without the influence of the
wall.

The equations for LES are formally derived by applying a low-pass
filter to the Navier--Stokes equations \citep{Leonard1975}.  The
common procedure is then to solve these filtered equations together
with a model for the SGS stresses, but no explicit filter form is
usually specified. Instead, the discrete differentiation operators and
limited grid resolution used to compute the LES solution are assumed
to act as an effective implicit filter
\citep{Lund2003,Carati2001,Bae_brief_2017,Bae_brief_2018}. The
approach, usually referred to as implicitly-filtered LES, yields a
velocity field that is considered representative of the actual
filtered velocity with filter size proportional to the grid resolution
\citep{Lund2003,Silvis2016}.  This lack of explicit filtering is
responsible for the aforementioned intimate relation between the grid
resolution and the LES equations \citep{Bose2014b}. Grid convergence
is only guaranteed in the limit of DNS-like resolution, and the LES
predictions may be sensitive in an intricate manner to the grid size
above such a limit.  This is a distinctive feature of
implicitly-filtered LES which entails important difficulties for
evaluating the quality of the solutions.

%

First studies aiming to assess the accuracy of SGS models include the
pioneering investigation by \citet{Clark1979}, who established the
numerical study of decaying isotropic turbulence as a reference
benchmark, although the grid resolutions and Reynolds numbers tested
were highly constrained by the computational resources of the time.
Since then, common benchmarks for LES have broadened to include simple
hydrodynamic cases such as forced or decaying isotropic turbulence
\citep{Metais1992}, rotating homogeneous turbulence
\citep{Kobayashi2001}, spatial or temporal mixing layers
\citep{Vreman1996, Vreman1997} and plane turbulent channel flow
\citep{Piomelli1988,Germano1991,Chung2010}, among others. See
\cite{Bonnet1998} for an overview of cases for LES validation.

The analysis of discretization errors in LES by
\citet{Ghosal1996,Kravchenko1997} and \citet{Chow2003} revealed that
the magnitude of the numerical errors can be comparable to those from
SGS modeling.  Recent developments in modeling and numerical error
quantification in isotropic turbulence by \citet{Meyers2003} also
showed that the partial cancellation of both sources can lead to
coincidentally accurate results. Along the same line,
\citet{Meyers2007} studied the combined effect of discretization and
model errors, and a further series of works resulted in the
error-landscape-methodology framework reviewed by \citet{Meyers2011},
where it is stressed that the determination of the quality of LES
based on one single metric alone may produce misleading results.  The
performance of SGS models in the presence of walls is even more
erratic. \citet{Meyers2007b} investigated the grid convergence
behavior of channel flow DNS at resolutions typically encountered in
SGS model testing.  They observed a non-monotonic convergence of the
skin friction and turbulence intensities with grid-refinements,
suggesting that the robustness of SGS models should be tested for a
range of Reynolds numbers and resolutions in order to avoid incidental
coincidences with DNS results. At much higher Reynolds numbers,
\citet{Sullivan2011} examined the numerical convergence of LES in
time-dependent weakly sheared planetary boundary layers. They assessed
the convergence of the second-order statistics, energy spectra, and
entertainment statistics, and concluded that LES solutions are
grid-independent provided that there is adequate separation between
the energy-containing eddies and those near the filter cut-off scale.
\citet{Stevens2014} showed the ability of LES to reproduce accurately
second and higher-order velocity moments for grid resolutions fine
enough to resolve 99\% of the LES kinetic energy. The convergence of
SGS models in complex geometries has been explored in a lesser degree,
but some noteworthy efforts are the pulsatile impinging jet in
turbulent cross-flow by \citet{Toda2014} and the full plane
calculations using the NASA Common Research Model by
\citet{Lehmkuhl2016}.


A central matter among the convergence studies above is the search for
the most meaningful flow quantity to collapse the LES errors when the
grid size, Reynolds number, and model parameters are systematically
varied. \citet{Geurts2002} characterized the simulation errors in
terms of the subgrid-activity, defined as the the relative
subgrid-model dissipation rate with respect to the total dissipation
rate. \citet{Klein2005} studied the accuracy of single-grid estimators
for the unresolved turbulent kinetic energy to assess the quality of
LES, and evaluated the sensitivity of the LES results on the modeling
and numerical errors. Similarly, \citet{Freitag2006} presented a
method to evaluate error contributions by assuming that the numerical
and modeling errors scale as a power of the grid spacing and filter
width, respectively.  Other indices to estimate the quality of the LES
solution are the fraction of the total turbulent kinetic energy in the
resolved motions \cite{Pope2004}, the relative grid size with respect
to Kolmogorov or Taylor scales, or the effective eddy viscosity
compared to the molecular viscosity \cite{Celik2009}.  Alternative and
more sophisticated metrics are still emerging, for instance, the
Lyapunov exponent measurement proposed by \citet{Nastac2017} for
assessing the dynamic content and predictability of LES among others,
but there is a lack of consensus regarding which should be the most
meaningful metric to quantify errors in a general set-up, if any.


In the present work, we study the error scaling of SGS models based on
the eddy viscosity assumption in the outer region of wall-bounded
turbulence at moderately high Reynolds numbers. Our goal is to
characterize the errors as a function of grid resolution and Reynolds
number, and to find the physical length-scale dictating the relative
size of the grid that is relevant for error quantification.  For that
purpose, we perform a theoretical estimation of the error scaling for
the mean velocity profile, turbulence intensities, and kinetic energy
spectra. Our results are numerically corroborated by LES of turbulent
channel flows using a wall model that acts as a surrogate of the
near-wall dynamics by supplying the exact mean wall-stress.
This numerical set-up is motivated by previous DNS
  and LES studies showing that the features of the outer flow are
  reproduced with reasonably good fidelity even if the near-wall layer
  is poorly represented
  \citep{Piomelli2002,Flores2006,Lee2013,Mizuno2013,Chung2014}.
Finally, it is important to remark that turbulent free shear flows
such as mixing layers, jets, and wakes are also tenable candidates for
studying shear-dominated flows away from walls. However, their large
scales are dynamically different to the large scale motions of
turbulent boundary layer flows typically relevant for external
aerodynamics, which is the focus of the present study.

The manuscript is organized as follows. In Section
  \ref{sec:necessity}, we illustrate the challenge of assessing the
  performance of SGS models in the outer layer of wall-bounded
  turbulence. We discuss the methodology and numerical setup to
assess the convergence of SGS models in Section \ref{sec:benchmark}.
The results for the errors in the mean velocity profile are presented
in Section \ref{sec:mean}, for the turbulence intensities in Section
\ref{sec:fluctuations}, and for the energy spectra in Section
\ref{sec:spectra}.  Finally, we conclude in Section
\ref{sec:conclusion}.

\section{The challenge of quantifying the performance of SGS models in the outer region of wall turbulence}
\label{sec:necessity}

\begin{figure}
\begin{center}
 \vspace{0.2cm}
 \includegraphics[width=0.85\textwidth]{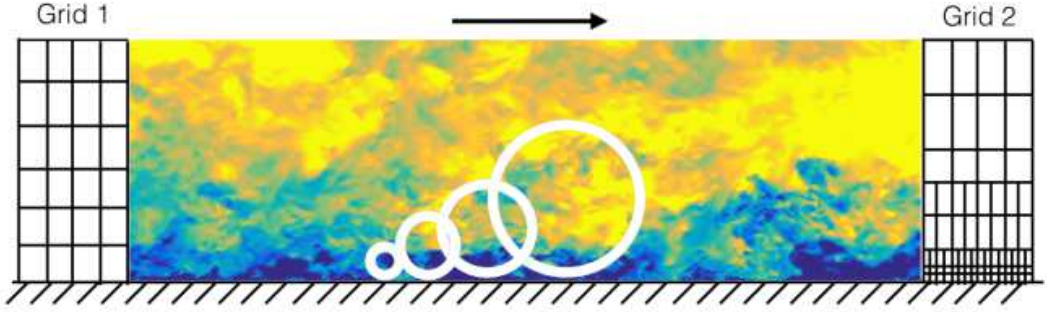}
\end{center}
\caption{Instantaneous streamwise velocity in a turbulent channel flow
  at $Re_\tau \approx 4200$ and sketch of wall-attached eddies of
  different sizes (white circles). Colors range from blue (low
  velocity) to yellow (high velocity). Grid 1 (left) depicts a uniform
  grid typical of WMLES. Grid 2 (right) represents a nested grid
  necessary for proper WRLES.
\label{fig:nested} }
\end{figure}

Most SGS models assume that a considerable fraction of the turbulent
kinetic energy (i.e., 80-90\% \cite{Pope2004}) is resolved by the
grid, and the Reynolds numbers and grid resolutions must comply with
this requirement in order to faithfully assess the performance of the
models.  In unbounded flows, such as isotropic turbulence, LES can be
performed at relatively coarse grid resolutions while still meeting
this condition. On the contrary, the scenario is not as favorable for
wall-bounded flows as discussed below.  The number of grid points $N$
to compute a turbulent boundary layer of thickness $\delta$ spanning a
wall-parallel area of $L_1 \times L_3$ is
\begin{equation}
N = \int_0^{L_1} \int_0^\delta \int_0^{L_3} \frac{\mathrm{d}x_1
  \mathrm{d}x_2 \mathrm{d}x_3}{\Delta_1 \Delta_2 \Delta_3},
\end{equation}
where $x_1$, $x_2$ and $x_3$ are the streamwise, wall-normal, and
spanwise directions, respectively, and $\Delta_1$, $\Delta_2$, and
$\Delta_3$ are the target grid resolutions in each direction which in
general are a function of space. The required number of grid points
can be expressed as $N \sim Re^\zeta$, where the exponent $\zeta$
depends on the sizes of the eddies expected to be accurately
represented by the grid. Estimations of $\zeta$ can be found in
\citet{Chapman1979} and \citet{Choi2012}. DNS aims to capture eddies
in the dissipative range, and hence $\Delta_i \sim \eta$ and $\zeta
\approx 2.6$, where $\eta$ is the Kolmogorov length-scale. To resolve
the energy-containing eddies as in traditional LES (also referred to
as wall-resolved LES or WRLES), $\Delta_i$ should scale as the
integral length scale, $\Delta_i \sim L_\varepsilon$, which yields
$\zeta \approx 1.9$. In the logarithmic region (log layer) of
wall-bounded flows, $L_\varepsilon$ grows linearly with $x_2$ and the
energy-containing eddies have sizes proportional to the distance to
the wall \citep{Jimenez2012,Larsson2016}. Consequently, the LES grid
must be accordingly reduced in all the spatial directions to resolve a
constant fraction of the turbulent kinetic energy, increasing the
computational cost. WRLES can be properly performed through nested
grids \citep{Sullivan1996,Kravchenko1999} such as the one depicted in
Fig.  \ref{fig:nested}. Otherwise, the near-wall grid resolution does
not suffice to capture the energy-containing eddies, and most SGS
models perform poorly \citep{Jimenez2000}.  Finally, if we target to
model only the outer flow motions as in wall-modeled LES (WMLES), the
grid requirements are such that $\Delta_i \sim \delta$, and $\zeta \le
1$ depending on the wall model approach.
%

Although WRLES has been practiced for a long time, actual WRLES is
scarce due to the complexity of its implementation and its associated
computational cost. In typical WRLES studies, only the wall-normal
resolution is properly refined according to the size of the
energy-containing eddies, while the wall-parallel directions remain
under-resolved. Most of the grid convergence studies in wall-bounded
LES mentioned in the introduction fall within this category.  The
consequence is that the majority of previous validation works are
performed at relatively low Reynolds numbers to make the calculations
computationally affordable \citep{Choi2012} and to avoid the errors of
under-resolving the wall-parallel directions. Under these conditions,
it is questionable whether SGS models are active enough to adequately
measure their performance in the outer layer of wall turbulence.

\begin{figure}
\begin{center}
 \vspace{0.1cm}
 \includegraphics[width=0.43\textwidth]{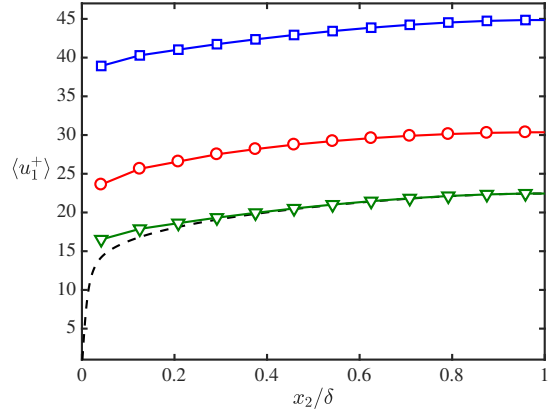}
 \end{center}
\caption{ Mean streamwise velocity profile, $\langle u_1^+ \rangle$,
  for a turbulent channel flow as a function of the wall-normal
  distance, $x_2$, scaled by the channel half-height, $\delta$. Lines
  and symbols are: (\textcolor{blue}{\mysquare}), no explicit SGS
  model with no-slip boundary condition at the wall (case NM950-NS);
  (\textcolor{red}{\mycircle}), dynamic Smagorinsky model with no-slip
  boundary condition at the wall (case DSM950-NS);
  (\textcolor{OliveGreen}{\mytriangledown}), no explicit SGS model
  with equilibrium wall model from \citet{Kawai2012} (case
  NM950-EQWM); (\dashed), DNS.  All cases are at friction Reynolds
  number $Re_\tau \approx 950$. More details are provided in Table
  \ref{table:cases_intro}. \label{fig:Umean} }
\end{figure}

To illustrate the low contribution of SGS models far from the wall and
their poor performance in the near-wall region, Fig. \ref{fig:Umean}
shows the mean streamwise velocity profile, $\langle \tilde{u}_1
\rangle$, for an LES turbulent channel flow as a function of the
wall-normal distance.  The details of the simulations are discussed in
Section \ref{sec:numerics} (see Table \ref{table:cases_intro}), but
for now, it is only important to remark that all cases were computed
using identical grids (with 13 points per boundary layer thickness)
and friction Reynolds number, $Re_\tau\approx950$.  Coarse DNS (no SGS
model and no wall model) provides the worst prediction (squares in
Fig. \ref{fig:Umean}). Ideally, a perfect SGS model would supply the
missing stresses at all distances from the wall. Indeed,
Fig. \ref{fig:Umean} shows that the solution improves by introducing
the dynamic Smagorinsky model (circles); however, the performance is
still poor and $\langle \tilde{u}_1 \rangle$ is far from the reference
DNS velocity profile.  In contrast, the agreement with DNS in the
outer layer ($x_2\gtrsim 0.2\delta$) is excellent when the equilibrium
wall model from \citet{Kawai2012} is employed (triangles), despite the
fact that there is no explicit SGS model in this case.
%

Note that for all cases, the shape of $\langle \tilde{u}_1 \rangle$
far from the wall is close to the DNS solution and barely affected by
the presence or lack thereof of an SGS model. The main source of error
comes from the inaccurate prediction of the wall friction velocity,
$u_\tau$, which translates into the overprediction of $\langle
\tilde{u}_1^+ \rangle = \langle \tilde{u}_1 \rangle/u_\tau$. This
suggests that the application of traditional SGS models alone is not
sufficient to provide the correct stress at the wall, problem that is
attenuated by means of a wall model. The result highlights the
importance of wall-modeling, but also shows that the validation of SGS
models in the outer layer of wall turbulence at low Reynolds numbers
or very fine grid resolutions could be meaningless due to the low
activity of the models themselves in this regime. On the contrary,
accurate quantification of SGS model errors could be achieved by
performing true WRLES using three-dimensional grid refinement as Grid
2 in Fig. \ref{fig:nested}.  However, we have mentioned above that the
latter is not a common practice, and most attempts at WRLES suffer
from the limitation demonstrated in Fig. \ref{fig:Umean}. Therefore,
many of the mismatches in the mean velocity profile between DNS and
LES reported in the literature are probably dominated by errors
accumulated near the wall. This calls for new numerical benchmarks
aiming to isolate LES errors in the outer flow from the region closest
to the wall. In the present work we propose a new benchmark to
overcome this limitation.

The behavior of SGS models close to the wall has been improved in
recent works such as the constrained LES by \citet{Chen2012}, the
integral length-scale approximation model by \citet{Rouhi2016}, and
the explicit algebraic model by \citet{Rasam2011}, among others. These
approaches reduce substantially the grid requirements by modifying the
SGS model near the wall while maintaining the no-slip boundary
condition.  Nevertheless, in the present work we focus on the error
analysis within the outer flow far from the wall, and the
aforementioned SGS models are not considered.\\

\section{Benchmark for the outer region of wall-bounded turbulence} 
\label{sec:benchmark}

\subsection{Exact mean wall-stress turbulent channel flows} 
\label{sec:exact_stress}

We consider a plane turbulent channel flow with periodic boundary
conditions in the streamwise and spanwise directions. The
incompressible LES equations are obtained by applying a spatial filter
to the Navier--Stokes equations,
\begin{eqnarray}\label{eq:NS}
\frac{\partial \bar u_i}{\partial t} + \frac{\partial \bar u_i \bar u_j  }{\partial x_j}
+ \frac{\partial \tau_{ij}  }{\partial x_j}  = - \frac{1}{\rho}\frac{\partial \bar p}{\partial x_i} +
\nu \frac{\partial^2 \bar u_i}{\partial x_j\partial x_j}, 
\quad\frac{\partial \bar u_i}{\partial x_i}  = 0,
\end{eqnarray}
where $\bar u_i$ for $i=1,2,3$ are the streamwise, wall-normal and
spanwise filtered velocities, respectively, $\bar p$ is the filtered
pressure, $\tau_{ij} = \overline{u_i u_j} - \bar u_i \bar u_j$ is the
effect of the sub-filter scales on the resolved eddies, $\rho$ is the
flow density, and $\nu$ is the kinematic viscosity. The streamwise,
wall-normal and spanwise spatial directions are $x_i$ for $i=1,2,3$,
respectively, and the walls are located at $x_2=0\delta$ and
$x_2=2\delta$.  The objective of LES modeling is to approximate
$\tau_{ij}$ via the SGS tensor $\tau^{\mathrm{SGS}}_{ij}$. To
emphasize that an LES model is not exact, the resolved LES velocity is
denoted by $\tilde u_i$ and we expect that $\tilde u_i \approx \bar
u_i$ for an accurate SGS model.

We have discussed in Section \ref{sec:necessity} the necessity of
benchmarks for wall-bounded turbulence that are independent of the
strict near-wall resolution requirements. To attain this goal, the
no-slip boundary condition at the wall is replaced by a constant
wall-stress condition imposed through a Neumann boundary condition of
the form
\begin{equation}
\label{eq:Neumann}
\left.\frac{\partial \tilde{u}_1}{\partial n}\right|_w = \frac{\tau_w -
\left.\tau^{\mathrm{SGS}}_{12}\right|_w}{\nu},
\end{equation}
where $w$ denotes quantities evaluated at the wall, $n$ is the
wall-normal direction oriented towards the interior of the channel,
and $\tau_w$ is the mean wall stress known \emph{a priori} from
DNS. Equation (\ref{eq:Neumann}) can be thought of as a wall-model
supplying the exact mean wall stress.  Equation (\ref{eq:Neumann}) is
used here in the context of LES, but a similar shear-stress boundary
condition was used by \citet{Chung2014} to study Townsend's
outer-layer similarity hypothesis in DNS.

The set-up above is not intended to capture the near-wall dynamics,
and the small eddies close to the wall are prone to be misrepresented
when Eq. (\ref{eq:NS}) is discretized for coarse grid
resolutions. However, our focus is on the outer flow along the range
$0.2\delta<x_2<\delta$ \citep{Pope2000}, and previous studies have
revealed that the flow statistics and structure of this region are
relatively independently of the particular configuration of the eddies
closest to the wall, even if they are partially or completely
under-resolved.  Some examples are the roughness experiments in
channels and boundary layers \cite{Perry1977, Jimenez2004, Bakken2005,
  Flores2006, Flores2007}, and the idealized numerical studies by
\citet{Flores2006,Mizuno2013,Chung2014} and \citet{Lozano2018b}, among
others. In all these cases, the near-wall region was seriously
modified or directly bypassed, but the properties of the outer layer
remained essentially unaltered. This is also the case for WMLES, where
it has been shown that imposing the correct mean wall-stress is
sufficient to predict one-point statistics accurately \citep{Lee2013},
consistent with the approach in Eq. (\ref{eq:Neumann}). Therefore, the
correct representation of the outer layer dynamics remains uncoupled
from the inner layer structure, supporting the numerical experiment
presented here as a valid framework to assess LES errors far from the
wall.  The independence of the outer flow with respect to the
near-wall dynamics together with the existence of inner-outer scale
separation are the main assumptions of the current numerical
set-up. Nonetheless, it has been reported in previous works that some
flow configurations, such as boundary layers subjected to strong
spanwise wall-stress variations \citep{Chung2018}, may invalidate
these assumptions.

\subsection{Numerical experiments} 
\label{sec:numerics}

We perform a set of LES of plane turbulent channels driven by a
constant mass flow in the streamwise direction.  The simulations are
computed with a staggered, second-order, finite difference
\citep{Orlandi2000} and fractional-step method \citep{Kim1985} with a
third-order Runge-Kutta time-advancing scheme \citep{Wray1990}. The
code has been validated in previous studies in turbulent channel flows
\citep{Lozano2016_Brief,Bae2018} and flat-plate boundary layers
\citep{Lozano2018a}.  Periodic boundary conditions are imposed in the
streamwise and spanwise directions, while for the top and bottom walls
we use either the no-slip (NS) boundary condition or exact-wall-stress
(EWS) Neumann boundary condition from Eq. (\ref{eq:Neumann}).

Two SGS models are investigated: dynamic Smagorinsky model (DSM)
\citep{Germano1991,Lilly1992} and anisotropic minimum dissipation
(AMD) model \citep{Rozema2015}, which are regarded as representative
eddy-viscosity models with and without test filtering, respectively.
We also consider cases without an explicit SGS model (NM) where the
numerical truncation errors act as an implicit SGS model. Some of our
results have also been computed for the Vreman model
\citep{Vreman2004} (see \ref{appendixA}) whose performance was found
to be similar to AMD.

The size of the computational domain is $8\pi \delta\times
2\delta\times 3\pi \delta$ in the streamwise, wall-normal, and
spanwise directions, respectively. The grid resolutions are denoted by
$\Delta_1, \Delta_2,$ and $\Delta_3$ for each spatial direction, and
they range from $0.025\delta$ to $0.2\delta$, which correspond to 5 to
40 points per boundary layer thickness. The present grids are in
accordance with the typical grid resolutions encountered in WMLES of
real external aerodynamic applications, and follow the recommendations
by \citet{Chapman1979} for resolving the large eddies in the outer
region of wall turbulence. Four different friction Reynolds numbers
are considered, $Re_\tau = u_\tau \delta/\nu \approx 950,2000,4200$
and $8000$, where $u_\tau$ is the friction velocity at the wall. The
LES results are compared with reference DNS data from
\citet{Hoyas2006,Lozano-Duran2014}, and \citet{Yamamoto2018}.  All the
LES channel flow were run at least for $100\delta/u_\tau$ after
transients.

The list of cases used in Section \ref{sec:necessity} is given in
Table \ref{table:cases_intro}.  The simulations discussed for the
remainder of the paper are named following the convention [SGS
  model][$Re_\tau$]-[boundary condition]-[grid resolution], where the
grid resolutions are denoted by i1, i2, i3 and i4 for isotropic grids,
and by a1, a2 and a3 for anisotropic grids. The different grid
resolutions are provided in Table \ref{table:resolutions}. For
example, DSM4200-EWS-i2 is an LES channel flow with DSM at $Re_\tau
\approx 4200$, EWS boundary condition, and grid resolutions $\Delta_1
= \Delta_2 = \Delta_3 = 0.1\delta$.
%
\begin{table}
\begin{center}
\setlength{\tabcolsep}{8pt}
\begin{tabular}{l c c c c c c}
Case         & SGS model  & Wall condition & $Re_\tau$ & $\Delta_1/\delta$ & $\Delta_2/\delta$ & $\Delta_3/\delta$  \\ 
\hline
\hline
NM950-NS     & NM & NS   & \multirow{3}{*}{$932$}  &
\multirow{3}{*}{0.10}    & 
\multirow{3}{*}{0.080}    & \multirow{3}{*}{0.050}  \\
DSM950-NS    & DSM & NS   &  &  &  &   \\ 
NM950-EQWM   & NM & EQWM &  &  &  &  \\ 
\hline
\end{tabular}
\end{center}
\caption{\label{table:cases_intro} List of cases used in Section
  \ref{sec:necessity}. The second column contains the SGS model: no
  explicit SGS model (NM) or dynamic Smagorinsky model (DSM). The
  third column refers to the wall boundary condition: no-slip (NS) or
  Neumann boundary condition using the equilibrium wall-model from
  \citet{Kawai2012} (EQWM).  The fourth column indicates the friction
  Reynolds number.  $\Delta_1$, $\Delta_2$, and $\Delta_3$ are the
  streamwise, wall-normal, and spanwise grid resolutions
  respectively. }
\end{table}
%
\begin{table}
\begin{center}
\setlength{\tabcolsep}{8pt}
\begin{tabular}{l c c c}
Grid resolution label & $\Delta_1/\delta$ & $\Delta_2/\delta$ & $\Delta_3/\delta$ \\ 
\hline
\hline 
i1     & 0.20    & 0.20    & 0.20   \\ 
i2     & 0.10    & 0.10     & 0.10  \\ 
i3     & 0.050   & 0.050    & 0.050 \\ 
i4     & 0.025   & 0.025    & 0.025 \\ 
\hline 
a1     & 0.20   & 0.10   & 0.05     \\ 
a2     & 0.10   & 0.10    & 0.07    \\ 
a3     & 0.20   & 0.10    & 0.10    \\ 
\hline
\end{tabular}
\end{center}
\caption{\label{table:resolutions} Tabulated list of resolutions as a
  fraction of the channel half-height, $\delta$. The first column
  contains the label used for naming the LES cases computed with
  different grids. $\Delta_1$, $\Delta_2$, and $\Delta_3$ are the
  streamwise, wall-normal, and spanwise grid resolutions,
  respectively.}
\end{table}


\section{Error scaling of the mean velocity profile} 
\label{sec:mean}

We examine first the mean velocity as it is the figure of merit for
most LES studies.  We assume that $\langle u_1 \rangle \approx \langle
\bar u_1 \rangle$, where $\langle\cdot\rangle$ denotes average in
homogeneous directions and time, and the LES mean velocity is directly
compared with unfiltered DNS data. The approximation is reasonable for
quantities dominated by large-scale contributions, as it is the case
for $\langle u_1 \rangle$.  The error for the mean velocity profile is
systematically quantified as the average difference of the LES and DNS
solutions in the outer region as
\begin{equation} \label{eq:error} 
\mathcal{E}_m = \left[ \frac{ \int_{0.2\delta}^{\delta} \left(\langle
\tilde{u}_1\rangle - \langle u_1\rangle\right)^2 \mathrm{d}x_2} {
\int_{0.2\delta}^{\delta} \langle u_1\rangle^2 \mathrm{d}x_2 }
\right]^{1/2},
\end{equation}
where $\langle \tilde{u}_1 \rangle$ is obtained from LES, and $\langle
u_1 \rangle$ is evaluated from DNS data. This choice excludes the
nonphysical/under-resolved range $x_2<0.2\delta$ for the LES cases
using the exact-wall-stress approach as discussed in Section
\ref{sec:exact_stress}. For a channel flow driven by constant mass
flux, \textcolor{black}{$Q$,} and exact mean wall-stress, some
reference errors can be obtained from two extreme cases, i.e., a fully
turbulent profile defined by the flat velocity $\langle \tilde u_1
\rangle = Q/2\delta$, and the laminar solution represented by the
parabolic function $\langle \tilde u_1 \rangle = 3Q/4\delta
(2-x_2/\delta)x_2/\delta$, with errors equal to
$\mathcal{E}_{m,turb}\approx 0.06$ and $\mathcal{E}_{m,lam} \approx
0.26$, respectively, at $Re_\tau\approx 4200$.

In general, the error depends on the grid resolution and Reynolds
number,
\begin{equation}
\mathcal{E}_m = \mathcal{E}_m(\Delta_1,\Delta_2,\Delta_3, Re_\tau).
\end{equation}
If we further assume that $\mathcal{E}_m \sim \Delta^{\alpha_m}
Re_\tau^{\gamma_m}$, where $\Delta$ is a (yet to be defined) measure
of the grid size, the exponents $\alpha_m$ and $\gamma_m$ can be
theoretically estimated from the error equation and empirically
computed from numerical experiments. Both analysis are performed
below. Ultimately, we will conclude that LES is a viable approach for
computing the outer flow of wall-bounded flows if the empirical values
of the exponents are such that $\alpha_m>0$ and $\gamma_m \approx 0$.

\subsection{Theoretical estimations} 
\label{subsec:mean:theoretical}

We estimate the expected error behavior of $\mathcal{E}_m$ which serves
as a reference for the numerical results in the next section.  If we
assume that the scaling of the integrand in Eq. (\ref{eq:error}) with
$\Delta_i$ is roughly the same within $x_2/\delta\in[0.2, 1]$,
$\mathcal{E}_m$ can be approximated by
\begin{equation}\label{eq:error_theory_0}
\mathcal{E}_m \sim \langle \bar u_1 \rangle - \langle u_1 \rangle,
\end{equation} 
where the denominator of Eq. (\ref{eq:error}) is discarded as it is
only a normalization factor which does not depend on $\Delta_i$. Let
us consider the streamwise momentum equation 
\begin{equation}\label{eq:du1dt}
\frac{\partial u_1}{\partial t} = -\frac{\partial u_1 u_j}{\partial x_j}
-\frac{\partial p'}{\partial x_1} + \frac{u_\tau^2}{\rho\delta},
\end{equation}
where the viscous terms are neglected and the pressure gradient is
decomposed into mean and fluctuating contributions
$u_\tau^2/(\rho\delta)$ and $\partial p'/\partial x_1$,
respectively. In the following, we denote fluctuating quantities by
$(\cdot)'$. After multiplying Eq. (\ref{eq:du1dt}) by $u_1$ and
averaging in the homogeneous directions and time, the resulting
equation is
\begin{equation}\label{eq:u1_du1dt}
-\left \langle u_1\frac{\partial u_1 u_j}{\partial x_j} \right \rangle
-\left \langle u_1' \frac{\partial p'}{\partial x_1} \right \rangle 
+ \frac{u_\tau^2}{\rho\delta} \left \langle u_1 \right \rangle = 0.
\end{equation}
A similar equation can be obtained for the filtered streamwise
velocity, and after subtraction and manipulation of both equations we
obtain
\begin{equation}\label{eq:u1_du1dt_sub}
\langle \bar{u}'_1 \bar{u}'_2 \rangle \frac{\partial \langle \bar{u}_1 \rangle }{\partial x_2} 
-\langle u'_1 u'_2 \rangle \frac{\partial \langle u_1 \rangle }{\partial x_2} 
=
\frac{1}{2} \frac{\partial}{\partial x_2} 
\left\langle 
u'_1 u'_1  u'_2 -
\bar{u}'_1  \bar{u}'_1 \bar u'_2 -2 \tau_{12} \bar{u}_1
\right\rangle
+ \left\langle u'_1 \frac{\partial p' }{\partial x_1} 
- \bar u'_1 \frac{\partial \bar p'}{\partial x_1} \right\rangle.
\end{equation}
For a symmetric filter with well-defined, non-zero, second moment in
real space, the terms in right-hand side of (\ref{eq:u1_du1dt_sub}) can be
expressed as \cite{Yeo1988,Winckelmans2001}
\begin{eqnarray}
\bar{u}'_1  \bar{u}'_1 \bar u'_2 - u'_1 u'_1  u'_2
&=&
\frac{ \bar \Delta_i^2 }{2}
\left( \bar u'_1 \bar u'_1 \frac{\partial^2 \bar u'_2}{\partial x_i^2} 
+ 2 u'_1 u'_2 \frac{\partial^2 \bar u'_1}{\partial x_i^2} \right )
+ \mathcal{O}(\bar \Delta_1^4 + \bar \Delta_2^4 + \bar \Delta_3^4), \label{eq:uv_approx} \\
\bar u'_1 \frac{\partial \bar p'}{\partial x_1} -
u'_1 \frac{\partial p' }{\partial x_1}
&=&
\frac{ \bar \Delta_i^2 }{2}
\left( \bar u'_1 \frac{\partial^3 \bar p'}{\partial x_1 \partial x_i^2} 
+ \frac{\partial \bar{p}'}{\partial x_1} \frac{\partial^2 \bar u'_1}{\partial x_i^2} \right )
+ \mathcal{O}(\bar \Delta_1^4 + \bar \Delta_2^4 + \bar \Delta_3^4), \label{eq:p_approx} \\
\tau_{1j} &=&
 \bar \Delta_i^2  \frac{\partial \bar u_1}{\partial x_i} \frac{\partial \bar u_j}{\partial x_i} 
+ \mathcal{O}(\bar \Delta_1^4 + \bar \Delta_2^4 + \bar \Delta_3^4), \label{eq:SGS_exp}
\end{eqnarray}
where repeated indices imply summation, and $\bar \Delta_i$ signifies
the filter size in the $i$-th direction defined as the square root of
the second moment of the filter operator
\begin{equation}
\bar \Delta_i^2 = 
\int_{-\infty}^{\infty}\int_{-\infty}^{\infty}\int_{-\infty}^{\infty} x_i^2 
H(x_1,x_2,x_3) \mathrm{d}x_1\mathrm{d}x_2\mathrm{d}x_3,
\end{equation}
with $H$ the filter kernel.  In general, $\bar \Delta_i^2 = \bar c
\Delta_i^2$, where $\bar c$ is a coefficient that depends on the
particular filter shape, e.g., for a box filter $\bar c = 1/12$.
Eqs. (\ref{eq:uv_approx}), (\ref{eq:p_approx}), and (\ref{eq:SGS_exp})
are valid for filter kernels with Fourier transform of class
$C^\infty$, which is the case for most filters defined in real space
such as the Gaussian filter, tophat filters, and all discrete filters
\citep{Carati2001}.

For the rest of the discussion, we neglect terms of the order of
$\mathcal{O}(\bar \Delta_i^4)$ and assume that traditional SGS
models are a fourth order approximation to $\tau_{ij}$. The
simplification is useful for estimating the error scaling
independently of any particular SGS model. We further assume that, far
from the wall, the wall-normal derivative of $\bar{u}_1$ in
Eq. (\ref{eq:SGS_exp}) is well approximated using the fluctuating
velocity, $\partial \bar{u}_1 / \partial x_2 = \partial \langle
\bar{u}_1 \rangle / \partial x_2 + \partial \bar{u}'_2 / \partial x_2
\approx \partial \bar{u}'_1 / \partial x_2$, since in the log layer
the gradients can be estimated as $\partial \langle \bar{u}_1 \rangle
/ \partial x_2 \approx u_\tau /(\kappa x_2) \ll u_\tau/\Delta_2 \sim
\partial \bar{u}'_1 / \partial x_2$, where $\kappa$ is the von
K\'arm\'an constant.

Introducing Eqs. (\ref{eq:uv_approx}), (\ref{eq:p_approx}), and
(\ref{eq:SGS_exp}) into (\ref{eq:u1_du1dt_sub}), invoking the
simplifications above, and considering the exact mean wall-stress
assumption from Section \ref{sec:exact_stress}, the error in the mean
velocity profile at $x_2=x^o_2$ can be shown to scale as
\begin{equation}\label{eq:error_theory}
\mathcal{E}_m \sim 
\bar \Delta_i^2 \left\langle 
\frac{\bar u'_1 \bar u'_1}{4} \frac{\partial^2 \bar u'_2}{\partial x_i^2}
+ u'_1 u'_2 \frac{\partial^2 \bar u'_1}{\partial x_i^2} +
\frac{\partial \bar u'_1}{\partial x_i} \frac{\partial \bar u'_2}{\partial x_i} u'_1
+
\int_{0}^{x^o_2}
\left(
\frac{\bar u'_1}{2} \frac{\partial^3 \bar p'}{\partial x_1 \partial x_i^2}
+ \frac{1}{2}\frac{\partial \bar{p}'}{\partial x_1} \frac{\partial^2 \bar u'_1}{\partial x_i^2}
- \frac{\partial \bar u'_1}{\partial x_i} \frac{\partial \bar u'_j}{\partial x_i} 
  \frac{\partial \bar u'_1}{\partial x_j}
\right)
\mathrm{d}x_2
\right\rangle.
\end{equation}
Note that the filter sizes in Eq. (\ref{eq:error_theory}) are arranged
in the form $ \bar \Delta_i^2 \sim \Delta_i^2$, which motivates the
use of the $L_2$-norm of $(\Delta_1,\Delta_2,\Delta_3)$ as the
characteristic grid-size, $\Delta$$\sim$$\sqrt{\Delta_1^2 + \Delta_2^2
  + \Delta_3^2}$, as long as the error is measured according to
Eq. (\ref{eq:error}). Eq. (\ref{eq:error_theory}) also shows that
$\Delta_i$ does not provide a full description of the error, and that
a complete characterization would involve an effective grid size such
as
\begin{equation}
\Delta_{\mathrm{eff}} = \sqrt{ d_1\Delta_1^2 + d_2\Delta_2^2 + d_3\Delta_3^2 },
\end{equation}
where $d_k$, $k=1,2,3$, are complicated flow-dependent functions from
Eq.  (\ref{eq:error_theory}).

Equation (\ref{eq:error_theory}) can be further exploited to determine
the scaling of $\mathcal{E}_m$ with $\Delta$ by assuming $p' \sim
u'^2$ and approximating the dependence of $\bar u'_i$,
$\frac{\partial^2 \bar u'_j}{ \partial x_i \partial x_i}$ and
$\frac{\partial \bar{u}'_j}{\partial x_i}$ on $\Delta$. A rough
estimation is performed by assuming that the kinetic energy spectrum
follows $E_{K} \sim k^\beta$, with the wavenumber $k \sim 1/\Delta$,
and the isotropic velocity gradient $G=\partial u /\partial x$ at
scale $\Delta$ such that
\begin{equation}
\bar u'_{1,2}
\frac{\partial^2 \bar u'_{2,1}}{ \partial x_i \partial x_i} , 
\frac{\partial \bar{u}'_1}{\partial x_i} \frac{\partial \bar{u}'_2}{\partial x_i}
\sim G^2  \sim \frac{u^2}{\Delta^2} \sim \frac{kE_K}{\Delta^2}
 \sim \Delta^{-(\beta+3)},
\end{equation}
where the exponent $\beta$ depends on the regime the SGS models
operates: for the shear-dominated range $\beta=-1$ \citep{Perry1977}
and $G^2 \sim \Delta^{-2}$, whereas for the inertial range $\beta =
-5/3$ \citep{Kolmogorov1941} and $G^2 \sim \Delta^{-4/3}$.  Taking
into account the scaling above, the expected error in the LES mean
velocity profile from Eq. (\ref{eq:error_theory}) scales as
\begin{equation}\label{eq:E_scaling}
\mathcal{E}^s_m\sim \Delta^0, \quad \mathcal{E}^i_m\sim \Delta,
\end{equation}
for SGS models acting on the shear-dominated ($\mathcal{E}^s_m$) or
inertial ($\mathcal{E}^i_m$) regimes, respectively.
\textcolor{black}{The results from Eq. (\ref{eq:E_scaling}) indicate
  that no improvement in the error is expected for grid resolutions
  comparable to the scales in the shear-dominated region, whereas an
  approximately linear improvement can be anticipated for finer grids
  with sizes comparable to the scales in the inertial range. This
  suggest that capturing the energy injection mechanism from the mean
  shear is critical to achieve accurate LES results.}
\textcolor{black}{The estimations from Eq. (\ref{eq:E_scaling}) assume
  that $\Delta$ lies completely either in the inertial range or in the
  shear-dominated regime. However, the error defined by
  Eq. (\ref{eq:error}) accounts for a wide range of wall-normal
  distances in which $\Delta$ may change from one regime to the
  other. In such a case, $\mathcal{E}_m$ is expected to exhibit an
  intermediate scaling between $\mathcal{E}^i_m$ and
  $\mathcal{E}^s_m$. An $x_2$-dependent formulation of the error is
  presented in Section \ref{subsec:mean:length}.}

The scaling of $\mathcal{E}_m$ with $\Delta$ in Eq.
(\ref{eq:E_scaling}) can be estimated from simpler dimensional
arguments without going through Eq. (\ref{eq:error_theory}), but it
was beneficial to write the explicit equation of the error to obtain
additional information about its functional form. Additionally, it is
important to remark that the results from Eq. (\ref{eq:E_scaling})
should be understood as rough estimations since actual errors evolve
according to a non-linear equation and, hence, their rigorous
mathematical treatment is highly elusive. This consideration is also
applicable to the error estimations for the turbulence intensities and
energy spectra in later sections.


\subsection{Numerical assessment} 
\label{subsec:mean:numerical}

\begin{figure}[t]
\begin{center}
 \subfloat[]{\includegraphics[width=0.40\textwidth]{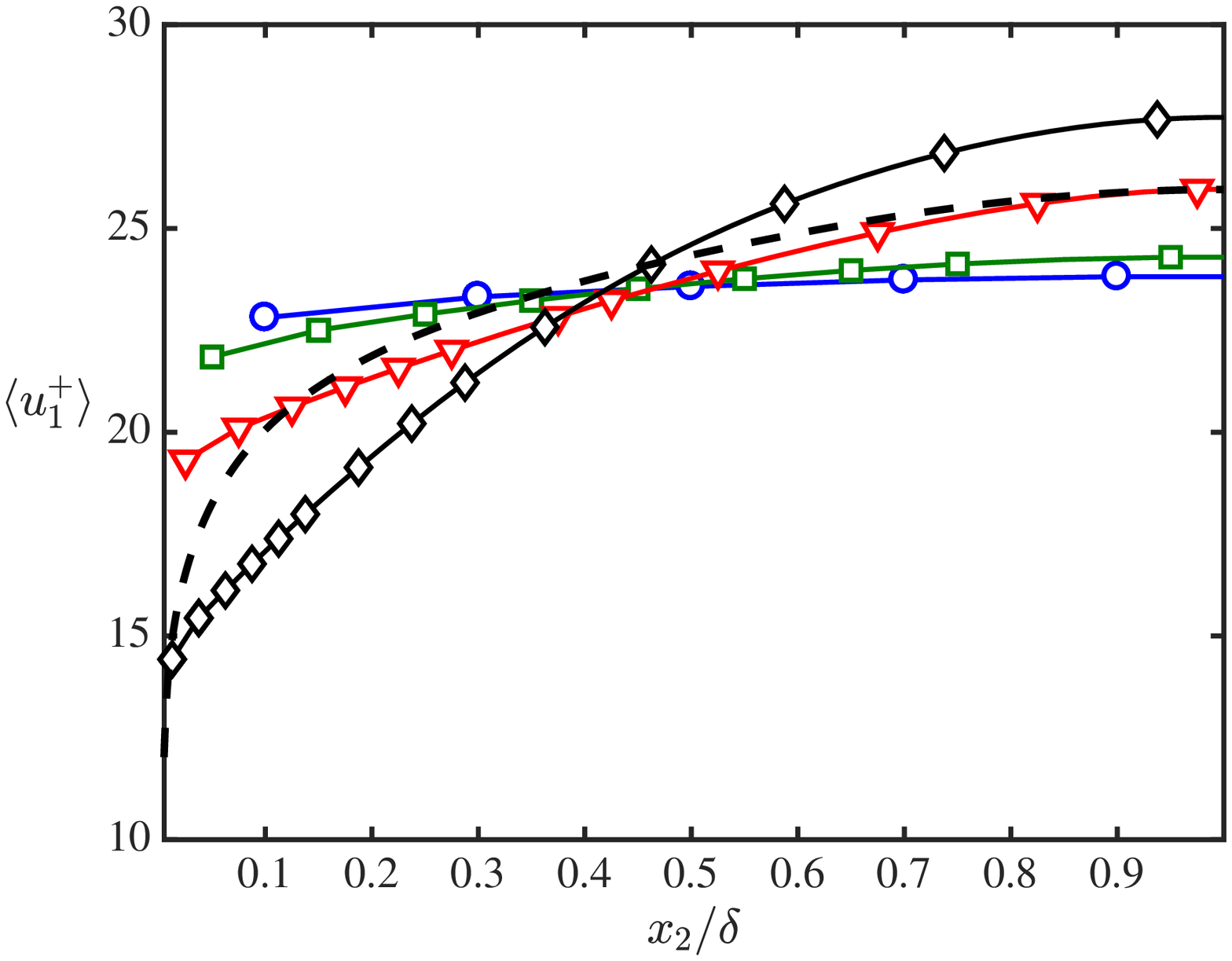}}
 \hspace{0.1cm}
 \subfloat[]{\includegraphics[width=0.40\textwidth]{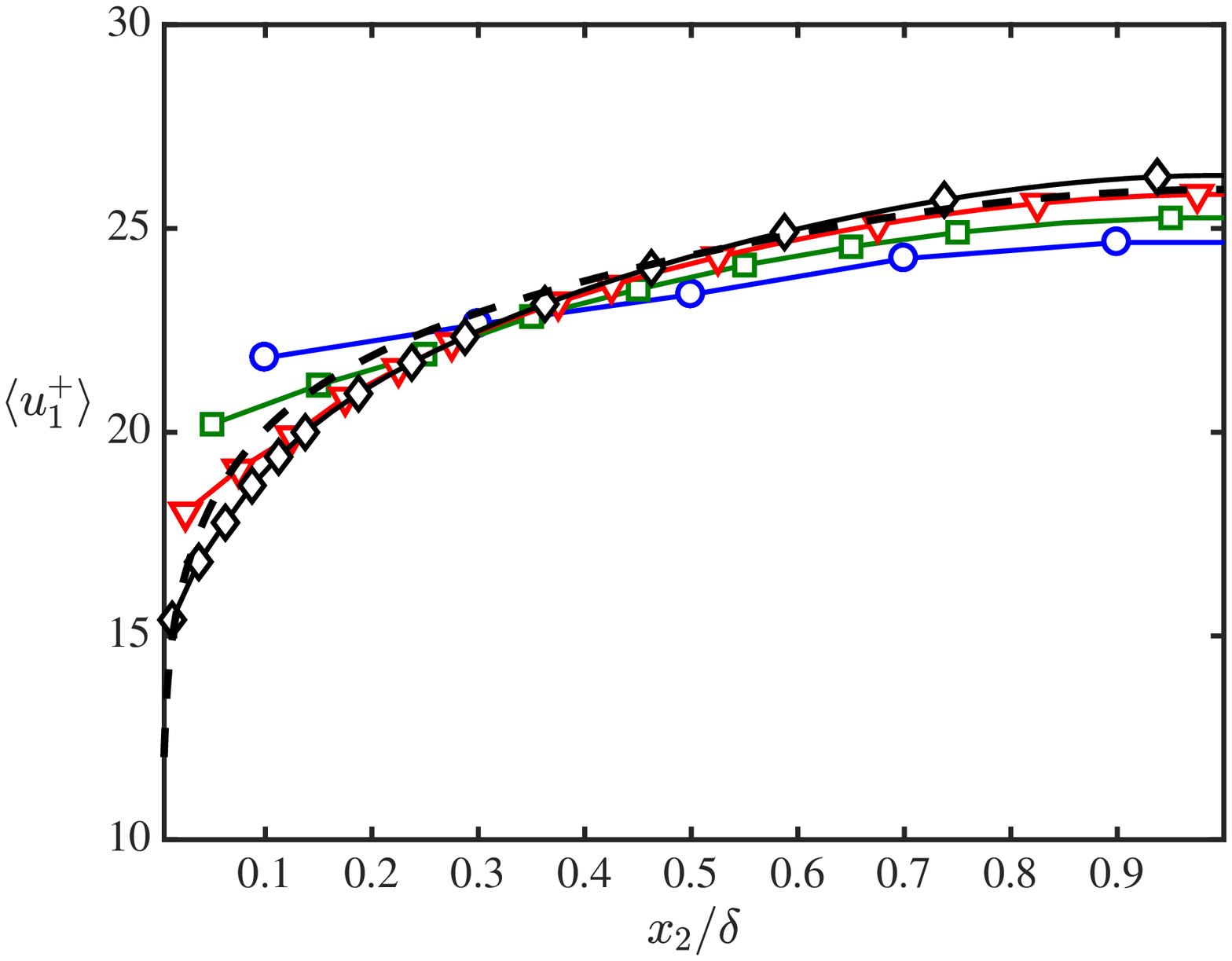}}
\end{center}
\caption{Mean streamwise velocity profile at $Re_\tau\approx 4200$ for
  (a) no explicit SGS model, and (b) DSM. Symbols are for grids i1
  (\textcolor{blue}{\mycircle}), i2
  (\textcolor{OliveGreen}{\mysquare}), i3
  (\textcolor{red}{\mytriangledown}), and i4 (\mydiamond) from Table
  \ref{table:resolutions}. The dashed line is
  DNS.\label{fig:Umean_examples} }
\end{figure}

Fig. \ref{fig:Umean_examples} shows the mean velocity profiles for a
selection of cases at $Re_\tau\approx4200$ and different grid
resolutions without SGS model (Fig.  \ref{fig:Umean_examples}a) and
with DSM (Fig.  \ref{fig:Umean_examples}b). As expected, $\langle
\tilde{u}_1\rangle$ converges to $\langle u_1\rangle$ as the grid is
refined for cases with DSM (equivalently for AMD), while the trend is
inconsistent for cases without explicit SGS model.
%

\begin{figure}[t]
\begin{center}
 \subfloat[]{\includegraphics[width=0.40\textwidth]{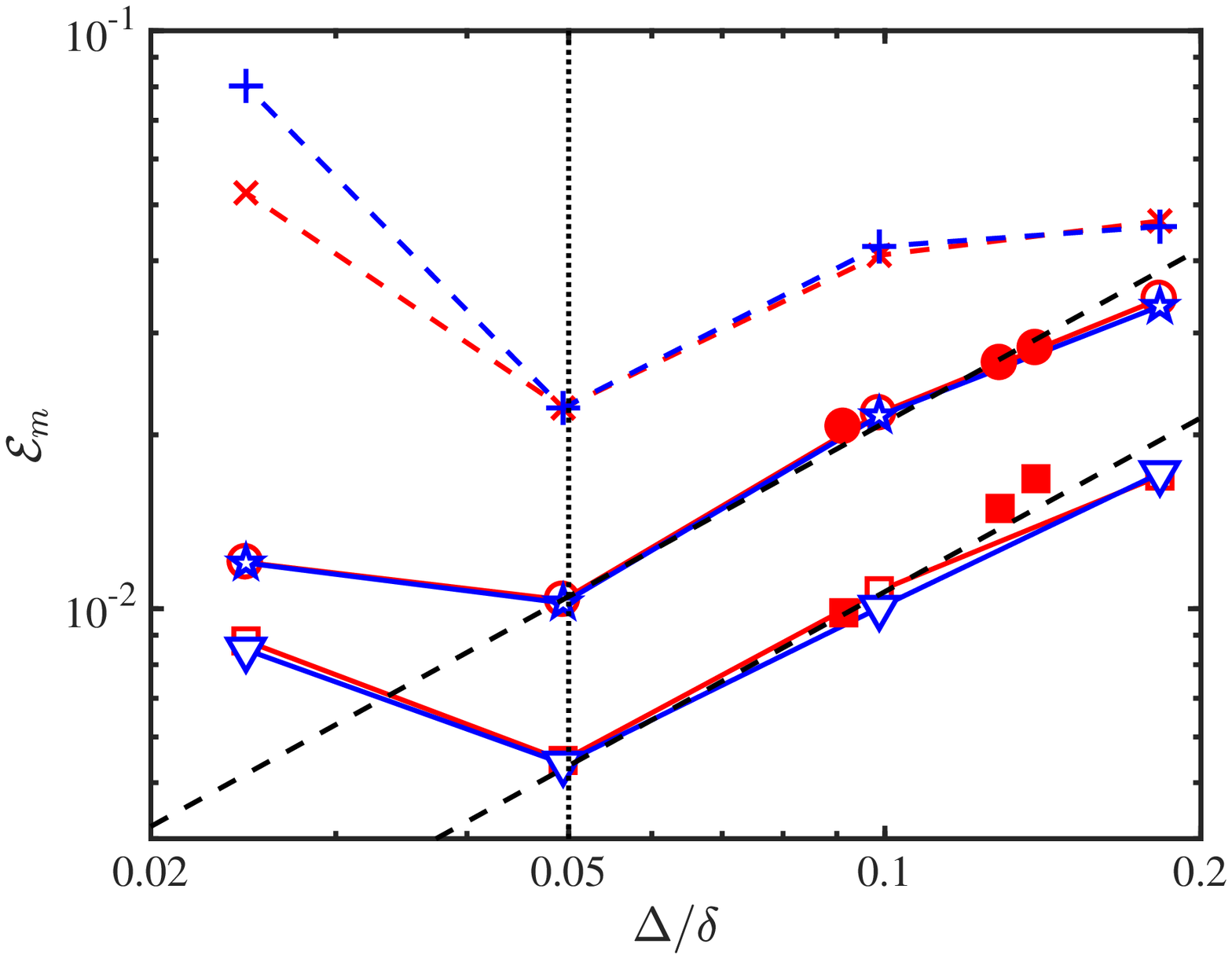}}
 \hspace{0.2cm}
 \subfloat[]{\includegraphics[width=0.40\textwidth]{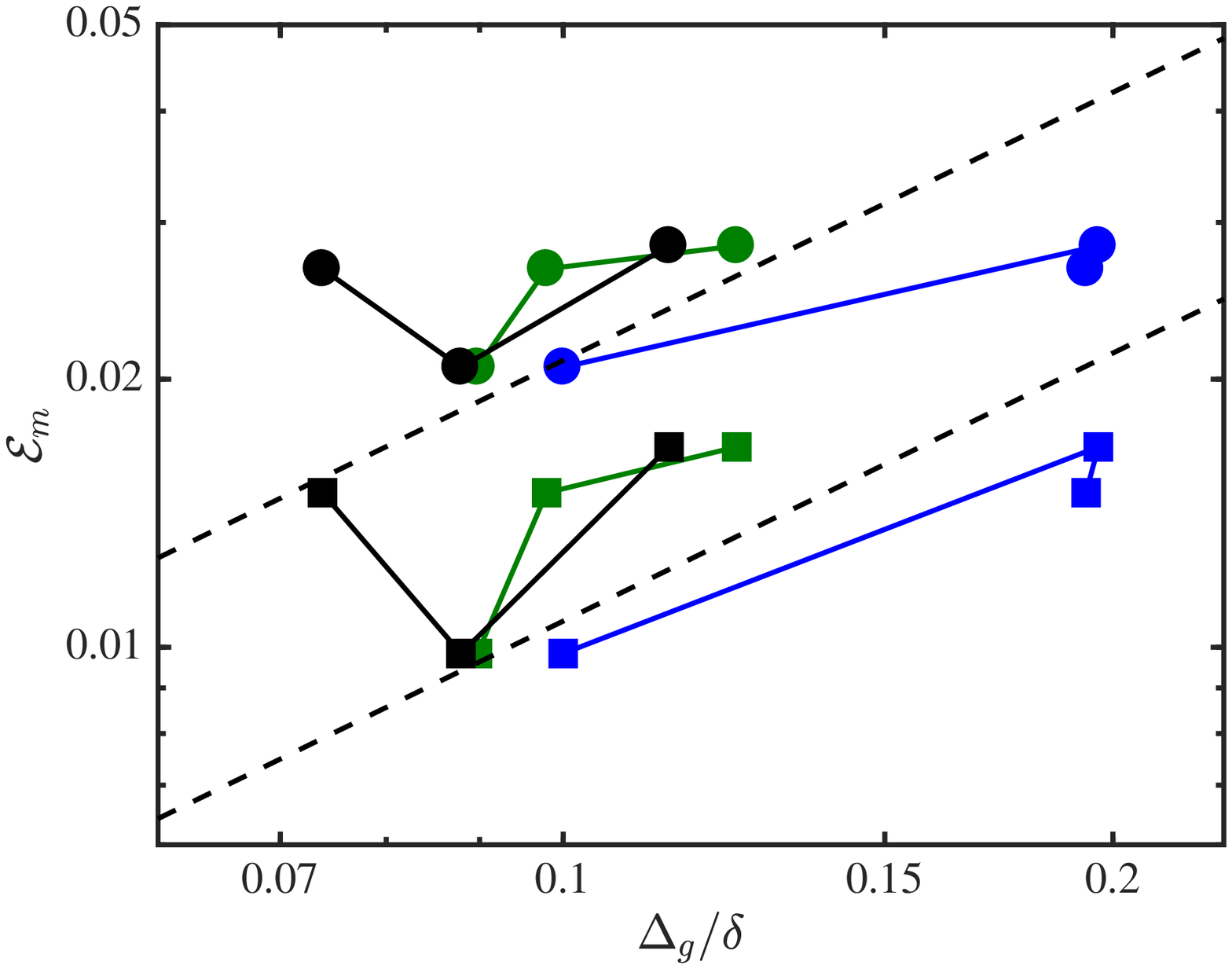}}
\end{center}
\caption{ (a) Error in the mean velocity profile as a function of the
  characteristic grid resolution $\Delta = \sqrt{(\Delta_1^2 +
    \Delta_2^2 + \Delta_3^2)/3}$. Colors are \textcolor{red}{red}, for
  cases at $Re_\tau \approx 4200$, and \textcolor{blue}{blue} for
  cases at $Re_\tau \approx 8000$. Symbols are (\mycircle) and
  (\mystar) for DSM, (\mytriangledown) and (\mysquare) for AMD,
  ($\times$) and ($+$) for no explicit SGS model. Open and closed
  symbols are for isotropic and anisotropic grids,
  respectively. Dashed lines are $\mathcal{E}_m = 0.107 \Delta/\delta$
  and $\mathcal{E}_m = 0.210 \Delta/\delta$, and the dash-dotted line
  is $\Delta/\delta = 0.05$. (b) Error in the mean velocity profile as
  a function of alternative characteristic grid sizes: $\Delta_g =
  \sqrt[3]{\Delta_1\Delta_2\Delta_3}$ (\textcolor{OliveGreen}{green}),
  $\Delta_g = \mathrm{max}(\Delta_1,\Delta_2,\Delta_3)$
  (\textcolor{blue}{blue}), and $\Delta_g =
  \sqrt{3/\left(1/\Delta_1^2+1/\Delta_2^2+1/\Delta_3^2\right)}$
  (black). The results are for cases with anisotropic grids at
  $Re_\tau \approx 4200$ with DSM (\mycircle), and AMD
  (\mysquare). For reference, panel (b) also includes the dashed lines
  from panel (a). \label{fig:Umean_error_exactWM} }
\end{figure}

The quantitative assessment of the $\mathcal{E}_m$ is shown in Fig.
\ref{fig:Umean_error_exactWM}(a) as a function of the characteristic
grid resolution $\Delta$, taken to be
\begin{equation}\label{eq:grid}
\Delta = \sqrt{\frac{\Delta_1^2 + \Delta_2^2 + \Delta_3^2}{3}},
\end{equation}
as motivated by Eq. (\ref{eq:error_theory}). Other grid definitions
were also inspected such as the cube root of the cell volume
\citep{Deardorff1970,Scotti1993}, the maximum of the grid sizes
\citep{Spalart1997}, or the square root of the harmonic mean of the
squares of the grid sizes (all reported in Fig.
\ref{fig:Umean_error_exactWM}b), among others. However, the best
collapse is found for the definition in Eq. (\ref{eq:grid}),
consistent with the discussion in Section
\ref{subsec:mean:theoretical}.  A survey of existing subgrid
length-scales can be found in \citet{Trias2017} but note that in the
current study we are discussing the most meaningful grid size to
characterize $\mathcal{E}_m$, which does not need to coincide with the
characteristic length-scale embedded in SGS models (i.e., $\tilde
\Delta$ in the Smagorinsky model $-2C_s \tilde \Delta^2 \sqrt{2\tilde
  S_{nm} \tilde S_{nm}}\tilde S_{ij}$, where $\tilde S_{ij}$ is the
resolved rate-of-strain tensor and $C_s$ is a constant).
%

\begin{figure}
\begin{center}
\includegraphics[width=1\textwidth]{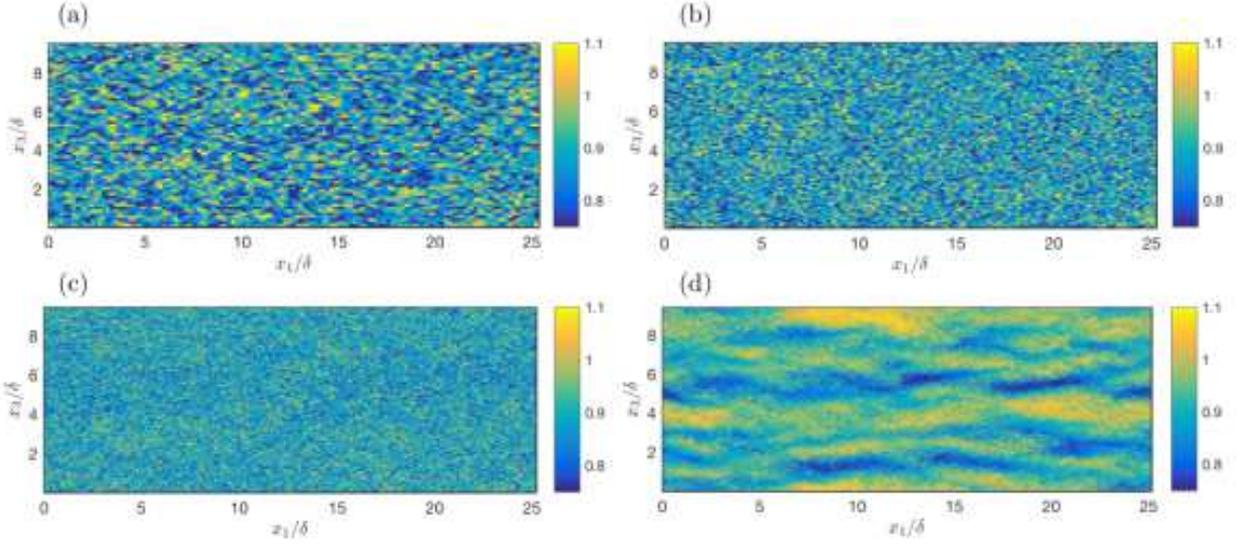}
\end{center}
\caption{ Instantaneous streamwise velocity in a wall-parallel plane
  at $x_2\approx 0.5 \delta$ for (a) NM4200-EWS-i1, (b) NM4200-EWS-i2,
  (c) NM4200-EWS-i3, (d) NM4200-EWS-i4. \label{fig:snapshots_panel} }
\end{figure}

For cases without SGS model, the errors are discernibly larger than
those calculated with DSM or AMD, especially for the finer grid
resolutions, and similar to those for fully turbulent flows
($\mathcal{E}_{m,turb}\approx 0.06$). Moreover, they follow a
non-monotonic behavior with $\Delta$, inconsistent with the
second-order prediction from the linear analysis of the spatial
discretization errors. This is expected, as the linear analysis holds
for $\Delta \rightarrow 0$, but it is no longer representative of
errors subjected to non-linear diffusion and convection for $\Delta
\sim \delta$. Visual inspection of the instantaneous streamwise
velocity fields for cases without SGS model in Fig.
\ref{fig:snapshots_panel} shows that there is a substantial change in
the flow topology at $\Delta \approx 0.05\delta$. For $\Delta >
0.05\delta$, the velocity field lacks the characteristic turbulence
features and exhibits instead a highly disorganized structure
(Figs. \ref{fig:snapshots_panel}a--c). On the other hand, clearly
defined streamwise velocity streaks emerge for $\Delta < 0.05\delta$
(Fig. \ref{fig:snapshots_panel}d). We can argue that these streaks are
nonphysical in the sense that they worsen the mean velocity profile
prediction as shown in Fig. \ref{fig:Umean_examples}(a) for
NM4200-EWS-i4.

For cases with SGS model and $\Delta > 0.05\delta$, the error follows
\begin{equation}\label{eq:error_ews}
\mathcal{E}_m \approx \epsilon \left( \frac{\Delta}{\delta} \right) Re_\tau^0,
\end{equation}
where $\epsilon$ is a model dependent constant. \textcolor{black}{Note
  that Eq. (\ref{eq:error_ews}) is obtained from the numerical
  evaluation of Eq. (\ref{eq:error}) using LES data, while
  Eq. (\ref{eq:E_scaling}) is the scaling analysis of
  Eq. (\ref{eq:error}) from theoretical considerations. The results
  show that the LES solution converges to the correct value free from
  viscous effects, $\mathcal{E}_m \sim Re_\tau^0$ (given a perfect
  wall model for the mean) as demanded by a proper LES far from the
  walls.  Our results also suggest that $\mathcal{E}_m \sim \Delta$,
  which agrees with the theoretical estimation of $\mathcal{E}^i_m$,
  i.e the expected error scaling when $\Delta$ is of the order of the
  length-scales in the inertial range}. Although both DSM and AMD
converge at the same rate with $\Delta$, the prefactor $\epsilon$ can
play an important role in the error magnitude and thus different
models may be preferred due to their lower $\epsilon$. The results in
\ref{appendixA} show that similar conclusions are drawn for the Vreman
model.

For $\Delta < 0.05\delta$, the errors depart from $\mathcal{E}_m \sim
\Delta$. This is probably a complicated non-linear effect which
involves the interplay between the numerical scheme and the flow
physics. Indeed, the observations from Fig. \ref{fig:snapshots_panel}
suggest that there is a competing effect of the improved prediction by
the SGS model versus the formation of nonphysical flow structures due
to discretization errors.  It is shown in Section
\ref{subsec:spectra:grid} that the grid resolution to resolve 90\% of
the turbulent kinetic energy is $\Delta^{\mathrm{min}}
\approx0.04\delta$ at $x_2 \approx 0.5\delta$ (same as in
Fig. \ref{fig:snapshots_panel}), that is very close to $\Delta \approx
0.05\delta$ for which the anomalous behavior of $\mathcal{E}_m$
appears. This transitional resolution $\Delta^{\mathrm{min}}$ is
related to the ability of the LES solution to support streaks without
SGS model, and we can hypothesize that it should have an impact on the
behavior of $\mathcal{E}_m$, even in the presence of an SGS
model. Additional tests included in \ref{appendixA} show that the
trend $\mathcal{E}_m \sim \Delta$ is recovered again for finer
grids. Although not inspected here, the convergence of $\langle \tilde
u_1 \rangle$ towards the DNS solution at even finer grids may entail
an intricate non-monotonic response as reported in
\citet{Meyers2007b}.

\subsection{Alternative metrics for error quantification} 
\label{subsec:mean:alternative}

\begin{figure}
\begin{center}
 \subfloat[]{\includegraphics[width=0.40\textwidth]{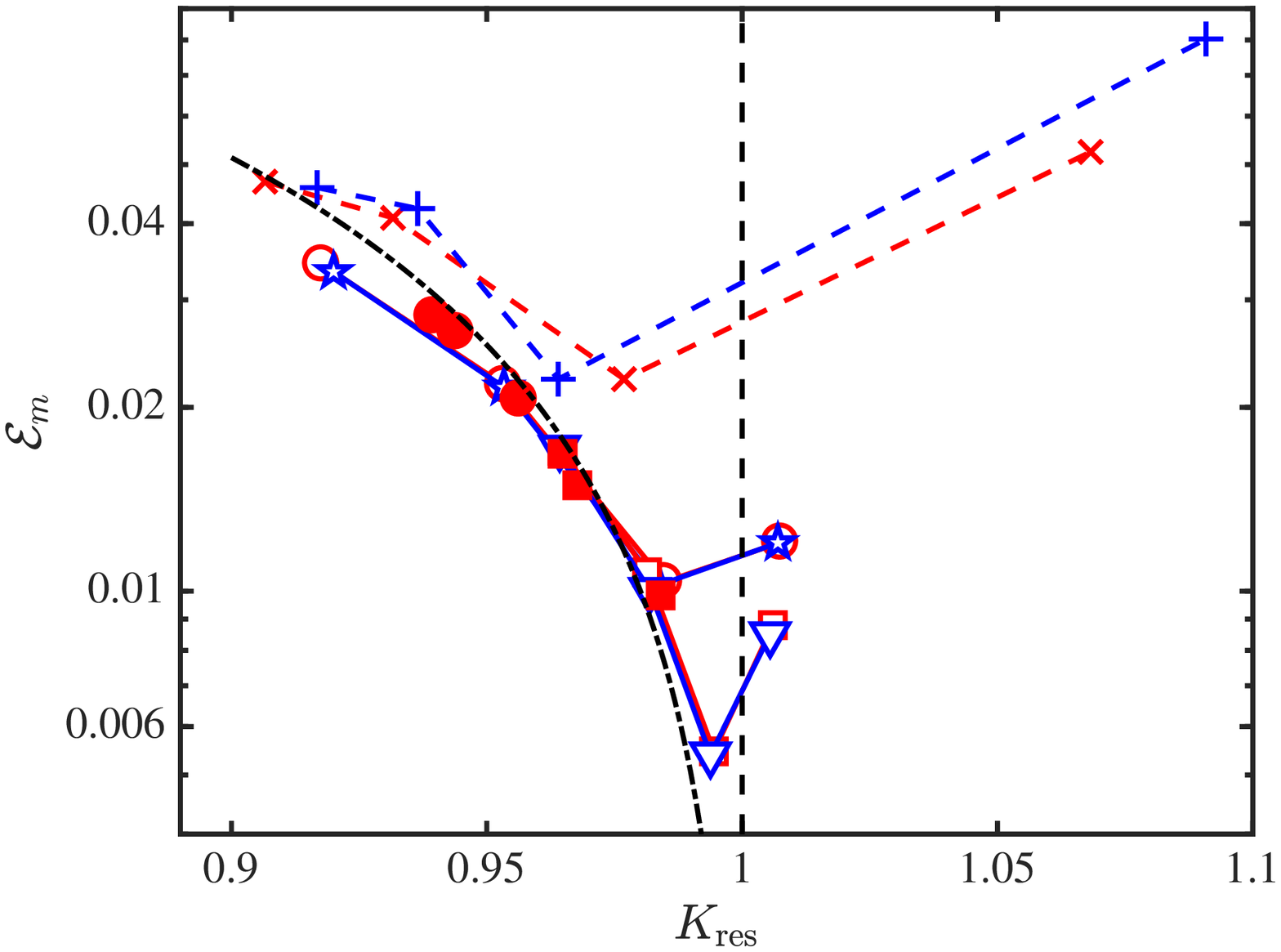}}
 \hspace{0.2cm}
 \subfloat[]{\includegraphics[width=0.40\textwidth]{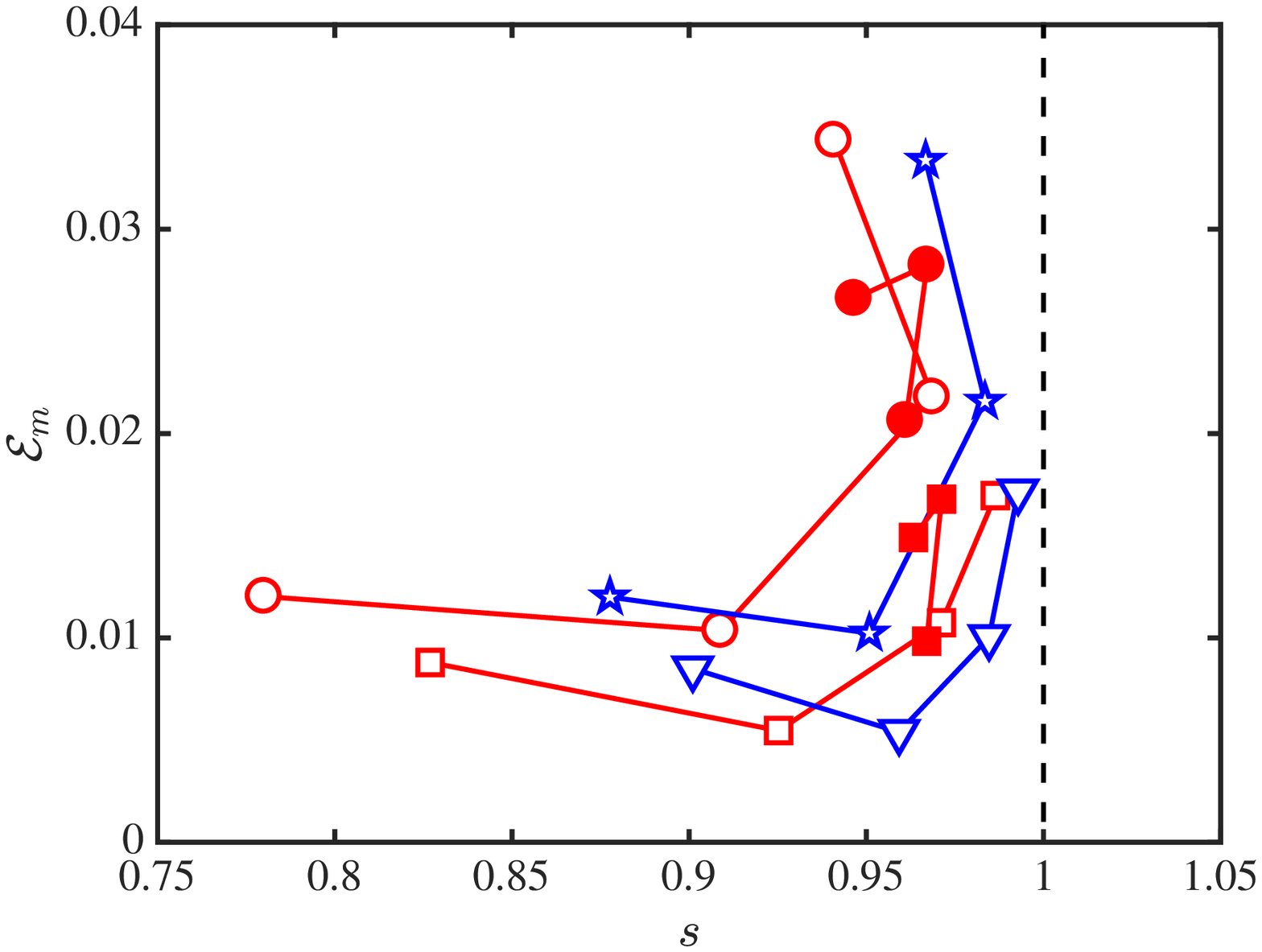}}
\end{center}
\caption{ Error in the mean streamwise velocity profile as a function
  of (a) the resolved total kinetic energy $K_{\mathrm{res}}$, and (b)
  SGS activity parameter $s$. Colors are \textcolor{red}{red} for
  cases at $Re_\tau \approx 4200$, \textcolor{blue}{blue} for cases at
  $Re_\tau \approx 8000$. Symbols are (\mycircle) and (\mystar) for
  DSM, (\mytriangledown) and (\mysquare) for AMD, ($\times$) and ($+$)
  for no explicit SGS model. Open and closed symbols are for isotropic
  and anisotropic grids, respectively. The dash-dotted line in (a) is
  $K_{\mathrm{res}} \approx (1 - \mathcal{E}_m)^2$. The vertical
  dashed lines are $K_{\mathrm{\mathrm{res}}}=1$ in (a) and $s=1$ in
  (b). \label{fig:Umean_error_exactWM_others} }
\end{figure}

Alternative metrics to functionally quantify $\mathcal{E}_m$ are the
resolved total kinetic energy,
\begin{equation}
K_{\mathrm{res}} = \frac{\langle \tilde u_i \tilde u_i \rangle}{\langle u_i u_i \rangle},
\end{equation}
and the SGS activity parameter \citep{Geurts2002,Meyers2003},
\begin{equation}\label{eq:s}
s = \frac{ \langle 2\nu_t\tilde S_{ij} \tilde S_{ij} \rangle}
{ \langle 2\nu_t\tilde S_{ij} \tilde S_{ij} + 2\nu\tilde S_{ij} \tilde S_{ij} \rangle},
\end{equation}
where $\nu_t$ is the eddy viscosity. The results are shown in
Fig. \ref{fig:Umean_error_exactWM_others} for $K_{\mathrm{res}}$ and
$s$ averaged over the wall-normal range $[0.3\delta,\delta]$.  Despite
the coarse grid resolutions investigated in the present work, the
resolved kinetic energy remains above 90\% for all cases (Fig.
\ref{fig:Umean_error_exactWM_others}a) and emerges as an effective
metric to assess the errors in the mean profile even among different
SGS models.  The result is not surprising since $K_{\mathrm{res}}$ can
be easily related to $\mathcal{E}_m$ if we assume that $\langle u_1^2
\rangle / \langle u_i^2 \rangle \gg 1$ for $i=2,3$, and $\langle u_1^2
\rangle \approx \langle u_1 \rangle^2$. The former are usually
$\sim$100, while the last condition is reasonably well satisfied if
$u_1$ follows a normal distribution $\mathcal{N}(\mu,\sigma)$ with
mean $\mu$ and standard deviation $\sigma$ such that $\mu/\sigma \gg
1$, which is a fair approximation in high-Reynolds-number turbulent
channel flows. Under those conditions, the resolved kinetic energy can
be expressed as
\begin{equation}
K_{\mathrm{res}} \approx (1 - \mathcal{E}_m)^2,
\end{equation}
which shows an excellent agreement with the data in
Fig. \ref{fig:Umean_error_exactWM_others}(a). Therefore,
$\mathcal{E}_m$ and $K_{\mathrm{res}}$ are interchangeable metrics for
characterizing the errors in $\langle \tilde u_1\rangle$. Cases with
no explicit SGS model do not follow the trend, and $K_{\mathrm{res}}$
can even exceed unity due to nonphysical velocity fluctuations whose
origin is discussed in Section \ref{sec:fluctuations}.  The same
effect is observed for cases with SGS models for the finest grid
resolution but in a lesser degree. The SGS activity is plotted in
Fig. \ref{fig:Umean_error_exactWM_others}(b). Increasing $s$ is
associated with increasing $\mathcal{E}_m$, although the results are
Reynolds number and SGS model dependent and do not collapse for
isotropic and anisotropic grids.  Equation (\ref{eq:s}) has still some
value as it does not make use of DNS data and it is a more realistic
estimator for practical applications for which the reference DNS
solution is not available.
%

\subsection{Relevant length-scale for local error quantification} 
\label{subsec:mean:length}

\begin{figure}[t]
\begin{center}
 \subfloat[]{\includegraphics[width=0.39\textwidth]{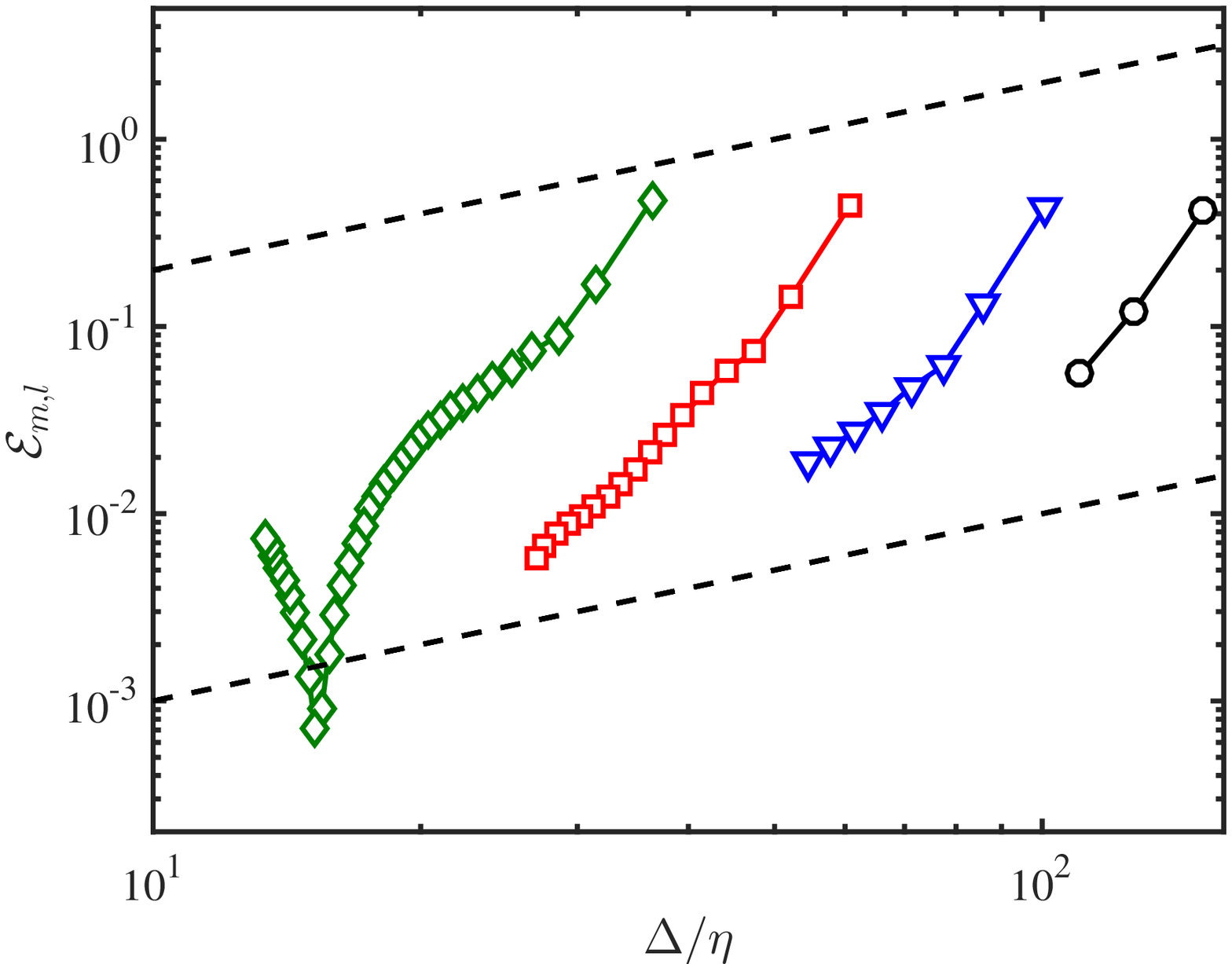}}
 \hspace{0.1cm}
 \subfloat[]{\includegraphics[width=0.39\textwidth]{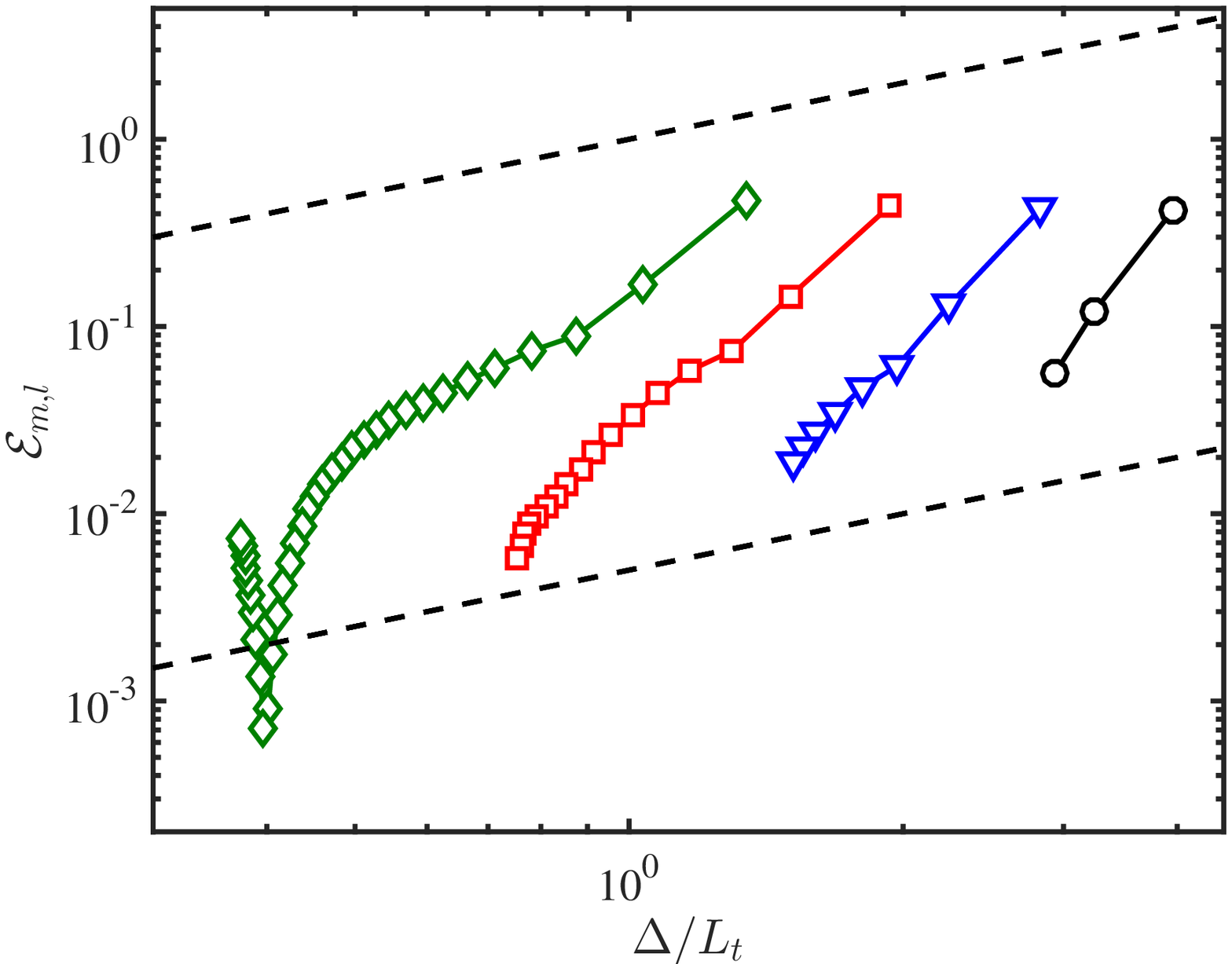}}
 \end{center}
\begin{center}
 \subfloat[]{\includegraphics[width=0.39\textwidth]{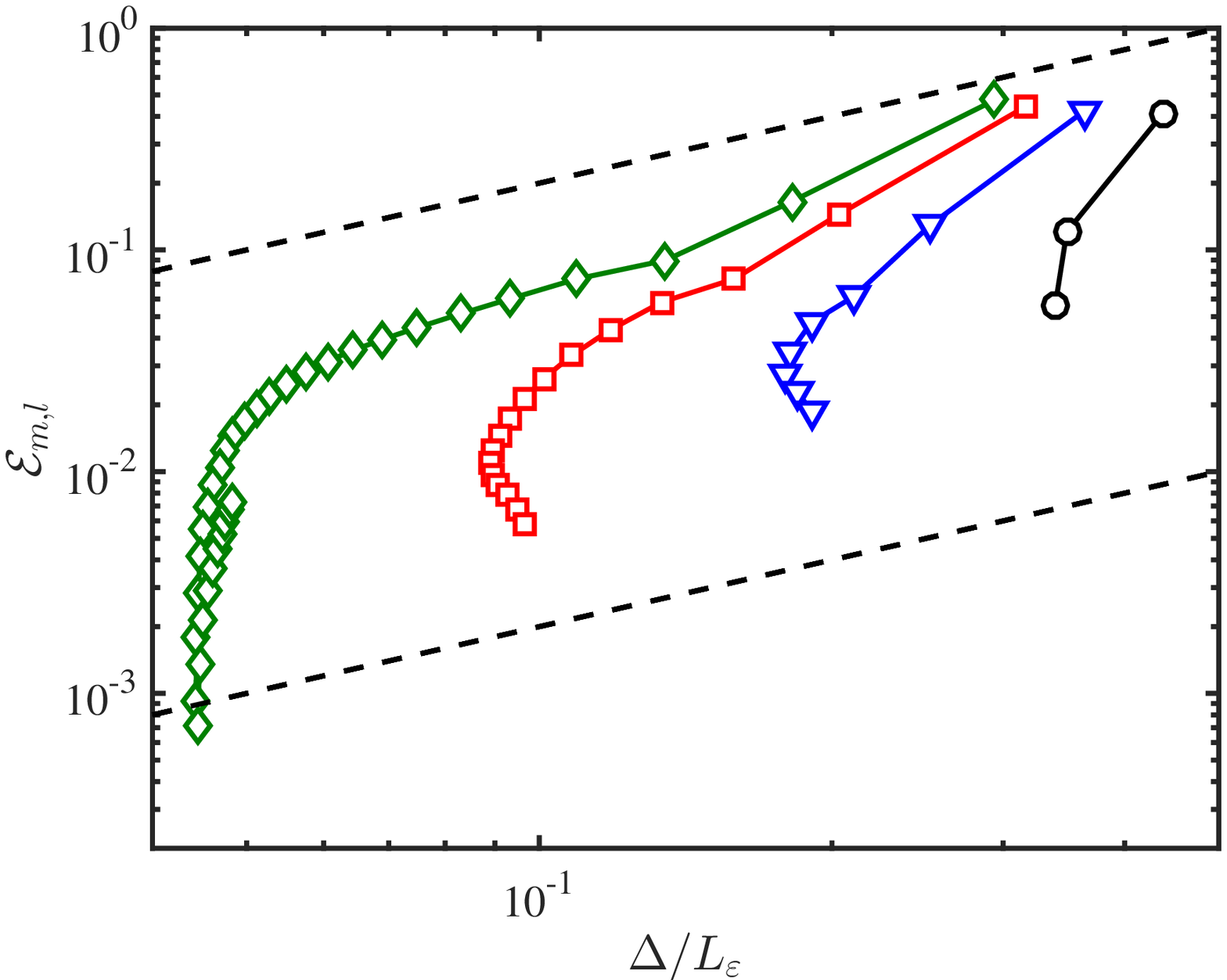}}
 \hspace{0.1cm}
 \subfloat[]{\includegraphics[width=0.39\textwidth]{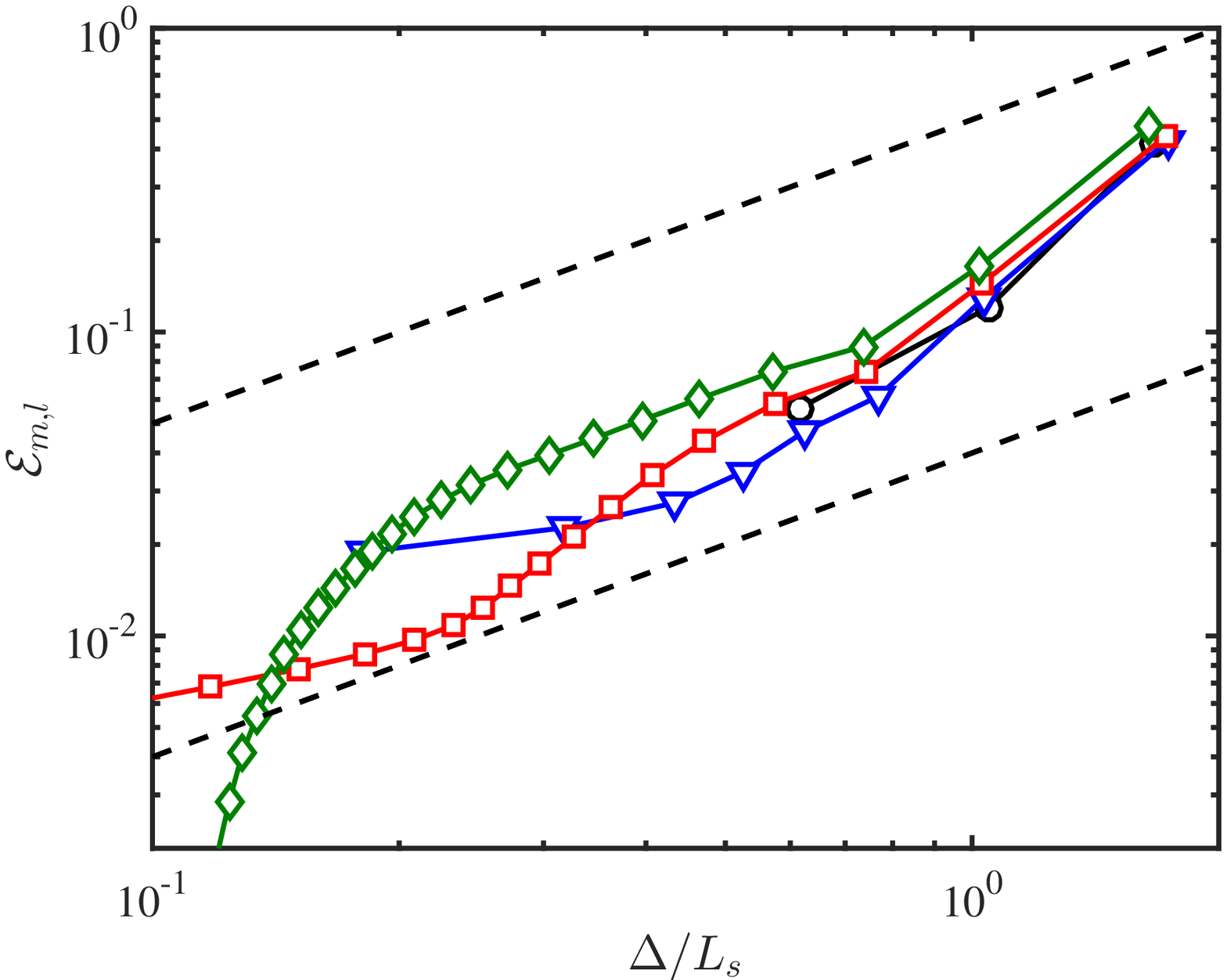}}
 \end{center}
\caption{ Local error in the mean velocity profile $\mathcal{E}_{m,l}$
  as a function of the grid size $\Delta$ normalized by (a) Kolmogorov
  scale $\eta$, (b) Taylor microscale $L_t$, (c) integral length-scale
  $L_\varepsilon$, and (d) shear length-scale $L_s$. Cases are
  AMD4200-EWS-i1 (\mycircle), AMD4200-EWS-i2
  (\textcolor{blue}{\mytriangledown}), AMD4200-EWS-i3
  (\textcolor{red}{\mysquare}), and AMD4200-EWS-i4
  (\textcolor{OliveGreen}{\mydiamond}). The dashed lines in (d) are
  $\mathcal{E}_{m,l} \sim \Delta/L_s$. \label{fig:Umean_error_local}}
\end{figure}

The error in the previous section is an integrated measure across the
entire outer layer and, consistently, the grid resolution is
non-dimensionalized by the boundary layer thickness $\delta$. However,
the length-scale of the energy-containing eddies is a function of the
wall-normal direction, and local errors at a given $x_2$ are expected
to vary accordingly. We investigate the physical length-scale relevant
for local error scaling and define the $x_2$-dependent error in the
mean velocity profile as
\begin{equation} \label{eq:error_local} 
\mathcal{E}_{m,l}(x_2) = \left[ \frac{ \frac{1}{2d}\int_{x_2-d}^{x_2+d} \left(\langle
\tilde{u}_1\rangle - \langle u_1\rangle\right)^2 \mathrm{d}x_2} 
{\frac{1}{0.8\delta}\int_{0.2\delta}^{\delta} \langle u_1\rangle^2 \mathrm{d}x_2 }
\right]^{1/2},
\end{equation}
where the integration limits, $x_2\pm d$, coincide with the grid
locations of $\tilde{u}_1$, and the integral is numerically performed
using the trapezoidal rule. Different candidates for the normalization
length-scale are tested, namely, the Kolmogorov scale
$\eta=(\nu^3/\varepsilon)^{1/4}$ \citep{Pope2000}, the Taylor
microscale $L_t=(15\nu\langle u'_i u'_i\rangle/\varepsilon)^{1/2}$
\citep{Tennekes1972}, the integral length-scale
$L_\varepsilon=(K/3)^{3/2}/\varepsilon$ \citep{Pope2000}, and the
shear length-scale $L_s=u_\tau (\partial \langle u_1\rangle/\partial
x_2)^{-1}$ \cite{Mizuno2011,Lozano2018b}, where $\varepsilon$ is the
rate of energy dissipation, and $K$ is the turbulent kinetic
energy. All the length-scales are computed for the reference DNS
data. The results for AMD4200-EWS-i1,i2,i3,i4 are shown in Fig.
\ref{fig:Umean_error_local}, and similar results are obtained for the
corresponding DSM cases. The best collapse is found for
$\Delta/L_s$. The local error lies below 10\% for $\Delta < L_s$, and
it drastically drops for $\Delta < 0.2L_s$, although theses ranges
should be understood as tentative estimates.  The largest errors are
obtained for $\Delta/L_s>1$, which corresponds to the grid points
closer to the wall.  
%

The scaling results for the local error $\mathcal{E}_{m,l}$ are
consistent with the excellent agreement in the global error
$\mathcal{E}_m$ when the grid resolutions are normalized by $\delta$
(Fig. \ref{fig:Umean_error_exactWM}).  The reason is that, at high
$Re_\tau$, the universal shape of the mean velocity profile in the
outer layer implies that the integrated effect of $L_s$ is
proportional to $\delta$. This can be easily seen by computing the
average value of $L_s$ for $x_2\in[0.2\delta, \delta]$ given by
$(L_s)_{\mathrm{avg}} = 1/0.8\delta \int_{0.2\delta}^{\delta} L_s(x_2)
\mathrm{d}x_2 \sim \delta$. Under the rough assumption that there is
no wake effect and the log layer is valid until the edge of the
boundary layer, then $(L_s)_{\mathrm{avg}} \approx 0.25 \delta$.

To conclude this section, we discuss one last interesting result
regarding the local error at the $n$-th off-wall grid point.
Considering that the $n$-th off-wall grid point is located at
$x_2=n\Delta$, and assuming that at high Reynolds numbers the $n$-th
point falls within the log layer (as expected in WMLES), then $L_s
\approx \kappa x_2$ and $\Delta/L_s \approx 1/(n\cdot\kappa) \approx
2/n$ independently of $\Delta$. Consequently, no improved predictions
are expected in $\langle \tilde u_1\rangle$ at the $n$-th off-wall
grid point as $\Delta$ is refined until the grid resolution reaches
the WRLES-like regime. A similar argument was provided by
\citet{Larsson2015} based on the size of wall-attached eddies across
the log layer.

\section{Error scaling of turbulence intensities} 
\label{sec:fluctuations}

In the previous section, we have measured the errors on $\langle
\tilde{u}_1 \rangle$ by assuming that LES and DNS are directly
comparable. The assumption is reasonable if the filtering operation
has a small impact on the mean of a variable $\phi$, that is,
$\langle\bar{\phi}\rangle \approx \langle\phi\rangle$, which is the
case for the mean velocity profile even at coarse filter
sizes. However, smaller-scale motions play a non-negligible role in
$\langle {u}_i'^{2} \rangle$, casting doubts on how to compare fairly
LES and DNS data. If LES is formally interpreted by means of a spatial
low-pass filter \citep{Deardorff1970,Leonard1975}, the meaningful
quantities to compare are the turbulence intensities of the filtered
DNS velocities. There are two caveats in order to carry on such
comparison. First, although numerical differentiation has a low-pass
filtering effect and the finite grid resolution prevents the formation
of small scales, the filter operator is not distinctly defined in
implicitly-filtered LES
\citep{Lund2003,Carati2001,Bae_brief_2017,Bae_brief_2018} and,
consequently, neither is the associated filter size. The second caveat
is probably more important: in real-world applications we are
interested in predicting DNS values, whereas their filtered
counterparts are of less practical importance. For these reasons, we
study the error scaling of the LES fluctuating velocities with respect
to unfiltered quantities.

In this section, we first argue that the physical mechanism regulating
the magnitude of the fluctuating velocities in implicitly-filtered LES
is not related to filtering, but rather to the requirement of
generating velocity gradients consistent with the statistically steady
state. Secondly, we study the theoretical and numerical convergence of
the LES turbulence intensities in wall-bounded flows.

\subsection{The mechanism controlling fluctuating velocities in implicitly-filtered LES} 
\label{subsec:fluctuations:mech}

\begin{figure}
\begin{center}
 \subfloat[]{\includegraphics[width=0.37\textwidth]{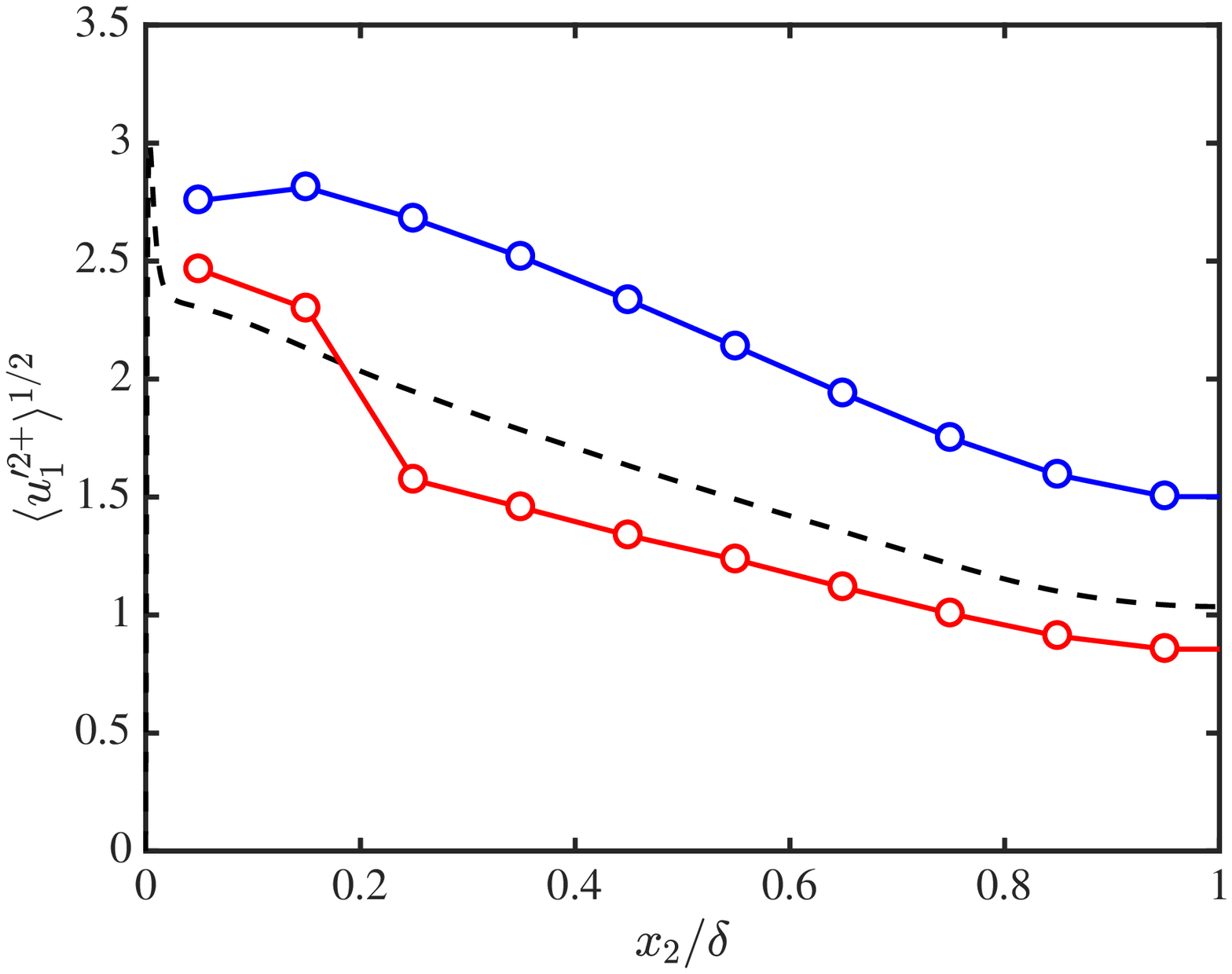}}
 \hspace{0.2cm}
 \subfloat[]{\includegraphics[width=0.45\textwidth]{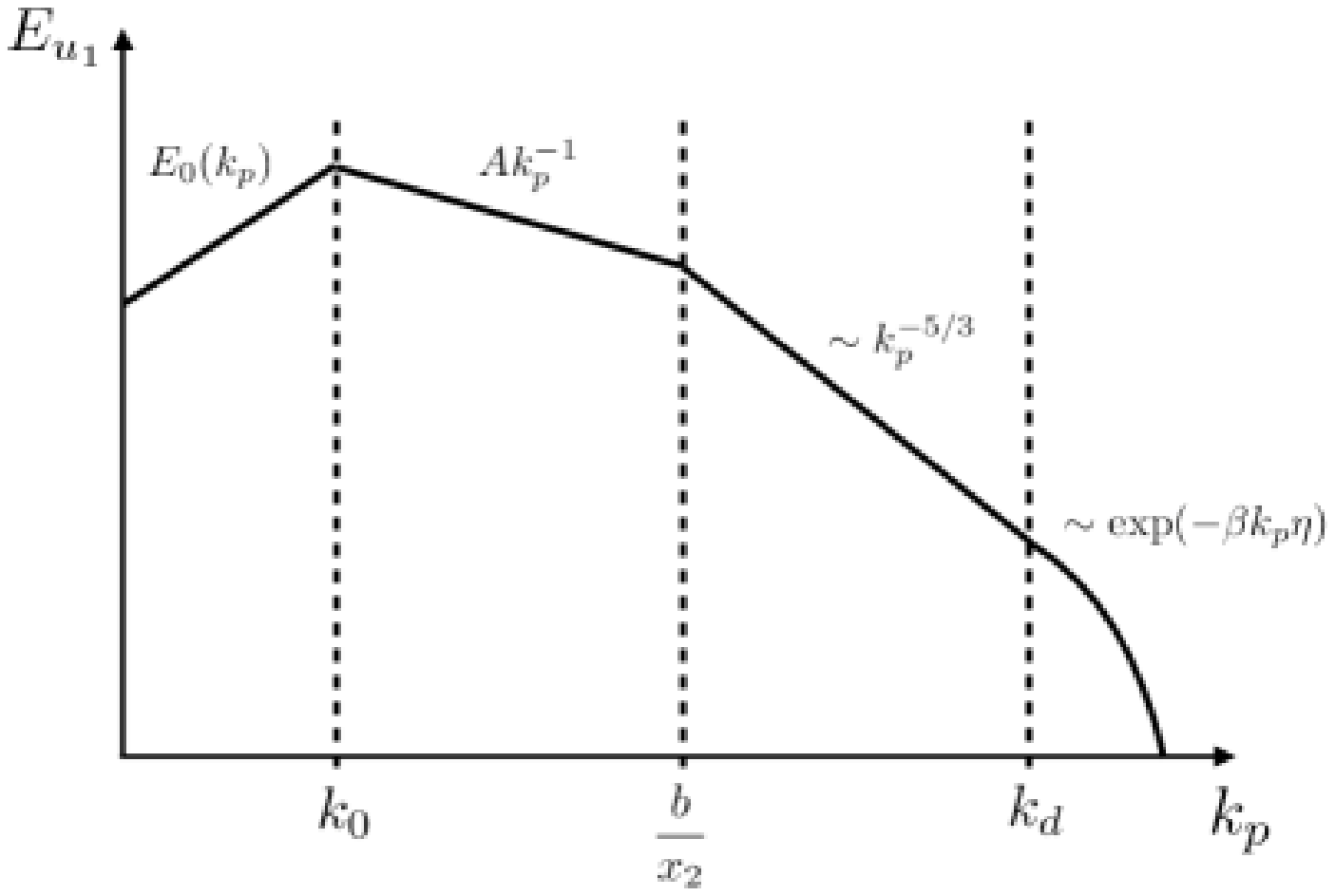}}
\end{center}
\caption{ (a) Streamwise r.m.s. velocity fluctuations for DNS at
  $Re_\tau\approx4200$ (\dashed), NM4200-EWS-i2
  (\textcolor{blue}{\mycircle}), and DSM4200-EWS-i2
  (\textcolor{red}{\mycircle}). (b) Model spectrum for the streamwise
  turbulence intensity. The parameters $A$, $b$, $k_0$, $k_d$ and
  $\beta$ are model constants.
  \label{fig:fluctuations_1} }
\end{figure}

Fig. \ref{fig:fluctuations_1}(a) shows the root-mean-squared (r.m.s.)
of the streamwise fluctuating velocity for DNS, LES without SGS model,
and LES with DSM. In the absence of model, the LES intensities are
over-predicted compared with DNS and, conversely, under-predicted with
DSM.  Similar results are obtained for the wall-normal and spanwise
velocity fluctuations.  The change in magnitude of the LES r.m.s.
fluctuating velocities can be understood through the energy equation
integrated over the channel flow domain $\mathcal{V}$ with volume $V$,
\begin{equation}\label{eq:int_energy}
 \frac{u_\tau^2 QV}{2\delta^2} = 
\int_{\mathcal{V}}  \left( \nu + \nu_t \right) 
\left( 
\frac{\partial \tilde u_i}{\partial x_k}\frac{\partial \tilde u_i}{\partial x_k} +
\frac{\partial \tilde u_i}{\partial x_k}\frac{\partial \tilde u_k}{\partial x_i} 
\right)
\mathrm{d}V.
\end{equation}
%
Eq. (\ref{eq:int_energy}) shows that the input power to maintain the
mass flow $Q$ must be dissipated by the viscous/SGS terms. In the DNS
limit ($\nu_t=0$) with fixed $\nu$, this is achieved by the velocity
gradients $\partial \tilde u_i/\partial x_k \sim \Delta u_c/l_c$,
where $\Delta u_c$ and $l_c$ are the characteristic velocity
difference and length of the smallest scales, respectively.  In LES
($\nu_t\neq 0$), the smallest available length-scale is limited by the
grid resolution $l_c \approx \Delta$. Thus, the two possible
mechanisms to maintain consistency with Eq. (\ref{eq:int_energy}) are
by $\nu_t>0$, or by augmenting $\Delta u_c$ (and hence the turbulence
intensities). If $\nu_t$ is large enough, $\Delta u_c$ is
under-predicted with respect to DNS as illustrated in
Fig. \ref{fig:fluctuations_1}(a). Conversely, if $\nu_t$ is small, as
in LES without explicit SGS model ($\nu_t = 0$), the result is an
increase of the turbulence intensities as shown in
Fig. \ref{fig:fluctuations_1}(a).  This illustrates how the mean LES
kinetic energy can exceed the mean DNS kinetic energy when SGS models
are not dissipative enough (as in
Fig. \ref{fig:Umean_error_exactWM_others}a), which may be problematic
for LES quality assessment.
%

In summary, the physical mechanism regulating the magnitude of the
fluctuating velocities in implicitly-filtered LES is related to the
necessity of generating dissipative terms of the correct magnitude
rather than by the (non-existent) filtering operation. Nevertheless,
the results above shows that even if implicitly-filtered LES is not
rigorously equivalent to the filtered Navier--Stokes equations, it
does hold some resemblance in the sense that the values of $\nu_t$
providing the correct mean velocity profile scaled by
  free stream, centerline, or bulk velocity, are accompanied by lower
r.m.s. velocities as it would be expected from the filtered DNS
velocity field.

\subsection{Theoretical estimations} 
\label{subsec:fluctuations:theoretical}

The metric adopted to measure errors in the turbulence intensities is
\begin{equation}\label{eq:error_ms_def}
\mathcal{E}_{f,i} = \left[ \frac{ \int_{0.2\delta}^{\delta} \left(\langle
\tilde u'^2_i \rangle - \langle u'^2_i \rangle\right)^2 \mathrm{d}x_2} {
\int_{0.2\delta}^{\delta} \langle u'^2_i \rangle^2 \mathrm{d}x_2 }
\right]^{1/2},
\quad i=1,2,3.
\end{equation}
For brevity, we occasionally omit the subscript $i$ when the
particular flow component is not relevant in the discussion. Our goal
is to estimate $\mathcal{E}_{f}$ as a function of $\Delta$.

In the log layer of wall-bounded turbulence at high Reynolds numbers,
the intensities of the unfiltered velocity fluctuations are known to
follow
\begin{eqnarray}\label{eq:aat_u}
\frac{\langle u'^2_1 \rangle}{u_\tau^2} = B_1 - A_1 \log \left( \frac{x_2}{\delta}\right), \quad
\frac{\langle u'^2_2 \rangle}{u_\tau^2} = B_2, \quad
\frac{\langle u'^2_3 \rangle}{u_\tau^2} = B_3 - A_3 \log \left( \frac{x_2}{\delta}\right), 
\end{eqnarray}
where the coefficients $B_i$ and $A_i$ are constants considered to be
universal for turbulent channel flows.  Eq. (\ref{eq:aat_u}) can be
derived by using the attached-eddy hypothesis \citep{Townsend1976} or
by dimensional analysis on the $k^{-1}$ spectrum of $u_1$ and $u_3$
\citep{Perry1977}, and the blocking effect of the wall for $u_2$.  The
hypothesis has been confirmed at high Reynolds number flows
\cite{Marusic2013,Hultmark2012} and it has also been observed in the
spanwise velocity even for relatively low Reynolds numbers
\cite{Jimenez2008,Sillero2013,Lozano2014,Lee2015}. An important
consequence of Eq. (\ref{eq:aat_u}) is that, at a given $x_2/\delta$,
the magnitude of the velocity fluctuations scaled by $u_\tau$ is
constant and independent of the Reynolds number.

We are now interested in the LES asymptotic high-Reynolds-number
limit for the filtered fluctuating velocities $\langle \bar u'^2_i
\rangle$,
\begin{eqnarray}\label{eq:aat_uf}
\frac{\langle \bar u'^2_1 \rangle}{u_\tau^2} = \bar B_1 - \bar A_1 f \left( \frac{x_2}{\delta}\right), \quad
\frac{\langle \bar u'^2_2 \rangle}{u_\tau^2} = \bar B_2, \quad
\frac{\langle \bar u'^2_3 \rangle}{u_\tau^2} = \bar B_3 - \bar A_3 f\left( \frac{x_2}{\delta}\right), 
\end{eqnarray}
where $\bar B_i$ and $\bar A_i$ are constants that depend on
$\Delta_i$, and $f$ is an unknown function such that $f(x_2)
\rightarrow \log(x_2)$ as $\Delta_i \rightarrow 0$. The exact
dependence of $\bar B_i$, $\bar A_i$ on $\Delta_i$, and the particular
shape of $f$ is expected to vary for different filter kernels. The
value of $\langle \bar u'^2_i \rangle$ may be estimated for a
symmetric filter with well-defined, non-zero second moment in real
space by considering \citep{Carati2001,Yeo1988}
\begin{eqnarray}
\langle \bar u'^2_i \rangle = \langle \overline{u'^2_i} \rangle - 
\Delta_k^2  \left\langle  \left( \frac{\partial \bar u'_i }{\partial x_k}  \right)^2 \right\rangle 
+ \mathcal{O}(\Delta_1^4 + \Delta_2^4 + \Delta_3^4).
\end{eqnarray}
If we further assume that $\langle \overline{u'^2_i} \rangle \approx
\langle u'^2_i \rangle$, then
\begin{eqnarray}\label{eq:log_theory}
\frac{\langle \bar u'^2_i \rangle}{u_\tau^2} = B_i -
A_i \log \left( \frac{x_2}{\delta}\right) 
- \Delta_k^2  \left\langle  \left( \frac{\partial \bar u'_i }{\partial x_k}  \right)^2 \right\rangle 
+ \mathcal{O}(\Delta_1^4 + \Delta_2^4 + \Delta_3^4),
\end{eqnarray}
where $A_2=0$. Eq. (\ref{eq:log_theory}) shows that the wall-parallel
turbulence intensities of the filtered field do not follow Eq. 
(\ref{eq:aat_u}), and the major contributor to the departure from the
classic log-law is the correction term on the right-hand side of
Eq. (\ref{eq:log_theory}). The error is then given by
\begin{equation}\label{eq:Ef_error}
\mathcal{E}_{f} \sim 
\Delta_k^2 \left\langle \left( \frac{\partial \bar u'_i}{\partial x_k} \right)^2 \right\rangle.
\end{equation}
Equation (\ref{eq:Ef_error}), together with the estimations for the
velocity gradient $G$ in Section \ref{subsec:mean:theoretical}, yields
\begin{equation}\label{eq:error_ms_scaling}
\mathcal{E}^s_{f} \sim \Delta^0, \quad \mathcal{E}^i_{f} \sim \Delta^{2/3},
\end{equation}
which predict a low convergence rate of the LES turbulence intensities
\textcolor{black}{for $\Delta$ comparable to the scales in
  shear-dominated regime ($\mathcal{E}^s_{f}$), and in the inertial
  range ($\mathcal{E}^i_{f}$)}.

A limitation of Eq. (\ref{eq:error_ms_scaling}) is that it does not
provide any insight into the explicit logarithmic dependence of
$\langle u'^2_1 \rangle$ and $\langle u'^2_3 \rangle$ with $x_2$.  An
alternative procedure to estimate $\mathcal{E}_{f}$ is to connect
Eq. (\ref{eq:aat_u}) and Eq. (\ref{eq:aat_uf}) by the spectrum of the
streamwise velocity,
\begin{equation}\label{eq:urms_spectra}
\frac{\langle u'^2_i \rangle}{u_\tau^2} = 2\int_{0}^{\infty} E_{u_i}(k_p,x_2)\mathrm{d}k_p,
\end{equation}
where $E_{u_i}$ is the two-dimensional spectrum for the $i$-th
velocity component as a function of $k_p^2 = k_1^2 + k_3^2$, where
$k_1$ and $k_3$ are the streamwise and spanwise wavenumbers,
respectively. Similarly,
\begin{equation}\label{eq:urms_f_spectra1}
\frac{\langle \bar u'^2_i \rangle}{u_\tau^2} = 2\int_{0}^{\infty} \bar E_{u_i}(k_p,x_2)\mathrm{d}k_p,
\end{equation}
where $\bar E_{u_i}(k_p,x_2)$ is the energy spectra of the filtered
velocities. We focus on the streamwise velocity component, but the
reasoning below is also applicable to the spanwise component.  To make
the problem tractable, we adopt the model spectrum for $E_{u_1}$ from
Fig.  \ref{fig:fluctuations_1}(b).
%
The four different piecewise domains of the model correspond to the
large-scale, shear-dominated \citep{Perry1977}, inertial
\citep{Kolmogorov1941}, and viscous regimes \citep{Kraichnan1959},
respectively. Evaluation of Eq. (\ref{eq:urms_spectra}) using the
model from Fig. \ref{fig:fluctuations_1}(b) results in
\begin{equation}
\frac{\langle u'^2_1 \rangle}{2u_\tau^2} \approx \mathrm{constant}  -
A\log\left( \frac{x_2}{b}\right),
\end{equation}
where the contributions from inertial and viscous regimes have been
neglected. The result is consistent with the logarithmic functional
dependence of the streamwise turbulence intensity from Eq.
(\ref{eq:aat_u}).  Under the severe assumptions that the filtering
operator resembles a sharp Fourier cut-off, and neglecting filtering
in the wall-normal direction,
\begin{equation}\label{eq:urms_f_spectra2}
\frac{\langle \bar u'^2_1 \rangle}{u_\tau^2} = 2\int_{0}^{\infty} \bar E_{u_1}(k_p,x_2)\mathrm{d}k_p
\approx 2 \int_{0}^{\pi/\Delta} E_{u_1}(k_p,x_2)\mathrm{d}k_p.
\end{equation}
The difference $\langle u'^2_1 \rangle - \langle \bar u'^2_1 \rangle$
definitory of the error in Eq. (\ref{eq:error_ms_def}) is
\begin{equation}\label{eq:error_int_spec}
\mathcal{E}_{f} \sim \int_{0}^{\infty} E_{u_1}(k_p,x_2) \mathrm{d}k_p - 
\int_{0}^{\pi/\Delta} E_{u_1}(k_p,x_2)\mathrm{d}k_p,
\end{equation}
and after integration we obtain
\begin{equation}\label{eq:error_ms_scaling_2}
\mathcal{E}^s_{f} \sim \log(\Delta/x_2), \quad \mathcal{E}^i_{f} \sim \Delta^{2/3}.
\end{equation}
When the filter cut-off lies within the $k_p^{-1}$ regime, Eq.
(\ref{eq:error_ms_scaling_2}) predicts a $\log(\Delta/x_2)$ correction
to the $\Delta^0$-dependence estimated in Eq.
(\ref{eq:error_ms_scaling}), although both cases imply a slow
convergence with $\Delta$. Regarding the behavior of $\langle \bar
u'^2_1 \rangle$, for $\Delta$ within the shear-dominated region,
\begin{equation}
\langle \bar u'^2_1 \rangle \approx \mathrm{constant} + \mathcal{O}(\log(\Delta)),
\end{equation}
and the LES streamwise and spanwise turbulence intensities will not
reproduce the asymptotic logarithmic profile. For the inertial range,
the prediction of Eq. (\ref{eq:error_ms_scaling_2}) coincides with the
one reported in Eq. (\ref{eq:error_ms_scaling}). In this case,
integration of the model spectrum yields
\begin{equation}
\langle \bar u'^2_1 \rangle \approx \mathrm{constant} - A\log(x_2) + \mathcal{O}(\Delta^{2/3}),
\end{equation}
and LES is expected to capture the classic logarithmic behavior in
$x_2$ with a correction of the order of $\Delta^{2/3}$.



\subsection{Numerical assessment} 
\label{subsec:fluctuations:numeric}

We aim to quantify the exponents $\alpha_f$ and $\gamma_f$ for
\begin{equation}
\mathcal{E}_{f} \sim \left(\frac{\Delta}{\delta}\right)^{\alpha_f} Re_\tau^{\gamma_f},
\end{equation}
from LES data and the range of grid resolutions of interest in the
present work.  The results reported in this section are strictly valid
for LES with DSM. Nevertheless, similar conclusions are drawn for AMD
and $x_2 > 0.3\delta$, where the turbulence intensities predicted by
AMD and DSM are almost indistinguishable. The results are also
compared with filtered DNS data (fDNS), but this is only done
qualitatively.  For that, we use a three-dimensional box-filter with
filter size equal to the LES grid resolution in each direction. The
choice of this particular filter shape and filter size is arbitrary,
and it is argued before that no specific form can be established
\emph{a priori} for implicitly-filtered LES.

\begin{figure}[t]
\begin{center}
 \subfloat[]{\includegraphics[width=0.38\textwidth]{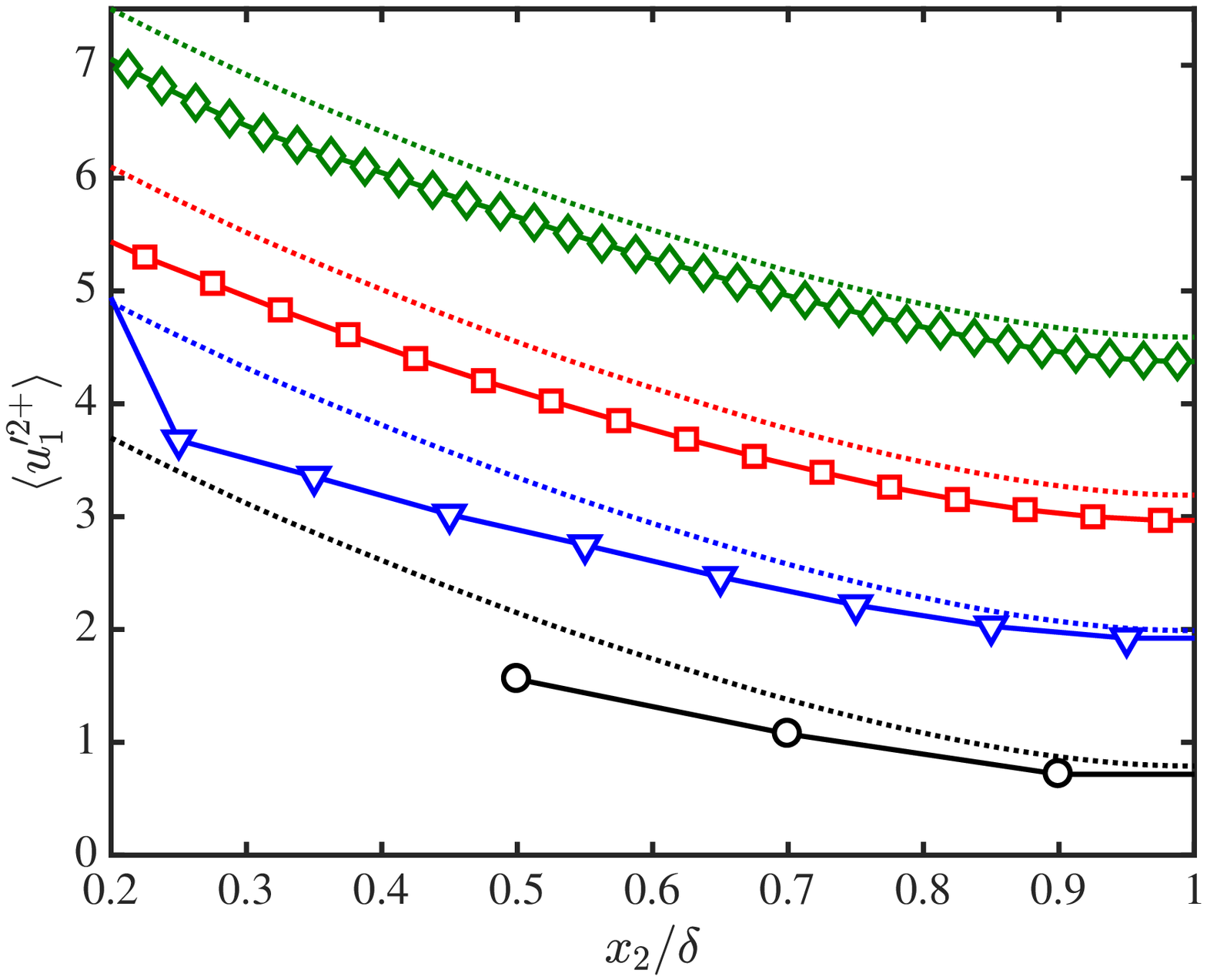}}
 \hspace{0.5cm}
 \subfloat[]{\includegraphics[width=0.38\textwidth]{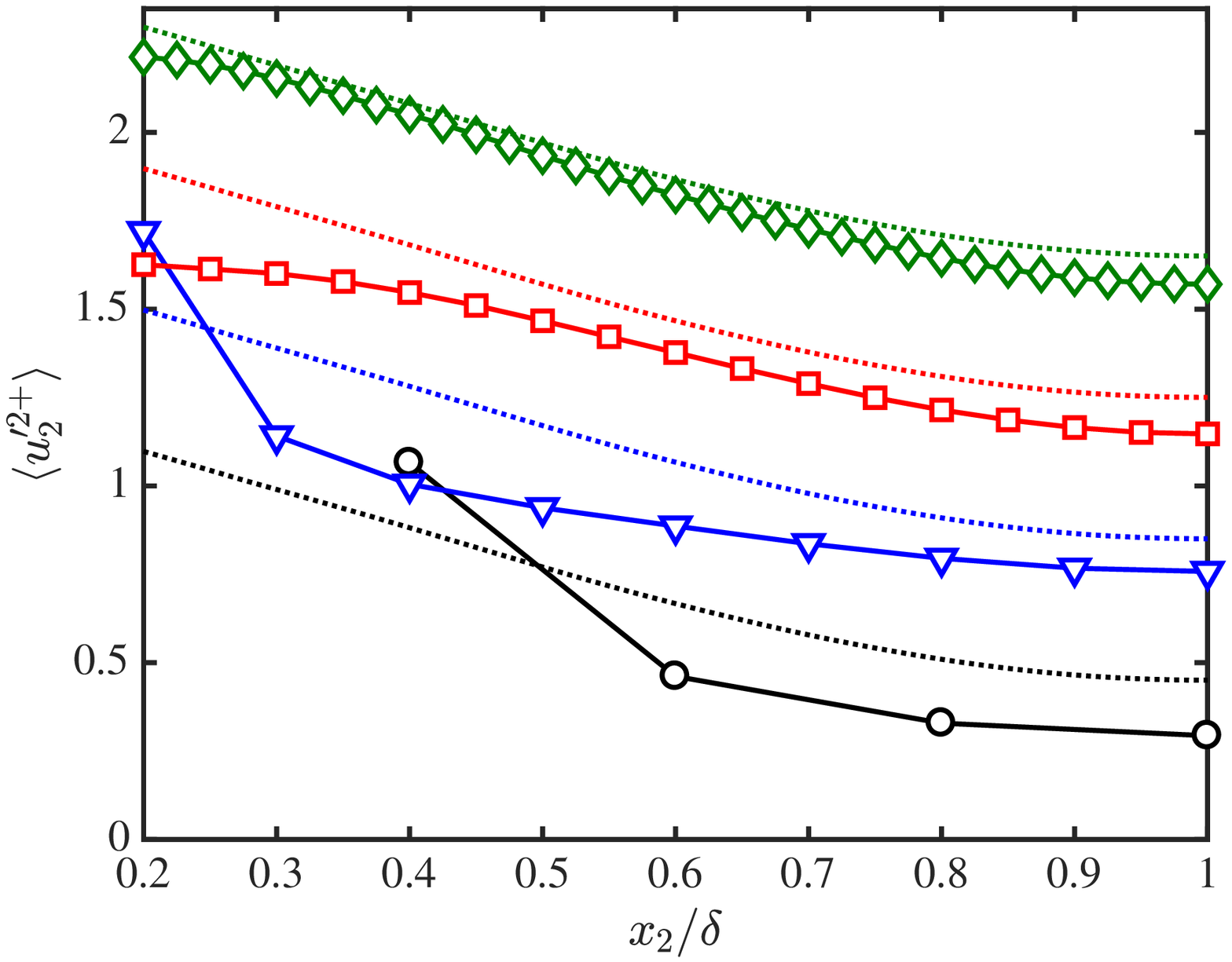}}
\end{center}
\begin{center}
 \subfloat[]{\includegraphics[width=0.38\textwidth]{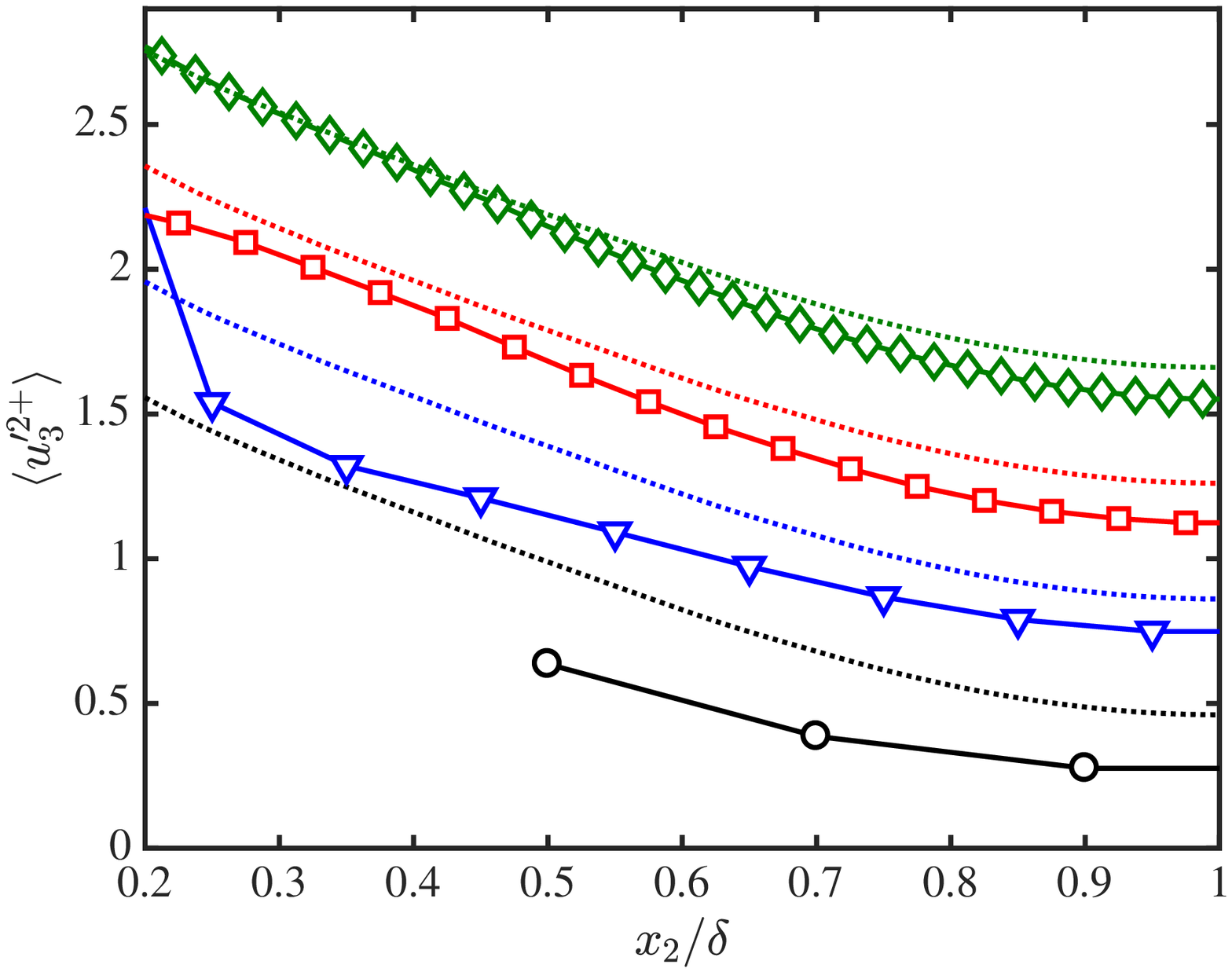}}
 \hspace{0.5cm}
 \subfloat[]{\includegraphics[width=0.38\textwidth]{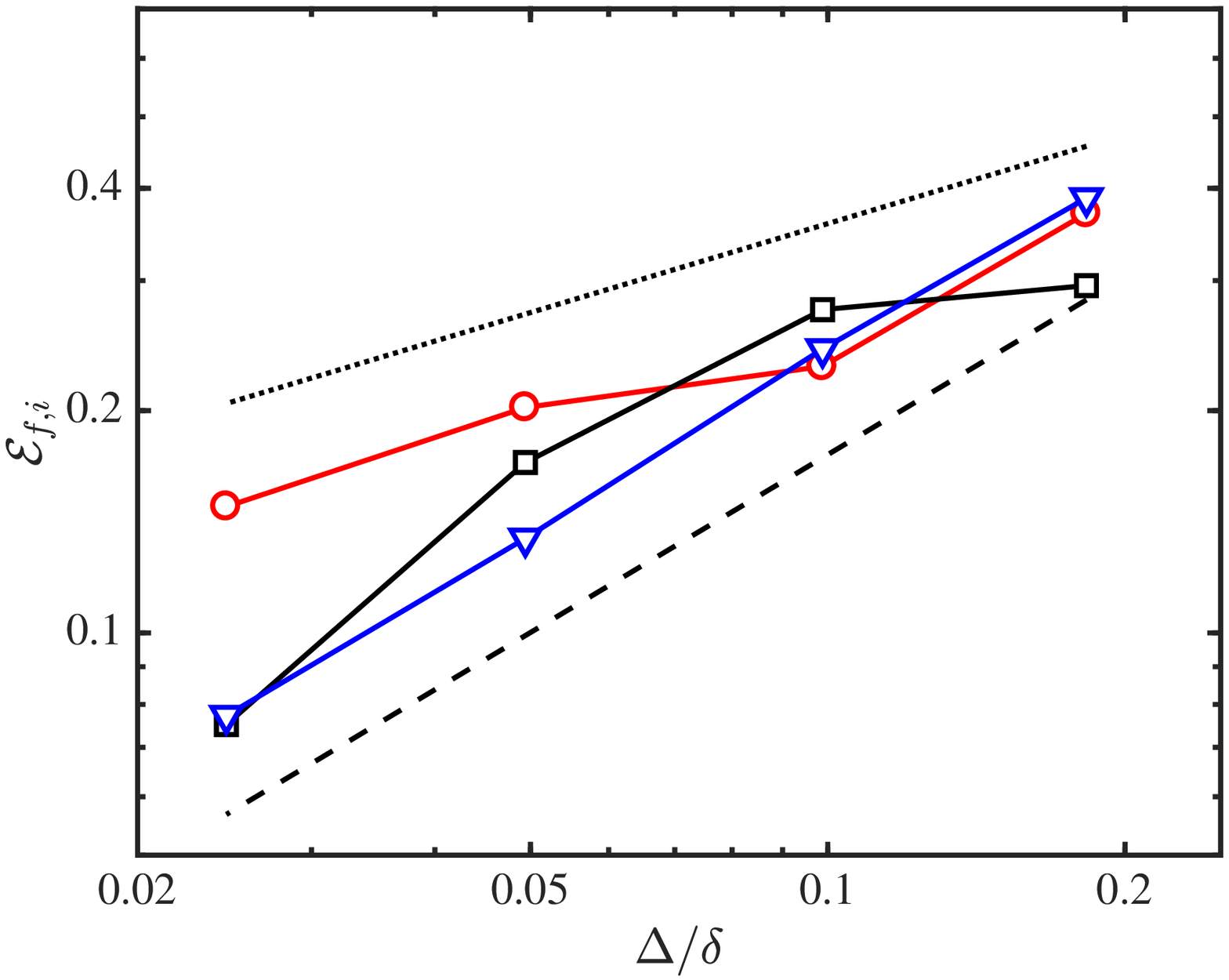}}
\end{center}
\caption{Streamwise (a), wall-normal (b), and spanwise (c) turbulence
  intensities as a function of the wall-normal distance for different
  grid resolutions at $Re_\tau \approx 2000$. Symbols are LES cases
  DSM2000-EWS-i1 (\mycircle), DSM2000-EWS-i2 (\mytriangledown),
  DSM2000-EWS-i3 (\mysquare), DSM2000-EWS-i4 (\mydiamond).  For
  clarity, cases DSM2000-EWS-i2, DSM2000-EWS-i3 and DSM2000-EWS-i4 are
  vertically shifted by $1.2$, $2.4$ and $3.8$ wall units,
  respectively. For comparison, each LES case is accompanied by DNS
  data (\dotted) vertically shifted by the same amount. The first two
  points closer to the wall for case DSM2000-EWS-i1 are omitted as
  they are contaminated by the nonphysical solution close to the wall.
  (d) Error in the streamwise $\mathcal{E}_{f,1}$
  (\textcolor{red}{\mycircle}), wall-normal $\mathcal{E}_{f,2}$
  (\mysquare), and spanwise $\mathcal{E}_{f,3}$
  (\textcolor{blue}{\mytriangledown}) turbulence intensities as a
  function of the characteristic grid resolution. The dashed and
  dotted lines are $\mathcal{E}_{f} \sim \Delta^{0.8}$ and
  $\mathcal{E}_{f} \sim \Delta^{0.4}$,
  respectively. \label{fig:fluctuations_grid}}
\end{figure}

Figs. \ref{fig:fluctuations_grid}(a)--(c) show the turbulence
intensities as a function of the wall-normal distance for DNS and LES
at $Re_\tau \approx 2000$ and various grid resolutions.  The main
observation from Fig. \ref{fig:fluctuations_grid}(a) is that the LES
turbulence intensities diverge from DNS as the grid is coarsened, and
the shape of the $\langle \tilde u'^2_i\rangle$ becomes distinctively
different from $\langle u'^2_i\rangle$.  Moreover, the effect is more
pronounced closer to the wall. Hence, the logarithmic behavior is not
captured by LES when $\Delta = \mathcal{O}(\delta)$, consistent with
the discussion in Section \ref{subsec:fluctuations:theoretical}. The
error between LES and DNS is quantified in
Fig. \ref{fig:fluctuations_grid}(d) and compared with the predictions
from Eq. (\ref{eq:error_ms_scaling}). The results show that
$\mathcal{E}_{f,1}$ converges as $\Delta^{0.4}$, whereas
$\mathcal{E}_{f,2}$ and $\mathcal{E}_{f,3}$ are well represented by
$\Delta^{0.8}$.
%
%

\begin{figure}[t]
\begin{center}
 \subfloat[]{\includegraphics[width=0.38\textwidth]{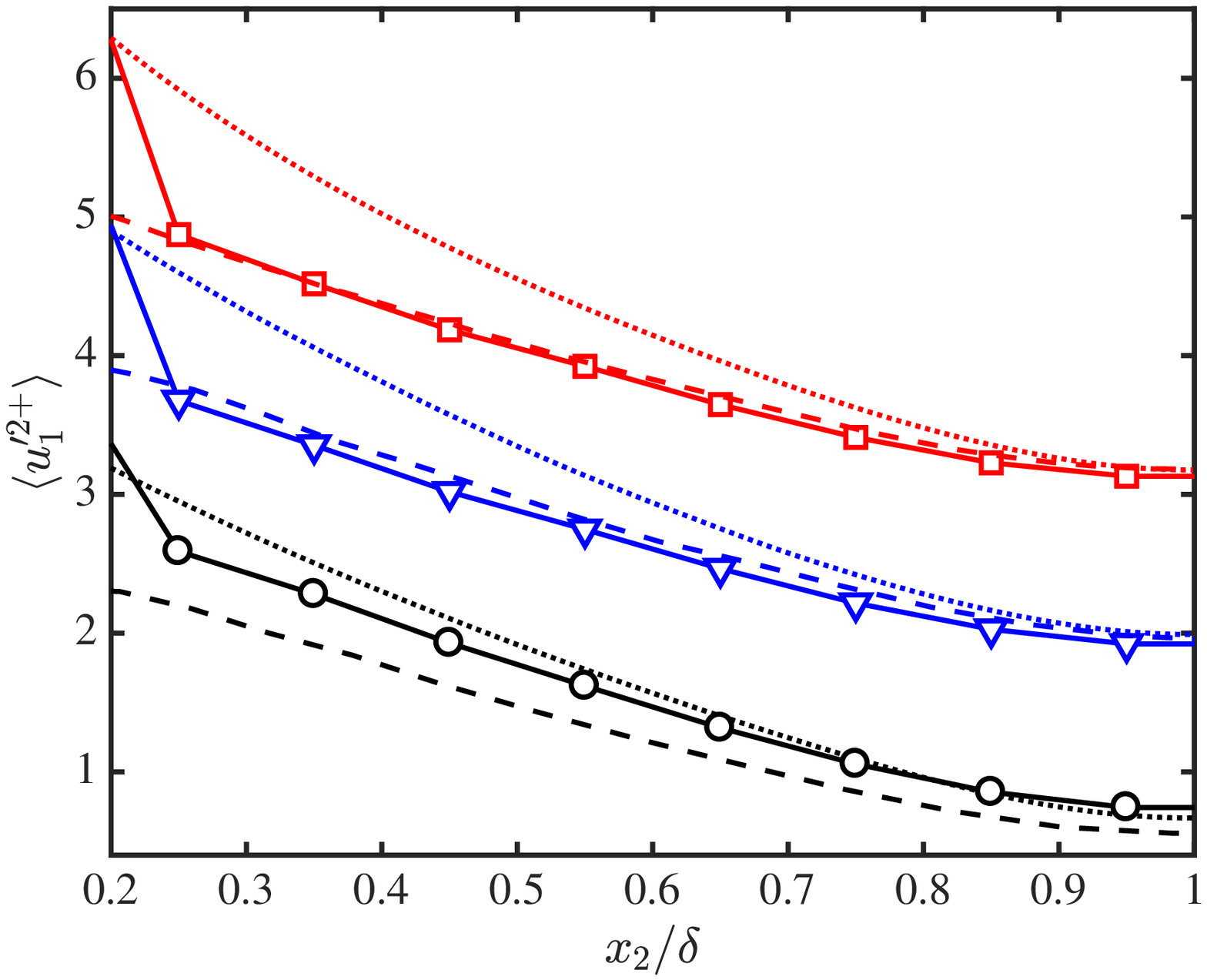}}
 \hspace{0.5cm}
 \subfloat[]{\includegraphics[width=0.38\textwidth]{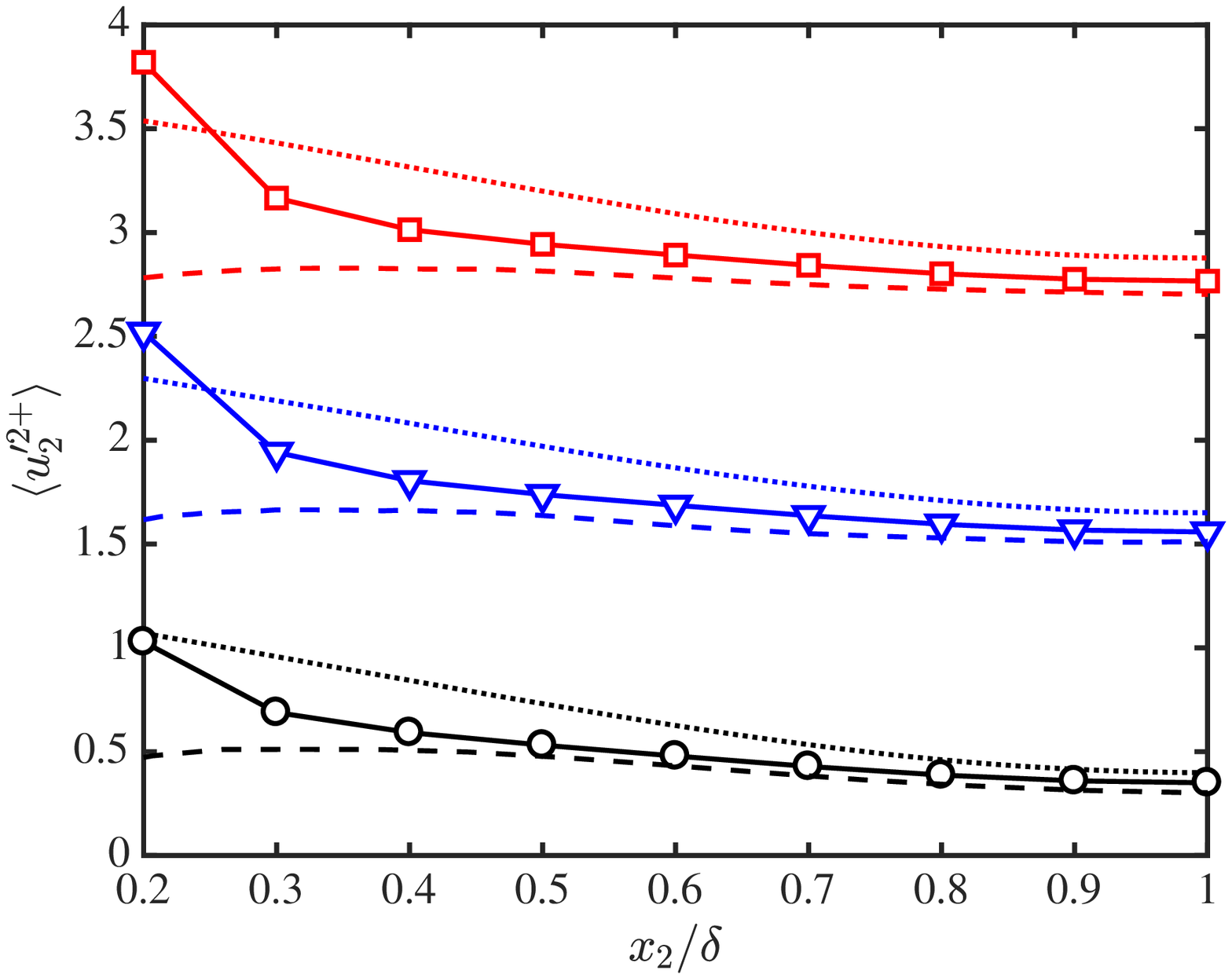}} 
\end{center}
\begin{center}
 \subfloat[]{\includegraphics[width=0.38\textwidth]{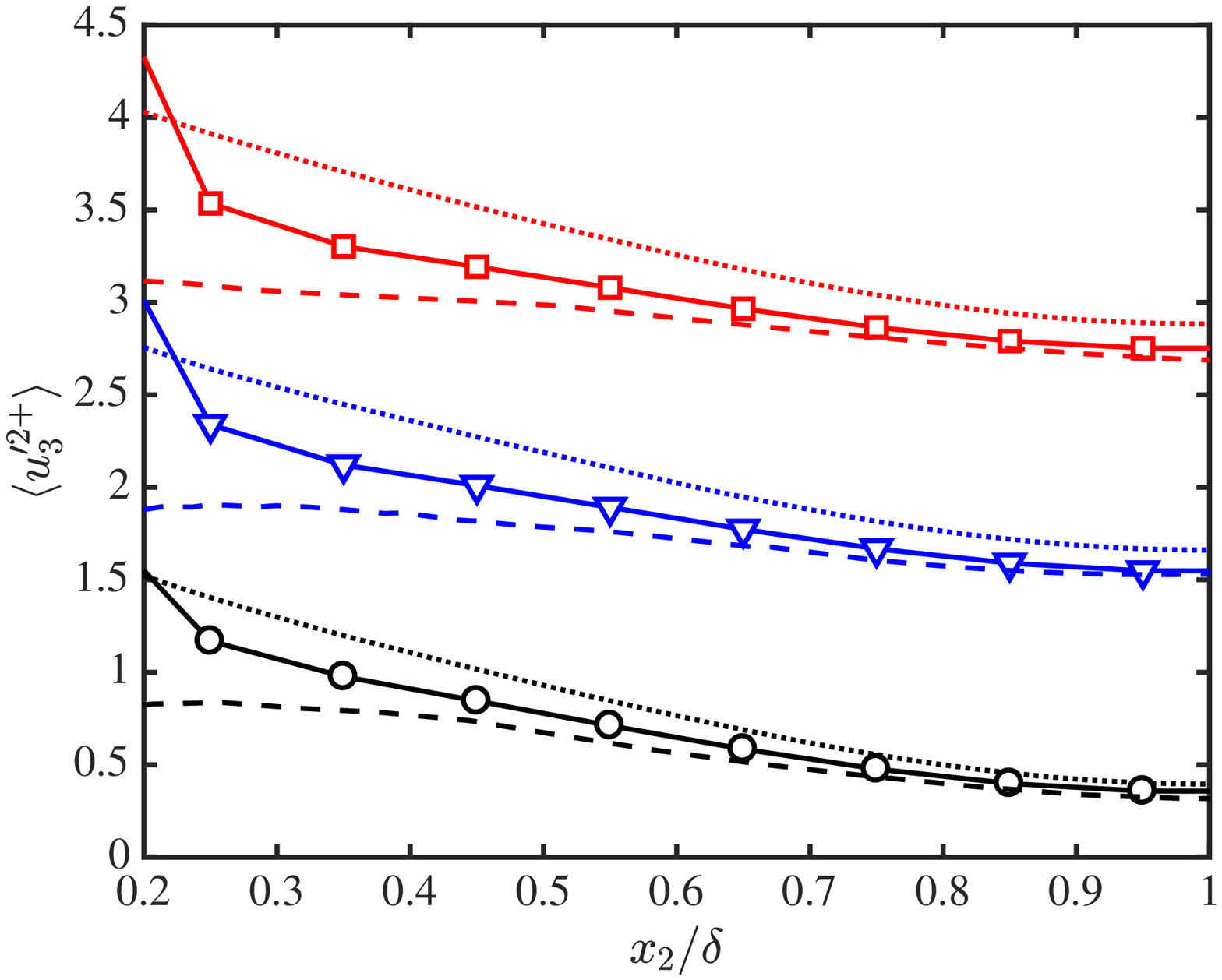}}
 \hspace{0.5cm}
 \subfloat[]{\includegraphics[width=0.38\textwidth]{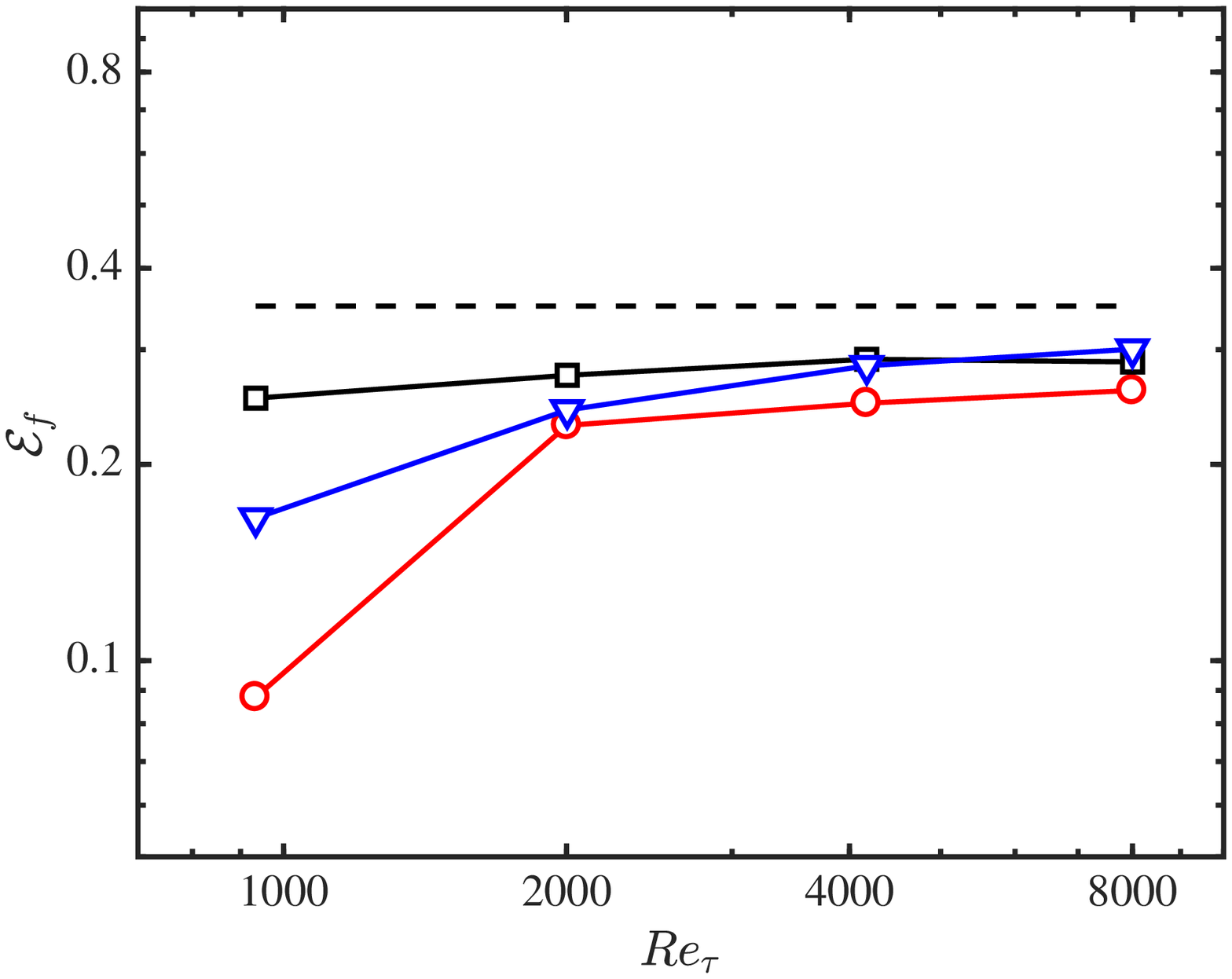}}
\end{center}
\caption{Streamwise (a), wall-normal (b), and spanwise (c) turbulence
  intensities as a function of the wall-normal distance for different
  Reynolds numbers with grid resolution i2. Symbols are LES cases
  DSM950-EWS-i2 (\mycircle), DSM2000-EWS-i2 (\mytriangledown), and
  DSM4200-EWS-i2 (\mysquare).  For clarity, cases DSM2000-EWS-i2 and
  DSM4200-EWS-i2 are vertically shifted by $1.2$ and $2.4$ wall units,
  respectively. For comparison, each LES case is accompanied by DNS
  data (\dotted) and box filtered DNS data (\dashed) vertically
  shifted by the same amount. (d) Error in the streamwise
  $\mathcal{E}_{f,1}$ (\textcolor{red}{\mycircle}), wall-normal
  $\mathcal{E}_{f,2}$ (\mysquare), and spanwise $\mathcal{E}_{f,3}$
  (\textcolor{blue}{\mytriangledown}) turbulence intensities as a
  function of the Reynolds number. The dashed line is $\mathcal{E}_{f}
  = 0.35$.
\label{fig:fluctuations_Retau} }
\end{figure}

The effect of the Reynolds number is evaluated in Fig.
\ref{fig:fluctuations_Retau}, which also includes comparisons with
fDNS. The grid resolution (or filter size) for the LES and fDNS cases
is set to i2 from Table \ref{table:resolutions} ($\Delta =
0.1\delta$), and $Re_\tau$ ranges from $\approx 950$ to $\approx
4200$. The dependence of $\mathcal{E}_{f,i}$ with $Re_\tau$ is weak,
and the error remains roughly constant for $Re_\tau > 950$, from where
we conclude that $\gamma_f \approx 0$. Therefore, the empirically
measured error for the LES turbulence intensities scales as
\begin{equation}\label{eq:error_fluc_empi}
\mathcal{E}_{f,1} \sim \left(\frac{\Delta}{\delta}\right)^{0.4} Re_\tau^0, \quad 
\mathcal{E}_{f,2/3} \sim \left(\frac{\Delta}{\delta}\right)^{0.8} Re_\tau^0,
\end{equation}
for $\Delta > 0.025\delta$. \textcolor{black}{The empirical results in
  Eq. (\ref{eq:error_fluc_empi}) corroborate that the correct
  representation of $\langle \bar u'^2_i \rangle$ is more demanding
  than that for the mean velocity profile, consistent with the
  analysis in Section \ref{subsec:fluctuations:theoretical}. The
  results are closer to the theoretical error prediction obtained for
  $\Delta$ comparable to the inertial length-scales
  ($\mathcal{E}^i_{f}\sim\Delta^{2/3}$), albeit the convergence rate
  for $\mathcal{E}_{f,1}$ is more moderate than for
  $\mathcal{E}_{f,2}$ and $\mathcal{E}_{f,3}$.  Nonetheless, we have
  discussed before that the error estimations from theoretical
  arguments presented above should be appraised as indicative of the
  actual non-linear error rather than as strict error laws. }
%

Fig. \ref{fig:fluctuations_Retau}(a) also shows that the LES
turbulence intensities are well approximated by fDNS, especially for
the highest Reynolds numbers and far from the wall. We have argued at
the beginning of the section that the filter operator is not
well-defined in implicitly-filtered LES, and the results here should
be interpreted only as an indication that the LES fluctuating
velocities are comparable to filtered DNS values when the filter size
is appropriately chosen.

\begin{figure}[t]
\begin{center}
 \subfloat[]{\includegraphics[width=0.37\textwidth]{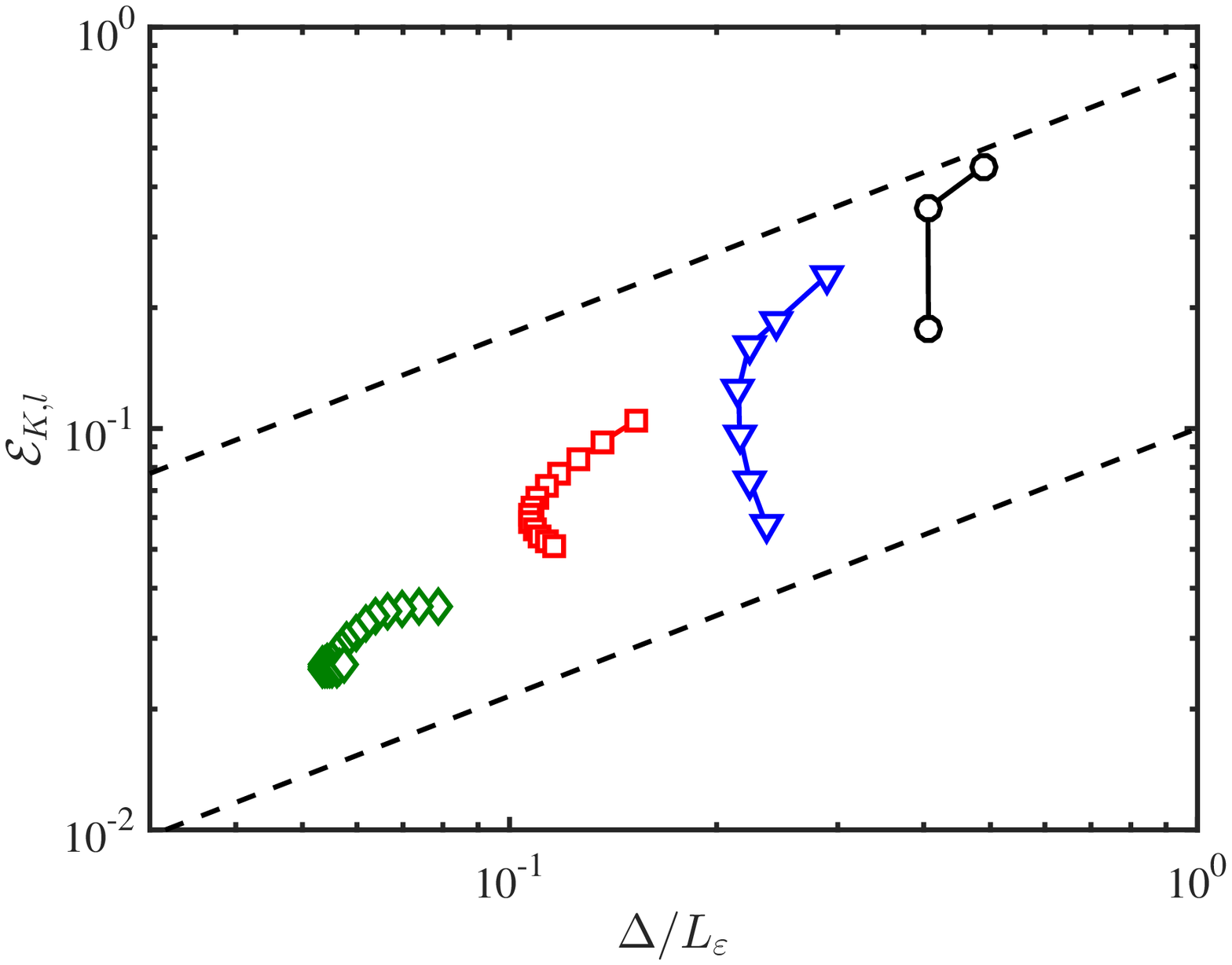}}
 \hspace{0.5cm}
 \subfloat[]{\includegraphics[width=0.37\textwidth]{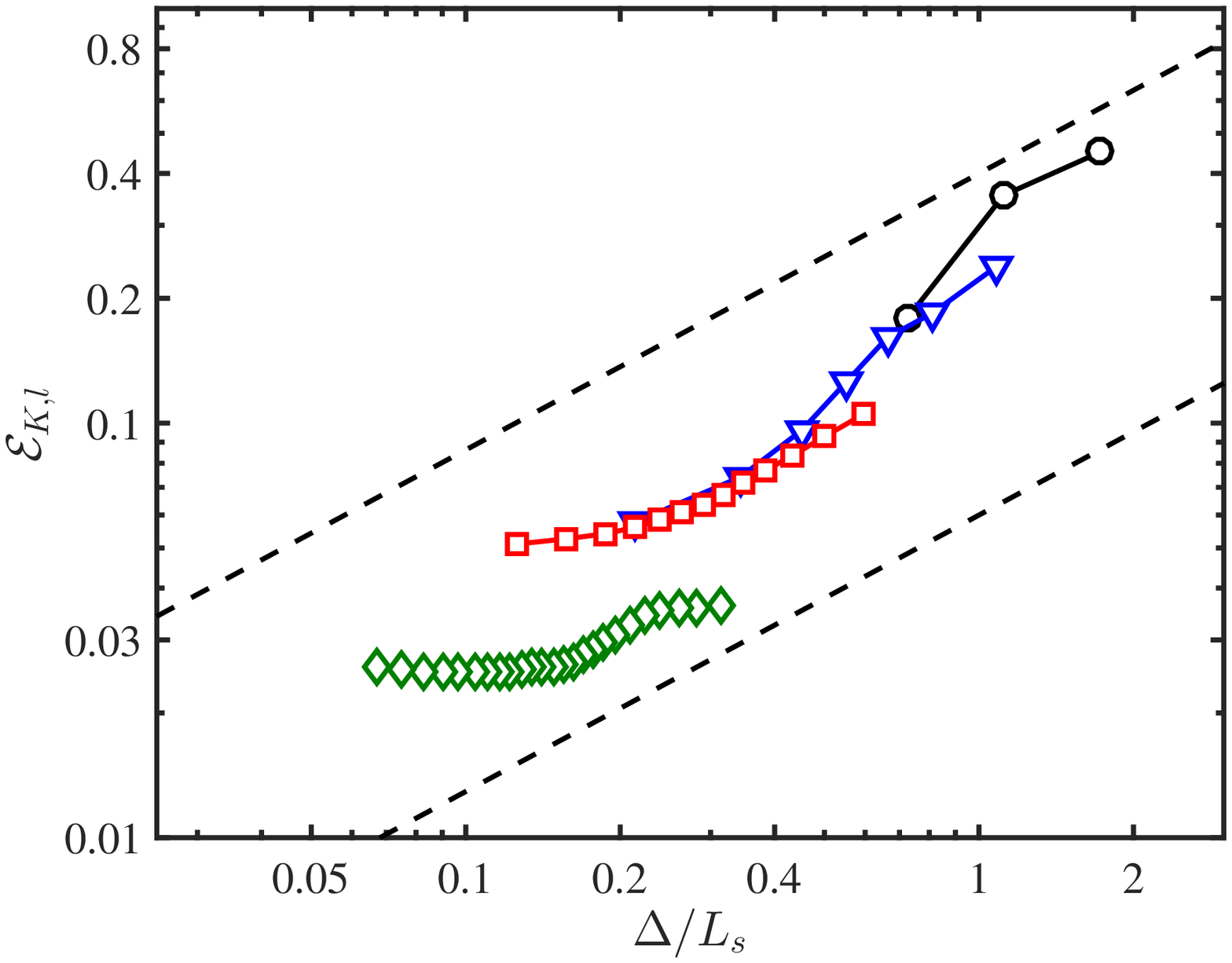}} 
\end{center}
\caption{ Local error in the turbulence kinetic
    energy $\mathcal{E}_{K,l}(x_2)$ as a function of $\Delta$
    normalized by (a) $L_\varepsilon$ and (b) $L_s(x_2)$. Colors
    represent different grid resolutions from Table
    \ref{table:resolutions}: i1 (black, DSM2000-EWS-i1), i2
    (\textcolor{blue}{blue}, DSM2000-EWS-i2), i3
    (\textcolor{red}{red}, DSM2000-EWS-i3), and i4
    (\textcolor{OliveGreen}{green}, DSM2000-EWS-i4). Dashed lines are
    (a) $\mathcal{E}_{K,l} = 0.2(\Delta/L_\varepsilon)^{2/3}$ and
    $\mathcal{E}_{K,l} = 1.2(\Delta/L_\varepsilon)^{2/3}$, and (b)
    $\mathcal{E}_{K,l} = 0.4(\Delta/L_s)^{2/3}$.
\label{fig:fluctuations_error_local} }
\end{figure}

Similarly to the mean velocity profile, the error from
Eq. (\ref{eq:error_ms_def}) can be re-evaluated locally along
different wall-normal bands to explore the relevant physical
scale-length to refer $\Delta$.  We define the local error for the
turbulent kinetic energy $K=(u'^2_1 + u'^2_2 + u'^2_3)/2$ (analogously
for LES) as
\begin{equation}\label{eq:error_tke}
\mathcal{E}_{K,l}(x_2) = 
\left[ 
\frac{ 
\frac{1}{2d}\int_{x_2 - d}^{x_2 + d} \left(\langle \tilde K \rangle - \langle K \rangle\right)^2 \mathrm{d}x_2} 
{\frac{1}{0.8\delta} \int_{0.2\delta}^{\delta} \langle K \rangle^2 \mathrm{d}x_2 }
\right]^{1/2},
\end{equation}
which is numerically computed as Eq. (\ref{eq:error_local}). Results
from Fig. \ref{fig:fluctuations_error_local} show that
$\mathcal{E}_{K,l} \sim (\Delta/L_s)^{2/3}$, and the shear
length-scale $L_s$ stands again as a sensible measure of the size of
the energy-containing eddies relevant for quantifying LES errors.  The
collapse obtained by scaling the grid resolution by $L_\varepsilon$,
$\eta$ and $L_t$ is less satisfactory, and the last two are not shown.
%

For completeness, we also consider the interpretation of $\langle u'_i
u'_j\rangle$ as a Reynolds stress tensor instead of as a velocity
variance. In the former case,
\begin{equation}\label{eq:R_DNS}
R^{DNS}_{ij} = \langle u'_i u'_j\rangle = \langle u_i u_j\rangle - \langle u_i\rangle \langle
u_j\rangle, 
\end{equation}
where the diagonal components of $R^{\mathrm{DNS}}_{ij}$ are the mean squared
DNS velocity fluctuations.  As argued in \citet{Carati2001}, assuming
$\langle\bar{\phi}\rangle \approx \langle\phi\rangle$,
\begin{equation}\label{eq:R_LES}
R^{\mathrm{DNS}}_{ij}  = \langle u_i u_j\rangle - \langle u_i\rangle \langle
u_j\rangle \approx \langle\overline{u_iu_j}\rangle -
\langle\bar{u}_i\rangle \langle\bar{u}_j\rangle \approx \langle\tilde{u}_i\tilde{u}_j\rangle +
\langle\tau_{ij}^{\mathrm{SGS}}\rangle -
\langle\tilde{u}_i\rangle\langle\tilde{u}_j\rangle = R^{\mathrm{LES}}_{ij}.
\end{equation}
Thus, the main difference between considering $\langle u'_i u'_j
\rangle$ as a stress rather than a velocity variance lies on the
contribution of the SGS tensor. An advantage of Eq. (\ref{eq:R_LES}) is
that $R_{ij}^{\mathrm{DNS}}$ and $R_{ij}^{\mathrm{LES}}$ are directly
comparable without prescribing a particular filtering
operation. However, the approach is also accompanied by a limitation
for the incompressible Navier--Stokes equations, where only the
traceless part of $\tau_{ij}^{\mathrm{SGS}}$ is modeled. Hence, in
order to allow for straight comparisons only the deviatoric
contributions of $R_{ij}^{\mathrm{DNS}}$ and $R_{ij}^{\mathrm{LES}}$
must be taken into consideration \citep{Winckelmans2002}. An error
analogous to Eq. (\ref{eq:error_ms_def}) can be defined using the
traceless counterparts of $R_{ij}^{\mathrm{DNS}}$ and
$R_{ij}^{\mathrm{LES}}$. The results, omitted for brevity, show that
the errors have a weak dependence on the grid resolution and follow
$\sim (\Delta/\delta)^{\alpha_f}$, with $\alpha_f<2/3$.

\section{Error scaling of the velocity spectra} 
\label{sec:spectra}

We consider the two-dimensional kinetic energy spectrum for the
unfiltered velocity field at a given wall-normal distance $x_2$,
$E_K(k_1,k_3,x_2)= \langle \hat u_i \hat u_i^\star \rangle_t/2$, where
$\widehat{(\cdot)}$ is the Fourier transform in the homogeneous
directions, $(\cdot)^\star$ denotes complex conjugate, and $ \langle
\cdot\rangle_t$ is average in time. Similarly, the kinetic energy
spectrum for the LES velocity is $\tilde E_K(k_1,k_3,x_2)=\langle
\widehat{\tilde u}_i \widehat{\tilde u}_i^\star\rangle_t/2$. The
magnitude of $E_K$ is given by
\begin{equation}
\langle u'^2_1 + u'^2_2 + u'^2_3\rangle = 
2\int_{0}^{\infty}\int_{0}^{\infty} E_K(k_1,k_3,x_2)\mathrm{d}k_1\mathrm{d}k_3,
\end{equation}
(analogously for $\tilde E_K$) and it was investigated in the previous
section. We are now interested in accuracy of LES to predict the
distribution of energy in the homogeneous scale-space at a given
wall-normal distance.  The error in the energy spectrum is defined as
\begin{equation}
\mathcal{E}_s(x_2) = 
\left[
\left\langle \left(\tilde E_K - E_K\right)^2 \right\rangle_{k_1,k_3}
\right]^{1/2},
\end{equation}
where $\langle \cdot \rangle_{k_1,k_3}$ denotes average over the
wall-parallel wavenumbers. Again, we are concerned with the error of
LES compared to unfiltered DNS.

\subsection{Theoretical estimations} 
\label{subsec:spectra:theoretical}

The effect of $\tau_{ij}$ on the distribution of energy can be
analyzed by considering the spectral kinetic energy equation for $\bar
E_K$ at a given wall-normal distance,
\begin{equation}\label{eq:budget_Ek}
\frac{\partial \bar E_K}{\partial t} = 
\hat P + \hat T + \hat \Pi + \hat D + \hat \varepsilon + \hat D_{\tau} + \hat \varepsilon_{\tau},
\end{equation}
where the first five terms on the right-hand are the production rate
of the turbulent kinetic energy ($\hat P$), turbulent transport $(\hat
T)$, pressure diffusion ($\hat \Pi$), viscous diffusion ($\hat D$),
and the molecular dissipation rate ($\hat \varepsilon$),
respectively. The explicit form of these terms can be found in
\citet{Mizuno2016}.  We focus on the contributions from
$\tau_{ij}$,
\begin{equation}
\hat \varepsilon_{\tau} = \mathbb{R} 
\left[  -\sqrt{-1} k_1 \langle \hat{\bar{u}}^{\star}_i \hat \tau_{i1} \rangle_t
        -\sqrt{-1} k_3 \langle \hat{\bar{u}}^{\star}_i \hat \tau_{i3} \rangle_t +  
        \left\langle \frac{\partial \hat{\bar{u}}^{\star}_i}{\partial x_2} \hat \tau_{i2} \right\rangle_t  \right], 
\quad \mathrm{and} \quad
\hat D_{\tau} = \mathbb{R} \left[ -\frac{ \partial \langle  \hat \tau_{i2} \hat{\bar{u}}^{\star}_i \rangle_t}{\partial x_2} \right],
\end{equation}
where $\mathbb{R}$ denotes real part. The term $\hat
\varepsilon_{\tau}$ is the dissipation rate of the spectral kinetic
energy by $\tau_{ij}$, while $\hat D_{\tau}$ is the wall-normal
turbulent transport by $\tau_{ij}$. A detailed equation for the
spectral error can be derived from Eq. (\ref{eq:budget_Ek}) although the
result is quite cumbersome. Instead, we assume by dimensional
arguments that the functional dependence of $\mathcal{E}_s$ on
$\Delta$ is proportional to the temporal integration of $(\hat
\varepsilon_{\tau}+\hat D_{\tau})$,
\begin{equation}
\mathcal{E}_s \sim \int_{0}^{t_c}  (\hat \varepsilon_{\tau} + \hat D_{\tau}) \mathrm{d}t \sim \Delta^2 G,
\end{equation}
where $t_c\sim G^{-1}$ is the characteristic
  time-scale for the evolution of the eddies of size $\Delta$, and $G$ is
  the characteristic velocity gradient. The estimated errors scale as
\begin{equation}
\mathcal{E}^s_s \sim \Delta, \quad \mathcal{E}^i_s \sim \Delta^{4/3}
\end{equation}
for grid resolutions comparable to the eddies in the shear-dominated
($\mathcal{E}^s_s$) or inertial ($\mathcal{E}^i_s$) range,
respectively.

\subsection{Energy-resolving grid resolution estimations} 
\label{subsec:spectra:grid}
%
\begin{figure}[t]
\begin{center}
 \subfloat[]{\includegraphics[width=0.38\textwidth]{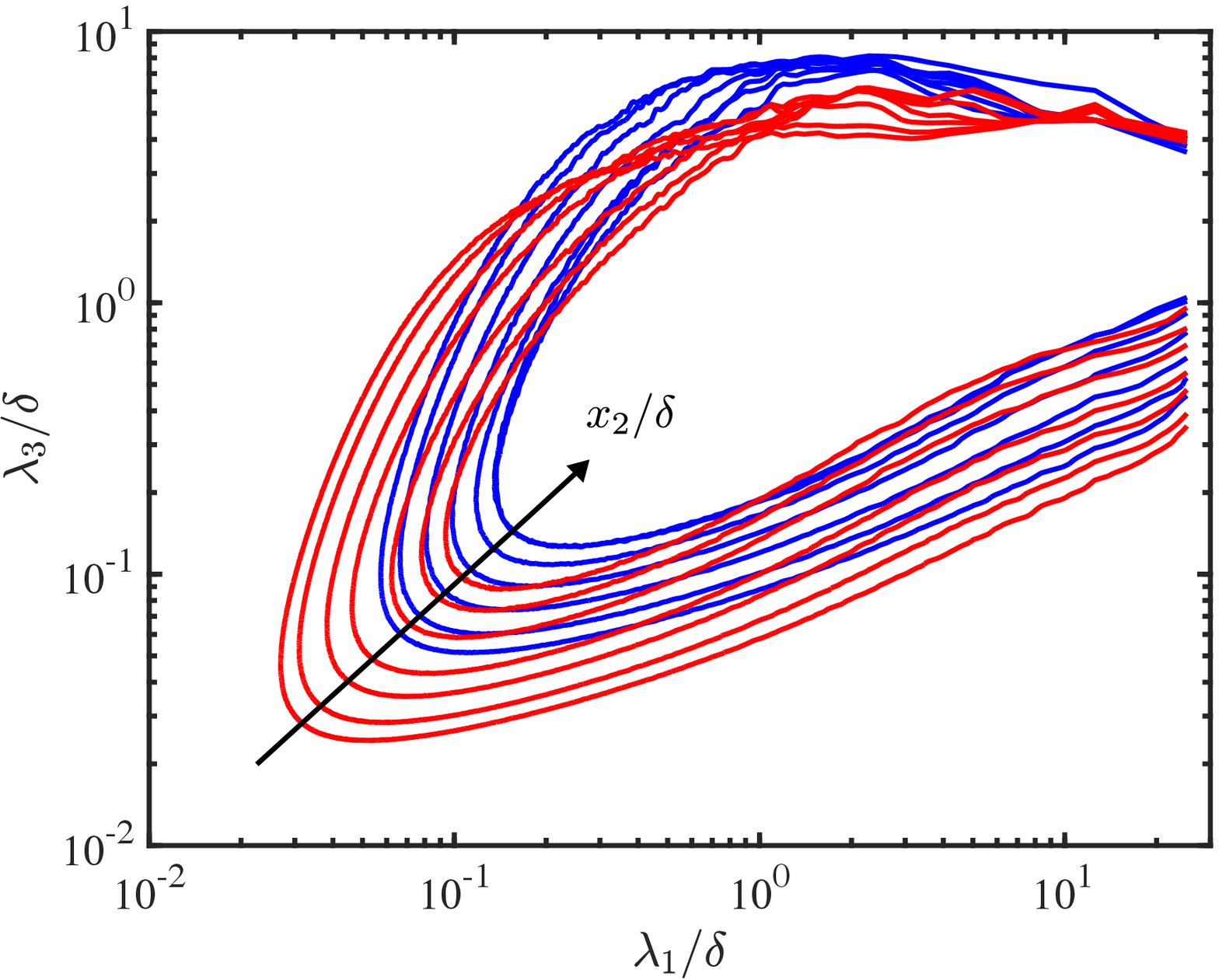}}
 \hspace{0.2cm}
 \subfloat[]{\includegraphics[width=0.35\textwidth]{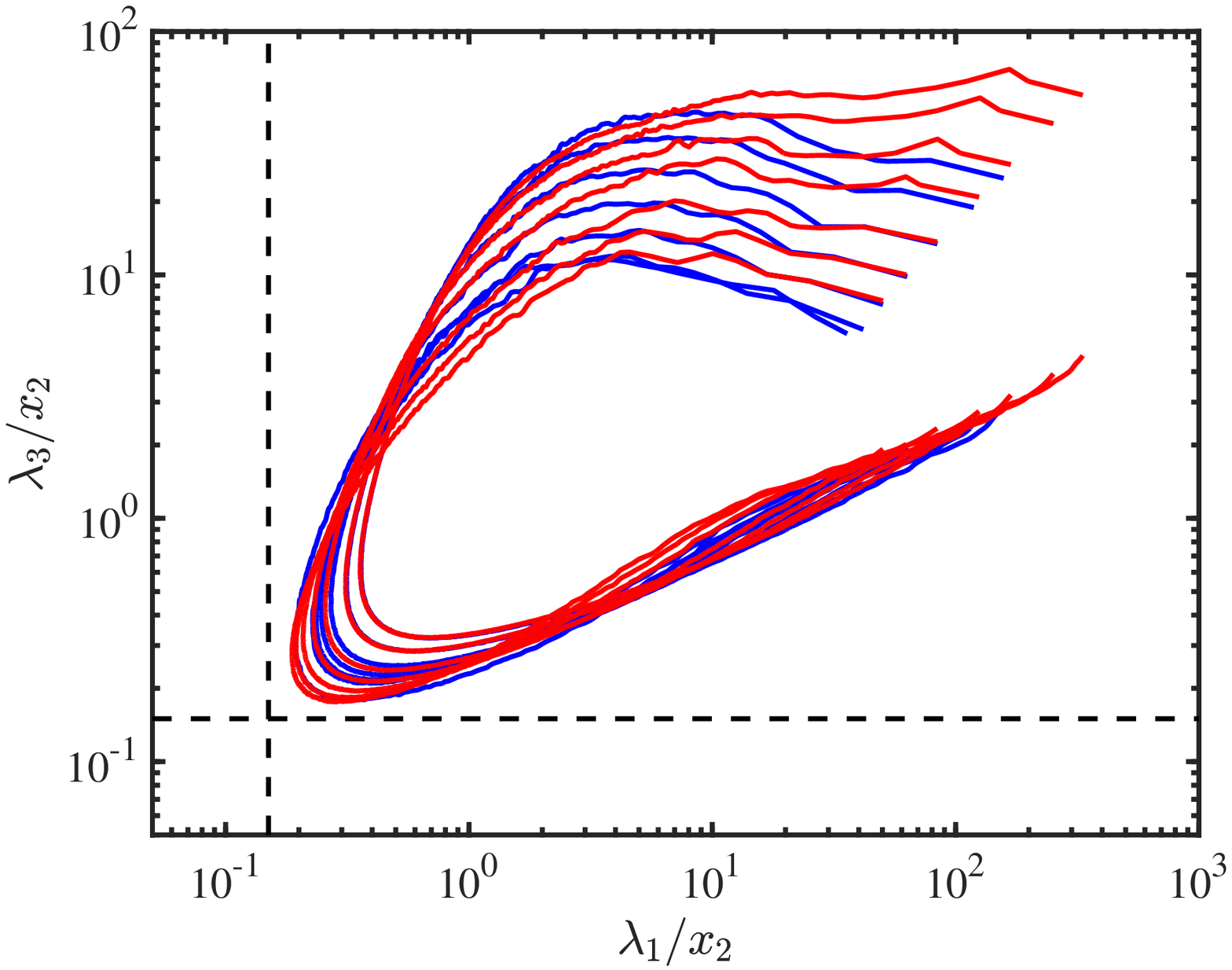}}
\end{center}
\caption{ Premultiplied two-dimensional kinetic energy spectra for DNS
  data as a function of the streamwise ($\lambda_1$) and spanwise
  ($\lambda_3$) wavelengths normalized by (a) $\delta$, and (b)
  wall-normal distance $x_2$. Colors are \textcolor{blue}{blue} for
  $Re_\tau\approx 950$ and \textcolor{red}{red} for $Re_\tau\approx
  2000$.  Different contours denote different wall-normal heights
  $x_2/\delta=0.16, 0.21, 0.30, 0.40, 0.50$ and $0.60$ for
  $Re_\tau\approx 950$, and $x_2/\delta=0.08, 0.10, 0.15, 0.20, 0.30,
  0.40$ and $0.50$ for $Re_\tau\approx 2000$. Contours contain 90\% of
  the turbulent kinetic energy. The straight dashed
    lines in (b) are $\lambda_1/x_2 = 0.15$ and $\lambda_3/x_2 =
    0.15$.\label{fig:spectra_scaling} }
\end{figure}

Prior to the numerical assessment of the error scaling, we estimate
the required LES grid resolution to resolve 90\% of the turbulent
kinetic energy at a given wall-normal distance. To that end, we use
the two-dimensional spectral energy density
$E_K(\lambda_1,\lambda_3,x_2)$ as a function of the streamwise and
spanwise wavelengths, namely $\lambda_1=2\pi/k_1$ and
$\lambda_3=2\pi/k_3$, respectively. Simple models describing the
two-dimensional energy spectral at moderate and high Reynolds numbers
have been proposed by \citet{DelAlamo2004} and \citet{Chandran2017},
respectively. However, both works focus on the energy bounds for the
large scales, whereas we are interested in the limiting length-scales
for the smaller energy-containing eddies; that is, we are seeking for
the minimum streamwise and spanwise grid spacing,
$\Delta_1^{\mathrm{min}}$ and $\Delta_3^{\mathrm{min}}$ such that
$E_K(\lambda_1>
2\Delta_1^{\mathrm{min}},\lambda_3>2\Delta_3^{\mathrm{min}},x_2)$
contains 90\% of the total turbulent kinetic energy.  Fig.
\ref{fig:spectra_scaling}(a) shows iso-contours of $E_K$ enclosing
90\% of the energy at difference wall-normal distances. As expected,
the size of the energy-containing eddies decreases as they get closer
to the wall.  As postulated by the attached-eddy hypothesis
\cite{Townsend1976} \citep[see also][for a review]{Marusic2019}, the
only relevant length-scales for the energy-containing motions spanning
along the log layer is $x_2$, which allows to write the energy spectra
as
\begin{equation}\label{eq:E_K}
 E_K=E_K(\lambda_1/x_2,\lambda_3/x_2). 
\end{equation}
 The proportionality of the sizes of eddies with the
   wall-normal distance was originally hypothesized as an asymptotic
   limit at very high Reynolds numbers and used in the classical
   derivation of the logarithmic velocity profile
   \citep{Millikan1938}, but it has been observed experimentally and
   numerically in spectra and correlations at relatively modest
   Reynolds numbers in pipes \citep{Morrison1969, Perry1975,
     Perry1977, Bullock1978, Kim1999, Guala2006, Bailey2008} and in
   turbulent channels and flat-plate boundary layers
   \citep{Tomkins2003, DelAlamo2004, Hoyas2006, Monty2007,
     Hoyas2008,Mizuno2013,Lozano2018b}. The performance of the scaling
   from Eq. (\ref{eq:E_K}) for DNS channel flows is shown in
   Fig. \ref{fig:spectra_scaling}(b) for various heights and Reynolds
   numbers. The results demonstrate the improved collapse of the
   kinetic energy spectra, and enables the estimation of energy bounds
   that are approximately valid at all the wall-normal distances
   within the outer layer. Taking $(\lambda_1)_{\mathrm{min}} = 2
 \Delta_1^{\mathrm{min}}$ and $(\lambda_3)_{\mathrm{min}} = 2
 \Delta_3^{\mathrm{min}}$, the \emph{a priori} minimum wall-parallel
 grid resolutions to resolve 90\% of the turbulent kinetic energy at
 $x_2$ are roughly given by
\begin{equation} \label{eq:spectra_res}
\left( \frac{\lambda_1}{x_2} \right)_{\mathrm{min}} = \frac{2\Delta^{\mathrm{min}}_1}{x_2} \approx 0.15, \quad 
\left( \frac{\lambda_3}{x_2} \right)_{\mathrm{min}} = \frac{2\Delta^{\mathrm{min}}_3}{x_2} \approx 0.15.
\end{equation}
The limit 0.15 was estimated from the dashed straight lines in
Fig. \ref{fig:spectra_scaling}(b), which bound the contours containing
90\% of the turbulent kinetic energy.  For example, to resolve 90\% of
the turbulent kinetic energy at $x_2 \approx 0.5\delta$, we require
$\Delta_1 = \Delta_3 \approx 0.04\delta$. These estimates were used in
Section \ref{subsec:mean:numerical} to explain the observations in
Fig. \ref{fig:snapshots_panel}. The grid resolution guidelines in
Eq. (\ref{eq:spectra_res}) imply that $\Delta_1 \approx \Delta_3$, in
contrast with the common choice of $\Delta_1 > \Delta_3$ among LES
practitioners, and usually argued in terms of the elongated streamwise
velocity streaks typical of wall-bounded flows. However, it is clear
from Fig. \ref{fig:spectra_scaling}(b) that the `nose' of the energy
spectra is located at $\lambda_1 \approx \lambda_3$, which justifies
the choice of $\Delta_1 \approx \Delta_3$. For coarser grid
resolutions aiming to resolve a lower fraction of the turbulent
kinetic energy, it is then reasonable to choose $\Delta_1 > \Delta_3$.

\subsection{Numerical assessment} 
\label{subsec:spectra:numeric}

\begin{figure}[t]
\begin{center}
\includegraphics[width=0.85\textwidth]{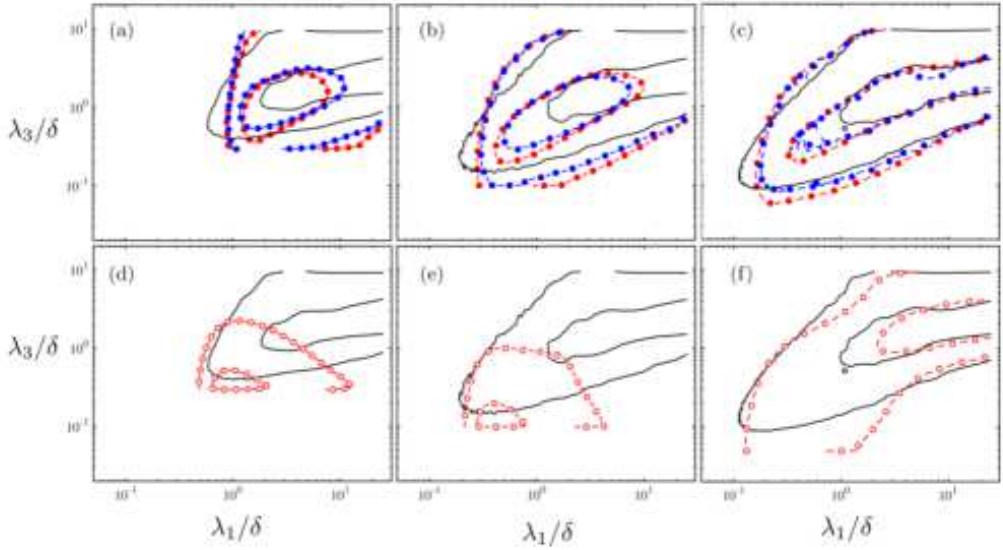}
\end{center}
\caption{ Premultiplied two-dimensional streamwise velocity spectra as
  a function of the streamwise ($\lambda_1$) and spanwise
  ($\lambda_3$) wavelengths at $x_2=0.75\delta$ for different grid
  resolutions from Table \ref{table:resolutions}. (a,d) i1 ($\Delta =
  0.2\delta$), (b,e) i3 ($\Delta = 0.05\delta$), (c,f) i4 ($\Delta =
  0.025\delta$). Colors and symbols are DSM
  (\textcolor{red}{\myclosedcircle}), AMD
  (\textcolor{blue}{\myclosedcircle}), and no explicit SGS model
  (\textcolor{red}{\mycircle}). Solid lines represent box filtered DNS
  data. Contours are 0.1 and 0.6 of the maximum. \label{fig:spectra}}
\end{figure}

Fig. \ref{fig:spectra} displays the premultiplied two-dimensional
spectra of the streamwise velocity for fDNS and LES (with DSM, AMD,
and no explicit SGS model). The filtered spectra was calculated from
box-filtered DNS data with a filter size
$\Delta_1\times\Delta_2\times\Delta_3$. The results show that both DSM
and AMD perform similarly, and that the LES spectra is representative
of the expected energy distribution for the filtered velocities
(Figs. \ref{fig:spectra}a-c), although the LES prediction tends to be
biased towards smaller scales for all grid resolutions.
%

For cases without explicit SGS model, the spectra is seriously
misrepresented for $\Delta>0.05\delta$ (Figs. \ref{fig:spectra}d-e),
with most of the energy piled up close to the smallest scales
supported by the grid. The physical interpretation of this effect was
provided in Section \ref{subsec:fluctuations:mech} in terms of the
necessary velocity gradients to comply with the conservation of
energy. Figs. \ref{fig:spectra}(d)-(e) are just the spectral depiction
of the same effect, i.e., the energy cascades towards the smallest
available scales until the resulting gradients can balance the input
power driving the channel. The distribution of energy changes
drastically for $\Delta<0.05\delta$, where large-scale streaks are now
a clear constituent feature of the flow (Fig.
\ref{fig:spectra}f). The result is consistent with the visualizations
in Fig. \ref{fig:snapshots_panel}(d), which shows a clear streaky
pattern in the streamwise velocity for $\Delta = 0.025\delta$, but a
notably different non-streaky structure for $\Delta>0.05\delta$.  The
existence of this critical grid resolution may be connected to the
grid requirements estimated in Section \ref{subsec:spectra:grid},
where it was concluded that $\Delta \approx 0.04 \delta$ in order to
capture at least 90\% of the turbulent kinetic energy at $x_2 \approx
0.5\delta$. This seems to be a necessary requirement to support the
development of streaks in the absence of SGS model, at least for the
particular numerical discretization adopted in this study.

\begin{figure}[t]
\begin{center}
 \subfloat[]{\includegraphics[width=0.39\textwidth]{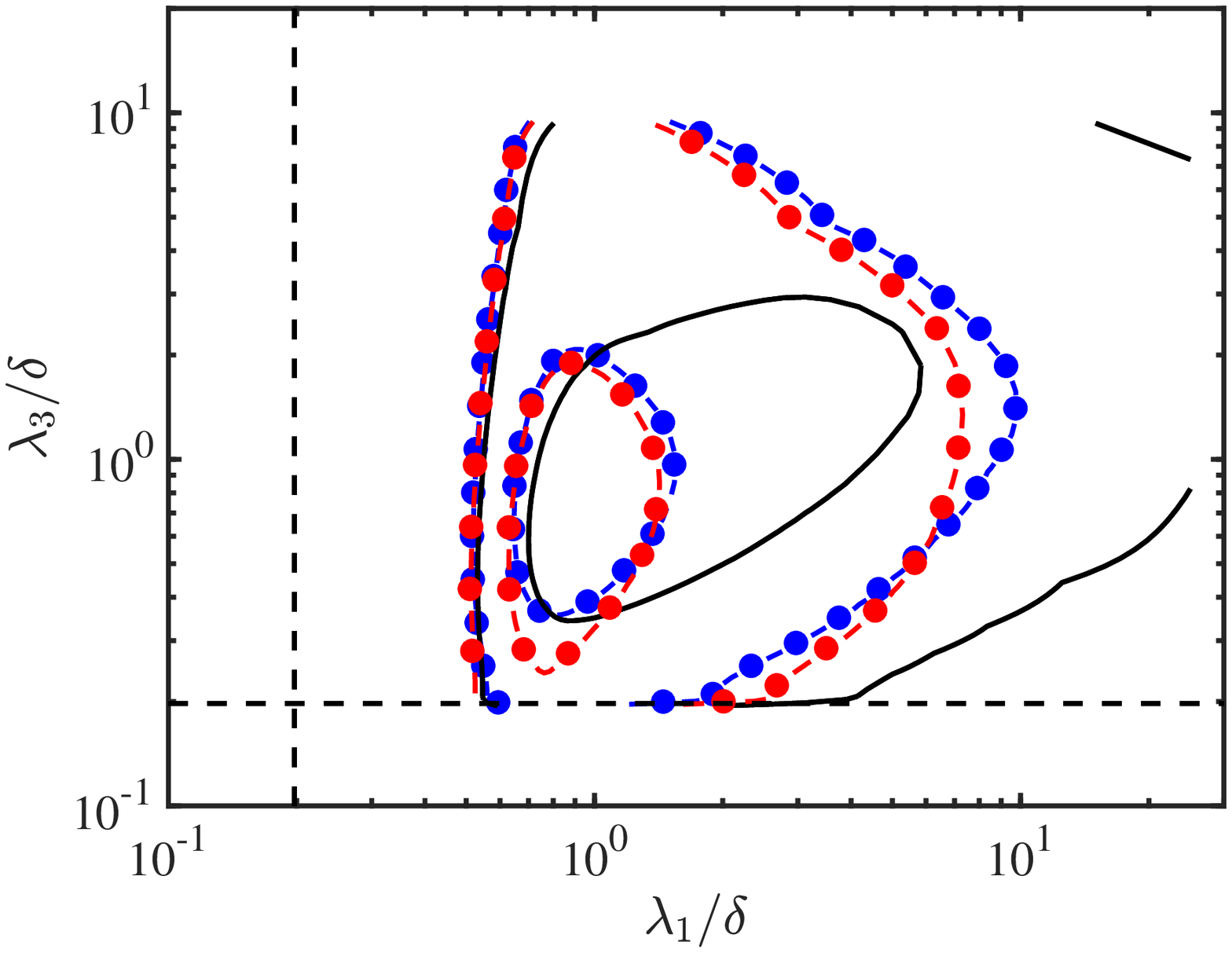}}
 \hspace{0.1cm}
 \subfloat[]{\includegraphics[width=0.39\textwidth]{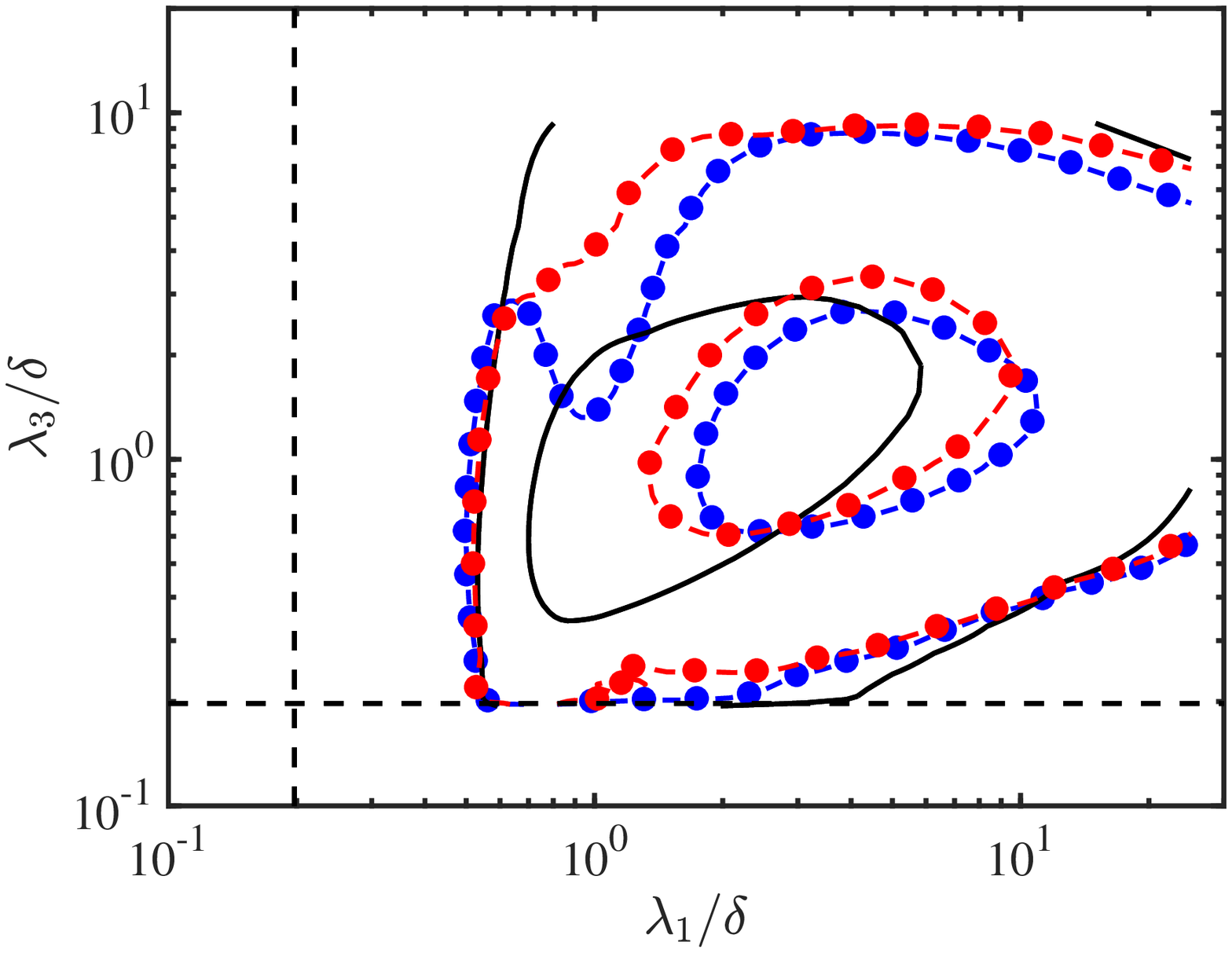}}
\end{center}
\caption{ Premultiplied two-dimensional spectra of the turbulent
  kinetic energy $\tilde E_K$ (\solid) compared with (a) the SGS
  dissipation rate of spectral kinetic energy $\hat
  \varepsilon_{\mathrm{SGS}}$ (closed circles), and (b) the SGS
  turbulent transport $\hat D_{\mathrm{SGS}}$ (closed circles) as
  function of the streamwise ($\lambda_1$) and spanwise ($\lambda_3$)
  wavelengths at $x_2=0.75\delta$. Symbols are
  (\textcolor{red}{\myclosedcircle}) for DSM2000-EWS-i2, and
  (\textcolor{blue}{\myclosedcircle}) for AMD2000-EWS-i2.  Contours
  are 0.1 and 0.6 of the maximum of $\tilde E_K$, $\hat
  D_{\mathrm{SGS}}$, and $|\hat \varepsilon_{\mathrm{SGS}}|$ for each
  quantity, respectively. The dashed lines are
  $\lambda_1/\delta=\lambda_3/\delta=0.2$.
\label{fig:spectra_model} }
\end{figure}

Two mechanisms are potentially responsible for the improvements
reported in Fig. \ref{fig:spectra} for cases with SGS model: the
dissipation of the energy piled up at the smallest LES scales by $\hat
\varepsilon_{\mathrm{SGS}}$, and the redistribution of energy in the
wall-normal direction by $\hat D_{\mathrm{SGS}}$. These are the LES
counterparts of $\hat \varepsilon_{\tau}$ and $\hat D_{\tau}$
discussed in Section \ref{subsec:spectra:theoretical} and their
spectra are plotted in Fig. \ref{fig:spectra_model}. The computed
values reveal that the main contributor is $\hat
\varepsilon_{\mathrm{SGS}}$ whose magnitude is roughly ten times
larger than that of $\hat D_{\mathrm{SGS}}$. Hence, the improved
predictions of the velocity spectra in Fig. \ref{fig:spectra}(a) and
(b) are mostly due to the removal of the excess of energy close to the
grid cut-off.
%

\begin{figure}
\begin{center}
 \subfloat[]{\includegraphics[width=0.39\textwidth]{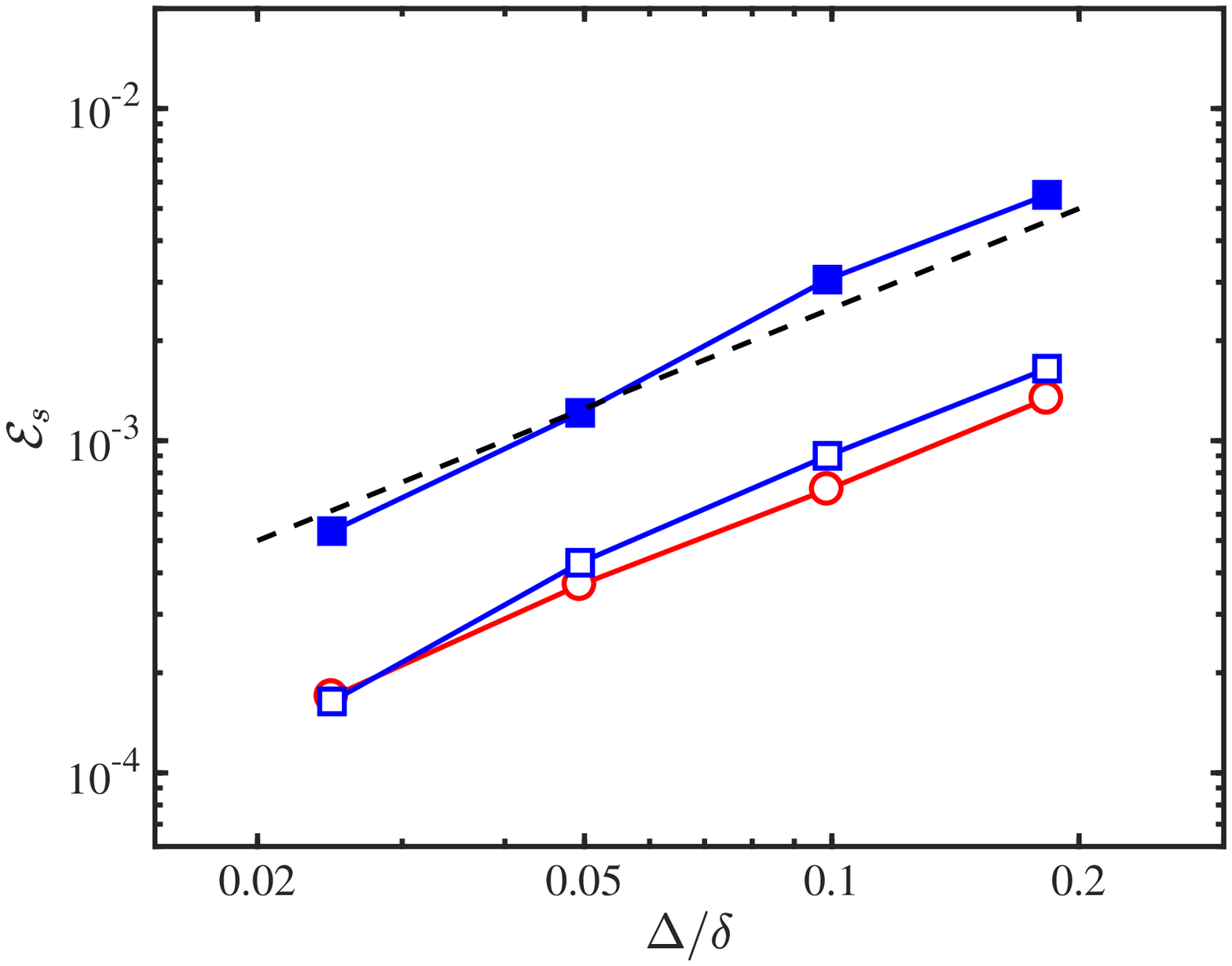}}
 \hspace{0.1cm}
 \subfloat[]{\includegraphics[width=0.39\textwidth]{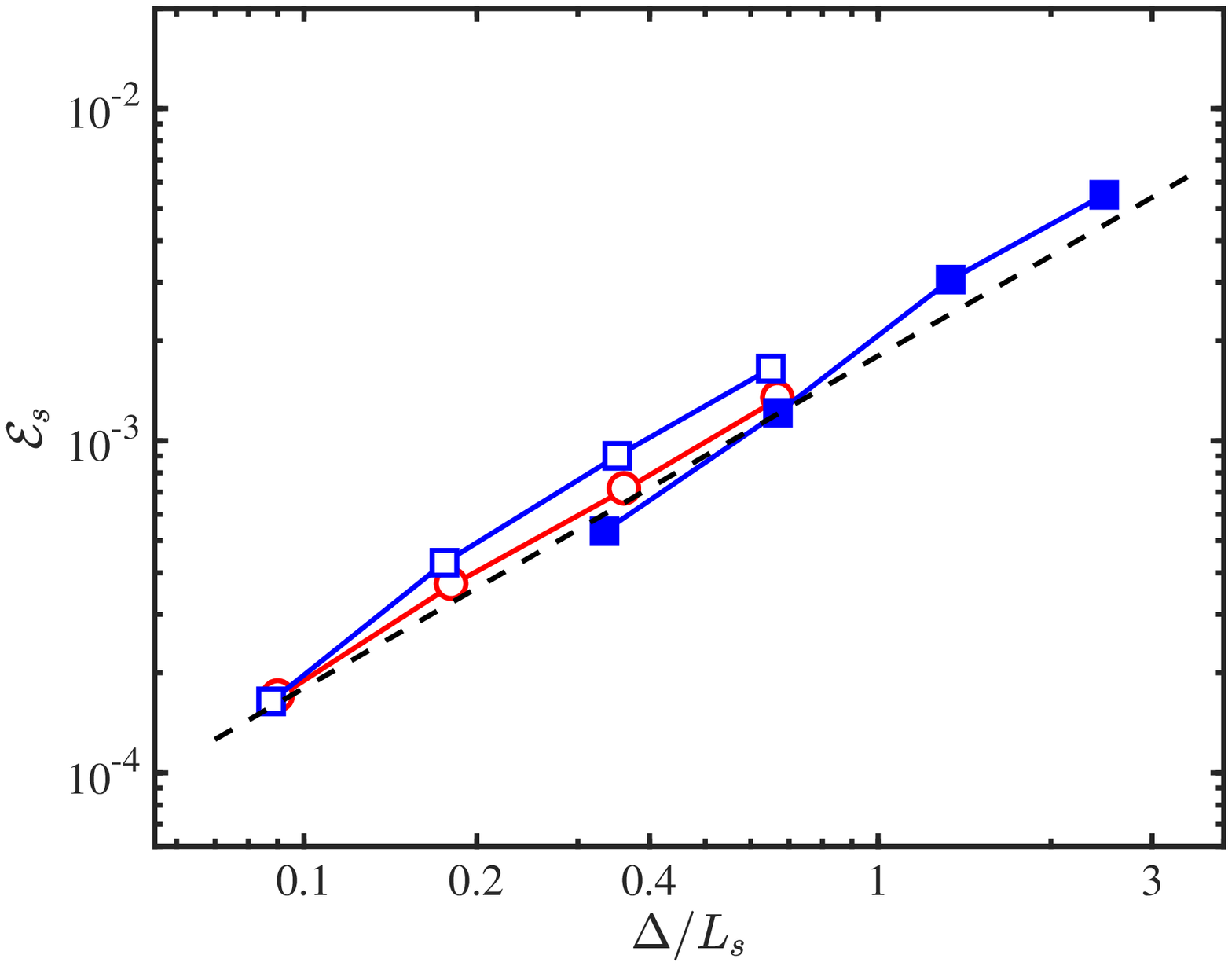}}
\end{center}
\caption{ Error in the kinetic energy spectra $\mathcal{E}_s$ as a
  function of the characteristic grid size $\Delta$ scaled by (a)
  $\delta$, and (b) $L_s(x_2)$. Open symbols are for $x_2=0.75\delta$,
  and closed symbols for $x_2=0.2\delta$.  Colors are
  \textcolor{red}{red} for $Re_\tau\approx 950$, and
  \textcolor{blue}{blue} for $Re_\tau\approx 2000$. Results are for
  LES with DSM. The dashed line is (a) $\mathcal{E}_s \sim
  (\Delta/\delta)^{4/3}$ and (b) $\mathcal{E}_s \sim
  (\Delta/L_s)^{4/3}$.
\label{fig:spectra_error} }
\end{figure}
%
Finally, the scaling of $\mathcal{E}_s$ is evaluated in Fig.
\ref{fig:spectra_error} using LES data (with respect to unfiltered DNS
data). Two wall-normal distances are considered, $x_2 = 0.2\delta$ and
$x_2 = 0.75\delta$.  The error scales as $\mathcal{E}_s \sim
\Delta^{4/3}$, consistent with estimations in Section
\ref{subsec:spectra:theoretical}, and are insensitive to variations in
the Reynolds number $\mathcal{E}_s \sim Re_\tau^0$. When the error is
expressed as a function of $\Delta/\delta$, $\mathcal{E}_s$ increases
for decreasing wall-normal distances due to the smaller eddy size
relative to $\delta$. Conversely, the errors collapse at different
$x_2$ locations when $\Delta$ is normalized by $L_s$, as shown for the
mean profile and turbulence intensities in previous sections.  In
summary, we conclude that the errors in the kinetic energy spectra
follow
\begin{equation}
\mathcal{E}_s \sim \left(\frac{\Delta}{L_s}\right)^{4/3} Re_\tau^0.
\end{equation}

\section{Conclusions} 
\label{sec:conclusion}


Large-eddy simulation has emerged as a fundamental tool for both
scientific research and industrial applications.  However, the
solutions provided by implicitly-filtered LES are grid-dependent, and
multiple computations are required in order to attain meaningful
conclusions. This brings the fundamental question of what is the
expected LES error scaling as a function of Reynolds number and grid
resolution, which has been the aim of the present investigation. In
particular, we have focused on the outer layer of wall-bounded flows
at moderately high Reynolds numbers with grid resolutions comparable
to the boundary layer thickness, as it is the typical scenario in
wall-modeled LES for external aerodynamics.

We have argued that LES of wall-bounded turbulence is challenging
since the energy-containing eddies are constrained to reduce their
characteristic size in order to accommodate the presence of the wall.
Proper wall-resolved LES calculations demand nested grid refinements
to capture these eddies, with a high computational overhead.  To make
the problem tractable, previous studies have quantified SGS errors in
WRLES at relatively low Reynolds numbers and unrealistically fine
grids.  In those conditions, most of the errors reported in the
literature are probably dominated by the near wall-region, where SGS
models are known to be deficient, while the contribution of SGS models
in the outer layer is negligible due to the fine grid resolution. For
example, we have shown that at $Re_\tau \approx 1000$ and 20 points
per $\delta$, the mean velocity profile in the outer layer is well
predicted by WMLES without any explicit SGS model. Given that SGS
models are mainly responsible for the outer flow in WMLES, it is
necessary to consistently isolate the errors in the bulk flow from
those in the near-wall region. It is only in this manner that we can
faithfully evaluate the behavior of SGS models.

To assess the performance of SGS models in the outer region
independently of the effect of the wall, we have designed a numerical
experiment, referred to as exact-wall-stress channel flows, where the
integrated effect of the near-wall region on the outer flow is
bypassed by supplying the exact mean stress at the wall. This
numerical experiment retains the same physics as the traditional
channel flow far from the wall, and hence is a suitable framework to
test boundary layer flows. We have considered two SGS models, i.e.,
dynamic Smagorinsky model and minimum dissipation model, that are
representative of eddy viscosity models with and without test
filtering, respectively.

We have investigated the error scaling of the mean velocity profile,
turbulence intensities, and kinetic energy spectra, with the grid
resolution and Reynolds number. The error is of the form
\begin{equation}
\label{eq:E_final}
\mathcal{E}_q \sim \left( \frac{\Delta}{L} \right)^{\alpha_q} Re_\tau^{\gamma_q}, 
\end{equation}
where $\Delta$ is the characteristic grid size, $L$ is length scale of
the energy-containing eddies, and $q$ denotes the quantity the error
$\mathcal{E}_q$ is referred to, i.e. $q=m$ for the mean velocity
profile, $q=f$ for the turbulence intensities, and $q=s$ for the
kinetic energy spectra. Our results show that $\Delta/L$ is an
intricate function of the flow state and grid resolution, but it is
well approximated by the $L_2$-norm of $(\Delta_1,\Delta_2,\Delta_3)$
divided by $\delta$ for quantities integrated over the outer layer,
and by the shear length-scale, $L_s$, for local errors as a function
of the wall-normal direction. The observation of $L_s$ as the relevant
physical length-scale to normalize $\Delta$ is consistent with its
ability to represent the size of the energy-containing eddies as
discussed by \citet{Lozano2018b}. For $Re_\tau>1000$, the errors are
independent of the viscous effects and $\gamma_q\approx0$, as expected
for WMLES. We have derived the theoretical values of $\alpha_q$ and
compared the results with the empirical estimations obtained by
numerical simulations. To be consistent with the current available
computational resources, we have only considered grid resolutions
which are a fraction of the boundary layer thickness. In these cases,
the corresponding LES filter cut-off lies either in the
shear-dominated regime or in the inertial range, and always far from
the viscous Kolmogorov region.  We have showed that the grid
resolution needs to be sufficient to resolve at least some fraction of
the shear-dominated eddies in order to obtain $\alpha_q > 0$. Overall,
our theoretical predictions match the numerical estimations, and we
detail below the results of Eq. (\ref{eq:E_final}) for the different
flow statistics investigated.

Errors in the mean velocity profile follow $\mathcal{E}_m \sim
\epsilon \Delta/\delta$, where $\epsilon$ is a SGS-model dependent
constant.  The local errors increase with the proximity of the wall,
and we have shown that the prediction at the $n$-th off-wall grid
point does not improve with grid refinement until the grid resolution
approaches the WRLES regime.

We have reasoned that the turbulence intensities in
implicitly-filtered LES are akin to those from filtered
Navier--Stokes, but the former are controlled by the necessity of
dissipating the energy input at the rate consistent with the
statistically steady state, while the latter are directly linked to
the filtering operation. In terms of convergence, the turbulence
intensities are more demanding than the mean velocity profile and
their error scales as $\mathcal{E}_f \sim (\Delta/\delta)^{\alpha_f}$
with $\alpha_f\approx 0.4 - 0.8$. Furthermore, in order to correctly
capture the classic wall-normal logarithmic dependence of the
streamwise and spanwise turbulence intensities, the grid resolution
must be comparable to the sizes of the eddies in the inertial range.

Errors in the wall-parallel kinetic energy spectra follow
$\mathcal{E}_s \sim (\Delta/L_s)^{4/3}$. We have pointed out that SGS
models affect the distribution of energy via two mechanisms, namely,
eddy-viscosity dissipation and wall-normal eddy-viscosity transport,
but the former is ten times larger in magnitude than the latter.  The
energy spectra from DNS was also utilized to estimate the LES grid
requirements to resolve 90\% of the turbulent kinetic energy as a
function of $x_2$, resulting in $\Delta_1 \approx \Delta_3 \approx
0.075 x_2$. For example, if we wish to accurately resolved 90\% of the
turbulent kinetic energy at $x_2 \approx 0.5\delta$, then
$\Delta_1=\Delta_3 \approx 0.04\delta$. If we further assume an
isotropic grid, the count yields $\sim$25 points per boundary layer
thickness.

In light of the present results, future efforts should be devoted to
enhancing the convergence rate of SGS models. This may be desired to
accelerate the convergence of the turbulence intensities in those
cases where their accurate prediction is of significant
importance. Examples are noise signature prediction, or particle laden
flows at certain Stokes numbers. Additionally, since our work relies
on a wall model providing the exact mean stress at the wall, we have
to emphasize the importance of developing and assessing the accuracy
of wall models as a pacing item to achieve practical LES.

\section*{Acknowledgments} 

This work was supported by NASA under the Transformative Aeronautics
Concepts Program, grant no. NNX15AU93A. The authors would like to
thank Prof. Parviz Moin, Dr. Sanjeeb T. Bose, Dr. Perry Johnson, and
Dr. Maxime Bassenne for their insightful comments.

\appendix
\section{Error in the mean velocity profile for Vreman model, finer grids, and WMLES} 
\label{appendixA}

\begin{figure}[t]
\begin{center}
 \subfloat[]{\includegraphics[width=0.40\textwidth]{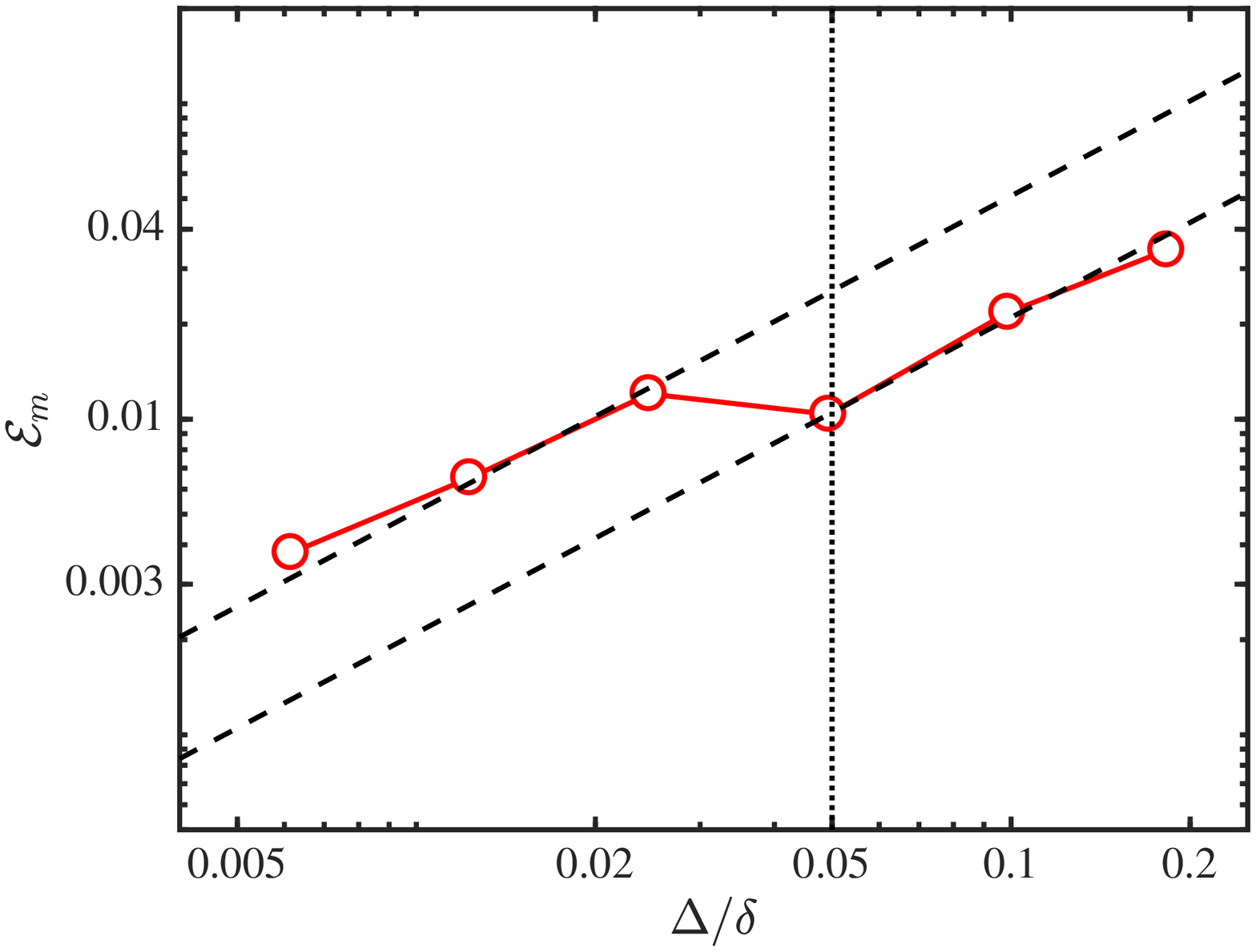}}
 \hspace{0.2cm}
 \subfloat[]{\includegraphics[width=0.40\textwidth]{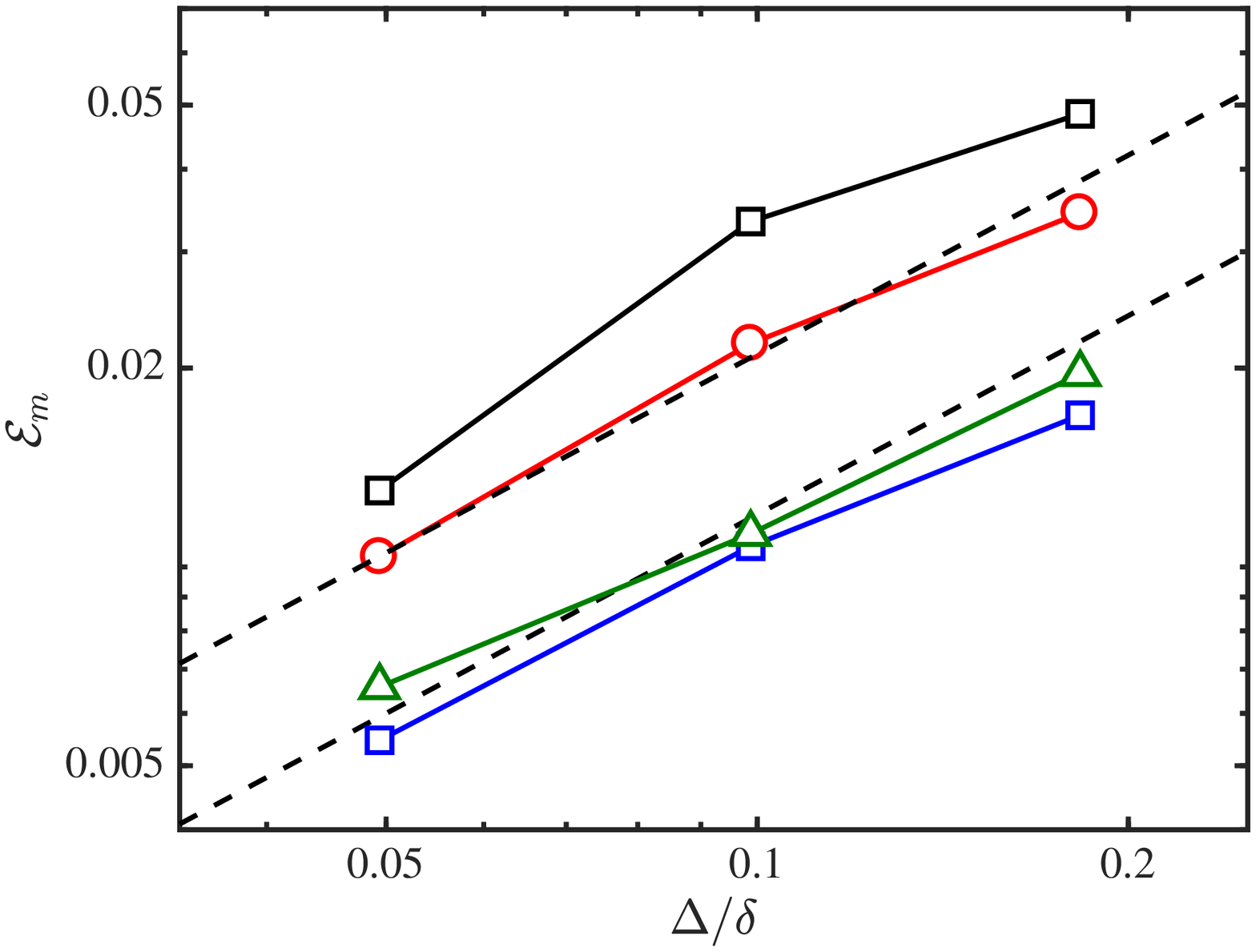}}
\end{center}
\caption{ Error in the mean velocity profile as a function of the
  characteristic grid resolution $\Delta = \sqrt{(\Delta_1^2 +
    \Delta_2^2 + \Delta_3^2)/3}$. (a) Assessment of finer grid
  resolutions for DSM4200-EWS-i1,i2,i3,i4,i5,i6. (b) Assessment of
  Vreman model and equilibrium wall model by \citet{Kawai2012}.
  Colors and symbols in (b) are (\textcolor{red}{\mycircle}) for
  DSM4200-EWS-i1,i2,i3, (\textcolor{blue}{\mysquare}) for
  AMD4200-EWS-i1,i2,i3, (\textcolor{OliveGreen}{\mytriangle}) for
  VRM4200-EWS-i1,i2,i3, and (\mysquare) for
  DSM4200-EQWM-i1,i2,i3. Dashed lines are $\mathcal{E}_m \sim
  \Delta/\delta$, and the dotted line is
  $\Delta=0.05\delta$. \label{fig:Umean_error_exactWM_appendix} }
\end{figure}

Cases with the Vreman model (VRM) are computed using the same
numerical set-up described in Section \ref{sec:numerics} with a Vreman
constant equal to $0.1$, $Re_\tau \approx 4200$, and grid resolutions
i1, i2 and i3. We use the same nomenclature as in Section
\ref{sec:numerics}. The error in the mean velocity profile is shown in
Fig. \ref{fig:Umean_error_exactWM_appendix}(a) and follows
$\mathcal{E}_m \sim \Delta/\delta$ with values comparable to those
obtained for AMD.

To test the effect of further grid refinements, two additional cases
are computed with isotropic grids equal to $\Delta = 0.0125\delta$
(denoted by i5) and $\Delta = 0.0063\delta$ (i6) for DSM at $Re_\tau
\approx 4200$. In order to alleviate the computational cost, the
streamwise and spanwise channel lengths are reduced to $2\pi\delta$
and $\pi\delta$, respectively. The error in the mean velocity profile
is plotted Fig. \ref{fig:Umean_error_exactWM_appendix}(a) which shows
that $\mathcal{E}_m \sim \Delta/\delta$ is recovered below
$\Delta\approx 0.05\delta$.

Finally, we also include results, where the EWS condition from Section
\ref{sec:exact_stress} is replaced by an actual wall model, namely,
the equilibrium wall model by \citet{Kawai2012} (EQWM). Cases are
computed at $Re_\tau \approx 4200$ for DSM and grid resolutions i1, i2
and i3. The linear scaling observed for cases with EWS deteriorates
slightly when the wall model is introduced, but remains close to
$\mathcal{E}_m \sim \Delta/\delta$ as shown in
Fig. \ref{fig:Umean_error_exactWM_appendix}(b).
%

\bibliographystyle{model1-num-names} 
\bibliography{convergence_LES}

\end{document}